\documentclass[longauth]{aa}

\usepackage{xcolor}  % acg
\usepackage{natbib}  % acg
\bibpunct{(}{)}{;}{a}{}{,} % to follow the A&A style

\usepackage[colorlinks = true,
            linkcolor = magenta,
            urlcolor  = magenta,
            citecolor = blue,
            anchorcolor = blue]{hyperref} %acg

\usepackage{graphicx}
\usepackage{txfonts}
\usepackage{chemformula} %art
\usepackage[version=4]{mhchem} %art
\let\ACMmaketitle=\maketitle
\renewcommand{\maketitle}{\begingroup\let\footnote=\thanks \ACMmaketitle\endgroup} %acg

\begin{document} 

   \title{An unusually low-density super-Earth transiting the bright early-type M-dwarf GJ 1018 (TOI-244)\thanks{Based on Guaranteed Time Observations collected at the European Southern Observatory (ESO) under ESO programs 108.2254.002, 108.2254.005, and 108.2254.006. Full Tables A.1 to A.5 are only available in electronic form at the Centre de Données astronomiques de Strasbourg (CDS).}}

   % Full Tables A.1 to A.5 are only available in electronic form at the Centre de Données astronomiques de Strasbourg (CDS).
   %\titlenote{This is a titlenote}

   %\subtitle{Subtitle Subtitle Subtitle Subtitle }

    \author{A.~Castro-Gonz\'{a}lez\inst{ \ref{CAB_villafranca} }
    \and
    O.~D.~S.~Demangeon\inst{\ref{CAUP}, \ref{dep_astro_porto}}
    \and
    J.~Lillo-Box\inst{\ref{CAB_villafranca} }
    \and
    C.~Lovis\inst{\ref{obs_geneva}}
    \and
    B.~Lavie\inst{\ref{obs_geneva}}
    \and
    V.~Adibekyan\inst{\ref{CAUP}, \ref{dep_astro_porto}}
    \and
    L.~Acuña\inst{\ref{max_planck}}
    \and
    M.~Deleuil\inst{\ref{LAM}}
    \and
    A.~Aguichine\inst{\ref{UCSC}}
    \and
    M.~R.~Zapatero~Osorio\inst{\ref{CAB_torrejon}}
    \and
    H.~M.~Tabernero\inst{\ref{CAB_torrejon}}
    \and
    J.~Davoult\inst{\ref{physikalisches_inst}}
    \and
    Y.~Alibert\inst{\ref{physikalisches_inst},\ref{space_hab_switz}}
    \and
    N.~Santos\inst{\ref{CAUP}, \ref{dep_astro_porto}}
    \and
    S.~G.~Sousa\inst{\ref{CAUP}, \ref{dep_astro_porto}}
    \and
    A.~Antoniadis-Karnavas\inst{\ref{CAUP}, \ref{dep_astro_porto}}
    \and
    F.~Borsa\inst{\ref{INAF_brera}}
    \and
    J.~N.~Winn\inst{\ref{Princeton_astro}}
    \and 
    C.~Allende Prieto\inst{\ref{IAC},\ref{uni_laguna}}
    \and
    P.~Figueira\inst{\ref{CAUP}, \ref{obs_geneva}}
    \and
    J.~M.~Jenkins\inst{\ref{NASA_ames}}
    \and
    A.~Sozzetti\inst{\ref{INAF_torin}}
    \and
    M.~Damasso\inst{\ref{INAF_torin}}
    \and
    A.~M.~Silva\inst{\ref{CAUP}, \ref{dep_astro_porto}}
    \and
    N.~Astudillo-Defru\inst{\ref{UCSC_chile}}
    \and
    S.~C.~C. Barros\inst{\ref{CAUP}, \ref{dep_astro_porto}}
    \and
    X.~Bonfils\inst{\ref{grenoble}}
    \and
    S.~Cristiani\inst{\ref{INAF_trieste}}
    \and
    P.~Di~Marcantonio\inst{\ref{INAF_trieste}}
    \and
    J.~I.~Gonz\'alez Hern\'andez\inst{\ref{IAC},\ref{uni_laguna}}
    \and
    G. Lo Curto\inst{\ref{ESO_chile}}
    \and
    C.~J.~A.~P.~Martins\inst{\ref{CAUP},\ref{centro_CAUP}}
    \and
    N.~J.~Nunes\inst{\ref{inst_lisboa}}
    \and
    E.~Palle\inst{\ref{IAC},\ref{uni_laguna}}
    \and
    F.~Pepe\inst{\ref{uni_geneva}}
    \and
    S.~Seager\inst{\ref{MIT_physics}, \ref{MIT_earth}, \ref{MIT_aeronautics}}
    \and
    A.~Su\'{a}rez~Mascare\~{n}o\inst{\ref{IAC},\ref{uni_laguna}}
    }
    
    \institute{Centro de Astrobiolog\'{i}a, CSIC-INTA, ESAC campus, 28692 Villanueva de la Ca\~{n}ada, Madrid, Spain\label{CAB_villafranca} \\\email{acastro@cab.inta-csic.es}
    \and
    Instituto de Astrof\'isica e Ci\^encias do Espa\c{c}o, Universidade do Porto, CAUP, Rua das Estrelas, 4150-762 Porto, Portugal\label{CAUP}
    \and
    Departamento de Fisica e Astronomia, Universidade do Porto, Rua do Campo Alegre, 4169-007 Porto, Portugal\label{dep_astro_porto}
    \and
    Observatoire de l'Universit\'e de Gen\`eve, 51 chemin Pegasi, 1290 Sauverny, Switzerland\label{obs_geneva}
    \and
    Max Planck Institute for Astronomy, Königstuhl 17, D-69117 Heidelberg, Germany\label{max_planck}
    \and
    Aix Marseille Univ, CNRS, CNES, LAM, 38 rue Frédéric Joliot-Curie, 13388 Marseille, France; Division Technique INSU, CS20330, 83507 La Seyne sur Mer cedex, France\label{LAM}
    \and
    Department of Astronomy and Astrophysics, University of California, Santa Cruz, CA, USA\label{UCSC}
    \and
    Centro de Astrobiolog\'ia, CSIC-INTA, Carretera de Ajalvir km 4, 28850 Torrej\'on de Ardoz, Madrid, Spain\label{CAB_torrejon}
    \and
    Physikalisches Institut, University of Bern, Gesellsschaftstrasse 6, 3012 Bern, Switzerland\label{physikalisches_inst}
    \and
    Center for Space and Habitability, University of Bern, Gesellsschaftstrasse 6, 3012 Bern, Switzerland\label{space_hab_switz}
    \and
    INAF – Osservatorio Astronomico di Brera, Via E. Bianchi 46, 23807 Merate (LC), Italy\label{INAF_brera}    
    \and
    Department of Astrophysical Sciences, Princeton University, Princeton, NJ 08544, USA\label{Princeton_astro}
    \and
    Instituto de Astrof{\'\i}sica de Canarias, E-38205 La Laguna, Tenerife, Spain\label{IAC}
    \and
    Universidad de La Laguna, Dept. Astrof{\'\i}sica, E-38206 La Laguna, Tenerife, Spain\label{uni_laguna}
    \and
    NASA Ames Research Center, Moffett Field, CA 94035, USA\label{NASA_ames}
    \and
    INAF - Osservatorio Astrofisico di Torin, Via Osservatorio 20, I-10025 Pino Torinese, Italy\label{INAF_torin}
    \and
    Departamento de Matemática y Física Aplicadas, Universidad Católica de la Santísima Concepción, Alonso de Rivera 2850, Concepción, Chile\label{UCSC_chile}
    \and
    Univ. Grenoble Alpes, CNRS, IPAG, F-38000 Grenoble, France\label{grenoble}
    \and
    INAF – Osservatorio Astronomico di Trieste, via G. B. Tiepolo 11, I-34143, Trieste, Italy\label{INAF_trieste}
    \and
    European Southern Observatory, Av. Alonso de Cordova 3107, Casilla 19001, Santiago de Chile, Chile\label{ESO_chile}
    \and
    Centro de Astrof\'{\i}sica da Universidade do Porto, Rua das Estrelas, 4150-762 Porto, Portugal\label{centro_CAUP}
    \and
    Instituto de Astrof\'isica e Ciências do Espa\c{c}o, Faculdade de Ci\^encias da Universidade de Lisboa,
    Campo Grande, PT1749-016 Lisboa, Portugal\label{inst_lisboa}
    \and
    Département d’Astronomie, Université de Genève, Ch. des Maillettes 51, 1290 Versoix, Switzerland\label{uni_geneva}
    \and
    Department of Physics and Kavli Institute for Astrophysics and Space Research, Massachusetts Institute of Technology, Cambridge, MA 02139, USA\label{MIT_physics}
    \and
    Department of Earth, Atmospheric and Planetary Sciences, Massachusetts Institute of Technology, Cambridge, MA 02139, USA\label{MIT_earth}
    \and
    Department of Aeronautics and Astronautics, MIT, 77 Massachusetts Avenue, Cambridge, MA 02139, USA\label{MIT_aeronautics}
    }

\date{Received 31 March 2023 / Accepted 2 May 2023}

% \abstract{}{}{}{}{} 
% 5 {} token are mandatory
 
  \abstract
   {Small planets located at the lower mode of the bimodal radius distribution are generally assumed to be composed of iron and silicates in a proportion similar to that of the Earth. However, recent discoveries are revealing a new group of low-density planets that are inconsistent with that description.}
   {We intend to confirm and characterize the TESS planet candidate TOI-244.01, which orbits the bright ($K$ = 7.97 mag), nearby ($d$ = 22 pc), and early-type (M2.5 V) M-dwarf star GJ 1018 with an orbital period of 7.4 days.}
   {We used Markov Chain Monte Carlo methods to model 57 precise radial velocity measurements acquired by the ESPRESSO spectrograph together with TESS photometry and complementary HARPS data. Our model includes a planetary component and Gaussian processes aimed at modeling the correlated stellar and instrumental noise.}
   {We find TOI-244 b to be a super-Earth with a radius of $R_{\rm p}$ = 1.52 $\pm$ 0.12 $\rm R_{\oplus}$ and a mass of $M_{\rm p}$ = 2.68 $\pm$ 0.30 $\rm M_{\oplus}$. These values correspond to a density of $\rho$ = 4.2 $\pm$ 1.1 $\rm g \cdot cm^{-3}$, which is below what would be expected for an Earth-like composition. We find that atmospheric loss processes may have been efficient to remove a potential primordial hydrogen envelope, but high mean molecular weight volatiles such as water could have been retained. Our internal structure modeling suggests that TOI-244~b has a $479^{+128}_{-96}$ km thick hydrosphere over a 1.17~$\pm$~0.09~$\rm R_{\oplus}$ solid structure composed of a Fe-rich core and a silicate-dominated mantle compatible with that of the Earth. On a population level, we find two tentative trends in the density-metallicity and density-insolation parameter space for the low-density super-Earths, which may hint at their composition.}
   {With a 8$\%$ precision in radius and 12$\%$ precision in mass, TOI-244~b is among the most precisely characterized super-Earths, which, together with the likely presence of an extended hydrosphere, makes it a key target for atmospheric observations.}


   \maketitle
%
%-------------------------------------------------------------------

\section{Introduction}

%--------------------------------------------------------------------

Over the last decades, exoplanet detection has been very successful.  Since the first discoveries \citep{1992Natur.355..145W,1995Natur.378..355M}, the number of known planets has increased year by year to the 5347 we know today\footnote{According to the NASA Exoplanet Archive \citep{2013PASP..125..989A}.}. This success has been achieved to a great extent thanks to the space-based \textit{Kepler} mission \citep[][2009-2014]{2010Sci...327..977B}, which, jointly with its extended \textit{K2} mission \citep[][2014-2018]{2014PASP..126..398H}, detected 62$\%$ of the known planets by means of the transit method \citep[e.g.,][]{2014A&A...562A.109L,2016ApJS..226....7C,2016ApJ...822...86M,2018AJ....155..136M,2018AJ....156..277L,2018AJ....156...78L,2020MNRAS.499.5416C,2021MNRAS.508..195D,2022AJ....163..244C}.

This plethora of planets has allowed us to perform statistical studies of many regimes of the parameter space, revealing different exoplanet populations. The \textit{Kepler} and \textit{K2} results definitively confirmed what \citet{2011arXiv1109.2497M} hinted: small planets (i.e., $R_{\rm p}$ < 4 $\rm R_{\oplus}$) are the most common ones within $\sim$1 AU of Sun-like stars \citep{2012ApJS..201...15H,2013ApJS..204...24B,2013PNAS..11019273P,2015ApJ...809....8B}. Also, \textit{Kepler} data revealed the existence of a bimodal radius distribution in the small planets sample \citep{2017AJ....154..109F}, which matches the radius gap previously predicted by photoevaporation numerical analysis \citep{2013ApJ...775..105O,2014ApJ...795...65J,2014ApJ...792....1L,2016ApJ...831..180C}. However, the physical cause of the distribution as well as the internal composition of the higher mode planets (2-4 $\rm R_{\oplus}$; gas dwarfs or water worlds) is still under debate \citep[e.g.,][]{2018MNRAS.476..759G,2019MNRAS.487...24G,2019PNAS..116.9723Z,2020A&A...643L...1V,2022Sci...377.1211L,2023arXiv230104321R}. In contrast, there is not such a debate as to the internal structure of the lower mode planets (1-2 $\rm R_{\oplus}$), whose densities point toward rocky-dominated compositions with a proportion of silicates and iron similar to that of the Earth \citep[33$\%$ Fe, and 67$\%$ $\rm MgSiO_{3}$ in mass; e.g.,][]{2019PNAS..116.9723Z}. 

Recent discoveries have revealed the existence of rocky planets inconsistent with Earth-like compositions. For example, planets K2-229 b \citep{2018NatAs...2..393S}, Kepler-107 c \citep{2019NatAs...3..416B}, K2-233 c \citep{2020A&A...640A..48L}, L 168-9 b \citep{2020A&A...636A..58A}, and K2-38 b \citep{2020A&A...641A..92T} have been found to have unusually high densities, similar to that of Mercury. These densities are thought to be caused by the presence of elevated iron content in the planetary cores, which is usually explained through the existence of high iron abundances in the initial protoplanetary disks \citep[e.g.,][]{2020MNRAS.493.4910S,2020ApJ...901...97A,2022A&A...662A..19J,2023A&A...670A...6B}, or through external factors such as mantle stripping caused by collisions during planetary formation \citep[e.g.,][]{2010ApJ...712L..73M}. On the other hand, planets TOI-561~b \citep{2021MNRAS.501.4148L,2021AJ....161...56W,2023AJ....165...88B}, L 98-59~c and d \citep{2021A&A...653A..41D}, HD 260655 c \citep{2022A&A...664A.199L}, and TOI-4481~b \citep{2023arXiv230106873P} have been found to have lower densities than expected for an Earth-like composition. These densities could be explained by a scarcity or total absence of iron in the planet structure, by the presence of a significant amount of volatile elements, or by a mixture of both. Interestingly, all these light planets are found to orbit metal-poor stars (i.e., [Fe/H] between -0.20 and -0.45 dex). This emerging density-metallicity correlation could be explained by the fact that the building blocks in the original protoplanetary disks of metal-poor stars have lower iron and higher water mass fractions than those expected for stars with solar metallicities, and hence, they are expected to form iron-poor and water-rich planets  \citep{2017A&A...608A..94S,2021Sci...374..330A}. Currently, interior structure models can shed some light on the composition of these low-density planets. However, they are limited by degeneracies that prevent us from determining their bulk compositions from mass and radius information alone. Future atmospheric observations through transmission spectroscopy will allow us to determine the volatile content of these planets, which will help us to uncover their nature.

The identification of rocky planets inconsistent with Earth-like compositions has only been possible recently thanks to the measuring of very precise planet masses. Today, only 24$\%$ of exoplanets have a true dynamical mass measured, and the percentage is reduced down to 9$\%$ for small planets (i.e., $R_{\rm p}$~<~4~$\rm~R_{\oplus}$). Measuring an accurate mass for transiting exoplanets is of crucial importance in order to perform internal structure analysis \citep[e.g.,][]{2020A&A...642A.121L,2021NatAs...5..775D} as well as atmospheric characterization through transmission spectroscopy. Recently, \citet{2019ApJ...885L..25B} conducted  retrievals on simulated transmission spectra from the James Webb Space Telescope, concluding that a 20$\%$ mass precision or better is required so that inferences of atmospheric properties are not limited by the mass precision of the planet. However, the percentage of small planets meeting this threshold is lower than 5$\%$.

In 2018, the launch of the TESS space telescope \citep{2015JATIS...1a4003R}, together with the start of operations of the ESPRESSO high-resolution echelle spectrograph \citep{2021A&A...645A..96P}, opened a new window to exoplanet characterization. TESS, by continuously monitoring a field of view (i.e., sector) of 24 x 96 degrees that changes every $\sim$27 days, is performing photometric observations of almost the entire sky, focusing on stars significantly brighter than those surveyed by the \textit{Kepler} mission \citep[e.g.,][]{2019ApJ...881L..19V,2020MNRAS.491.2982E}. ESPRESSO, with a spectral resolving power of 140\,000 over the 380-788 nm wavelength range, and mounted on the Very Large Telescope (VLT), is able to achieve an unprecedented instrumental radial velocity precision of 10 cm$\cdot\rm s^{-1}$. The powerful combination of TESS and ESPRESSO observations has allowed several exquisite planet characterizations \citep[e.g.,][]{2020A&A...642A..31D,2020A&A...642A.121L,2021A&A...653A..41D,2021A&A...648A..75S,2022A&A...665A.154B,2022arXiv221009713L}. 

In this work, we confirm and characterize the small ($R_{p}$ = 1.5 $\rm R_{\oplus}$) and close-in ($P$ = 7.49 d) transiting planet TOI-244 b \citep[recently validated in][]{2023AJ....165..134O} orbiting the bright ($K$ = 7.97 mag) and nearby ($d$ = 22 pc) M 2.5V star GJ 1018. Based on the transit signal detected from TESS data, we carried out an intensive radial velocity campaign with ESPRESSO in order to confirm its planetary nature, obtain a precise mass measurement, as well as to search for additional planets. In addition to TESS and ESPRESSO data, we used complementary spectroscopic and photometric data sets from HARPS and ASAS-SN in order to maximize the information for this system. 

In Sect.~\ref{sec:observations},  we describe the TESS, ESPRESSO, HARPS, and ASAS-SN observations. In Sect.~\ref{sec:stellar_charact}, we present our stellar characterization based on precise photometry and the ESPRESSO spectra. In Sect.~\ref{sec:analysis_results}, we describe our analysis of photometric and spectroscopic data and present the derived planetary parameters. In Sect.~\ref{sec:discussion}, we discuss the results, and we conclude in Sect.~\ref{conclusions}.

\section{Observations}
\label{sec:observations}

\subsection{TESS photometry}
\label{sec:obs_tess}

\begin{figure}
    \centering
    \includegraphics[width = 9cm]{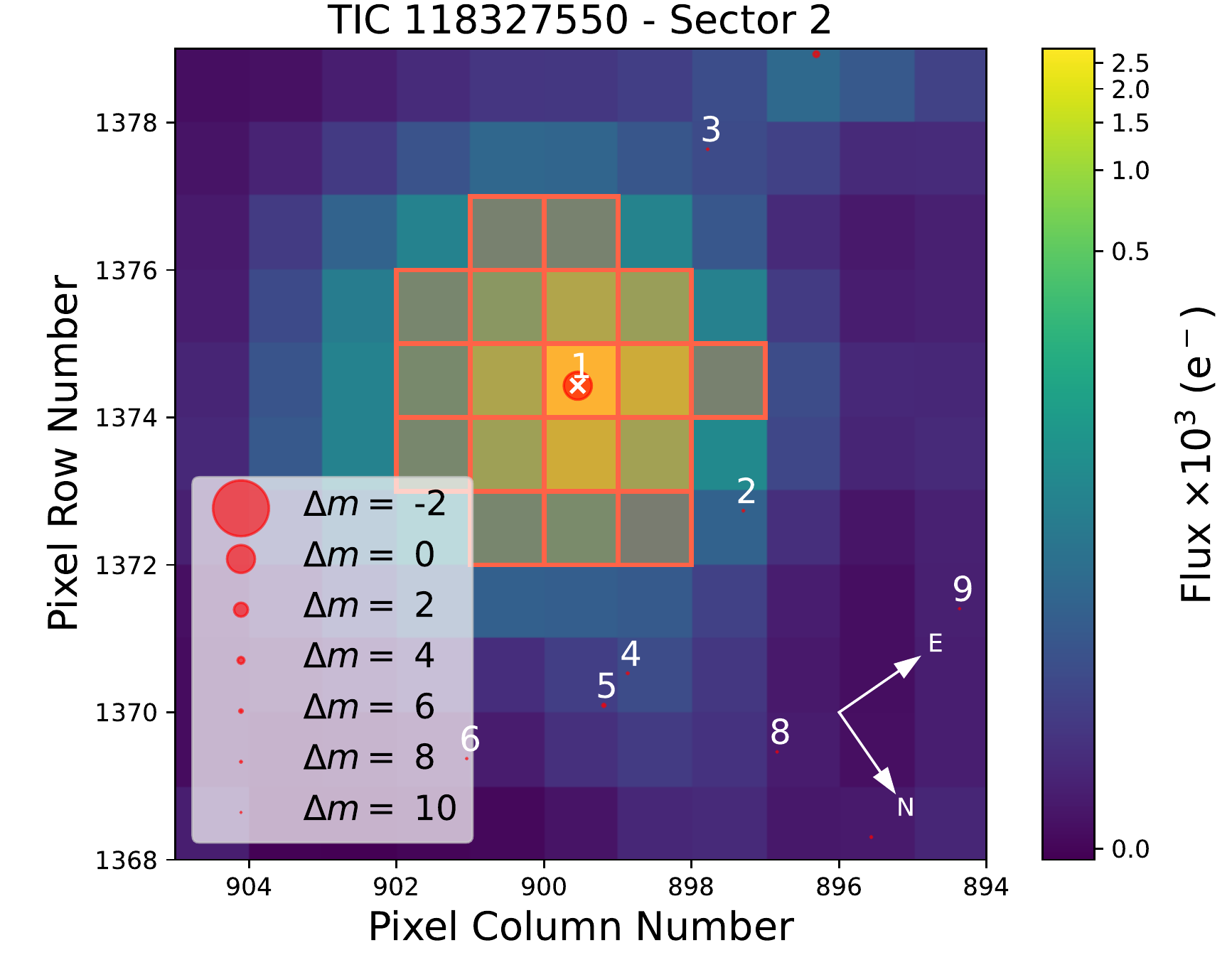}
    \caption{TESS target pixel file of GJ 1018. The orange grid is the selected aperture and the red circles correspond to the nearby \textit{Gaia} DR3 sources. Symbol sizes for \textit{Gaia} sources scale to their $G$ magnitudes. This plot has been prepared through \texttt{tpfplotter} \citep{2020A&A...635A.128A}.}
    \label{fig:tpfplotter}
\end{figure}

\begin{figure*}
    \centering
    \includegraphics[scale=0.51]{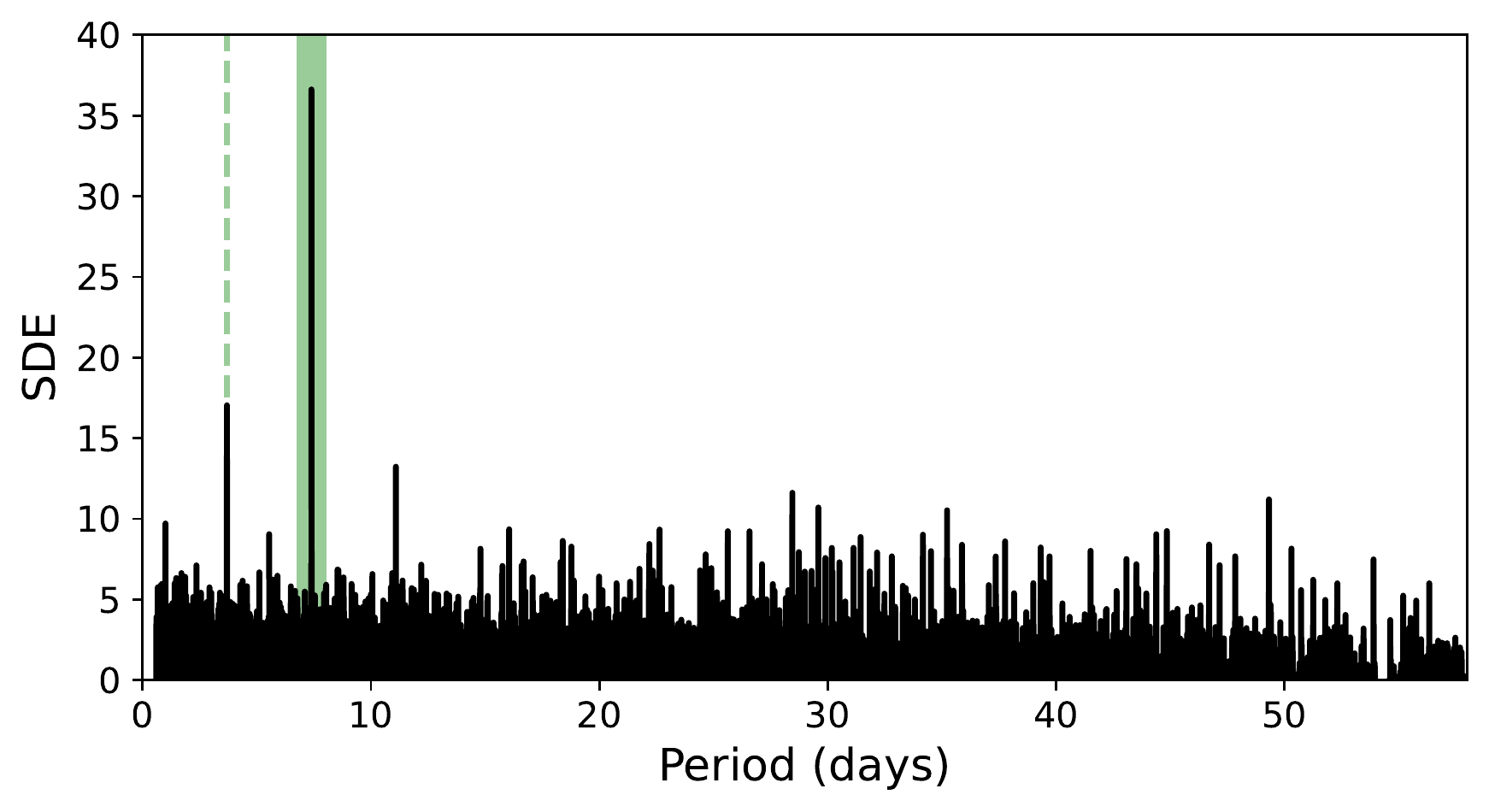}
    \includegraphics[scale=0.51]{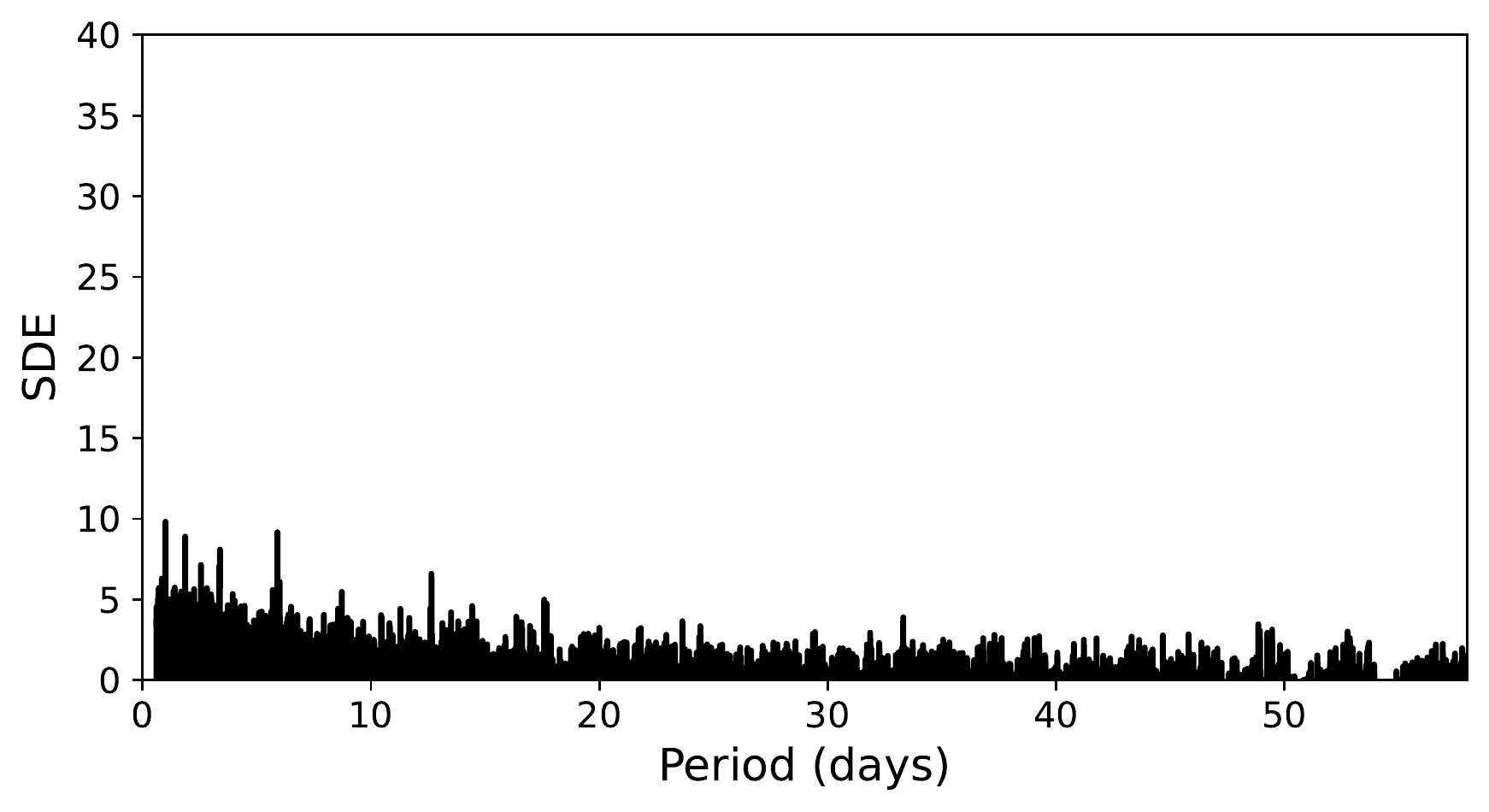}
    \caption{\textit{Left}: \texttt{TLS} periodogram of the complete and flattened TESS light curve of GJ 1018, where the highest peak (highlighted with the green broad line) corresponds to the TOI-244.01 transit signal. The green thin dashed line corresponds to the second harmonic. \textit{Right:} \texttt{TLS} periodogram of the TESS light curve after masking the TOI-244.01 signal. } 
    \label{fig:tls_periodograms}
\end{figure*}

The star GJ 1018 (TOI-244, TIC 118327550) was observed by TESS (camera $\#$2, CCD $\#$3) at a 2-min cadence in sector 2 (S2) from 22 August 2018 to 20 September 2018, and in S29 from 26 August 2020 to 22 September 2020, resulting in a total of 36$\,$229 target pixel files (TPFs) spanning a temporal baseline of two years. The central time of each sector shows a short gap ($\sim$1-2 days) due to the satellite repointing toward the Earth to downlink the data, resulting in a total duty cycle of 50 days.

The observations were processed by the Science Processing Operation Center (SPOC) pipeline \citep{2016SPIE.9913E..3EJ} and are publicly available in the TESS archive of the Mikulski Archive for Space Telescopes (MAST)\footnote{\url{https://mast.stsci.edu/portal/Mashup/Clients/Mast/Portal.html}}. The TESS SPOC data products include simple aperture photometry (SAP) and presearch data-conditioned simple aperture photometry (PDCSAP), being the latter the SAP processed by the PDC algorithm, which corrects the photometry of instrumental systematics that are common to all stars in the same CCD \citep{2012PASP..124.1000S,2012PASP..124..985S,2014PASP..126..100S}. The photometric aperture was automatically selected by the SPOC module Create Optimal Apertures (COA), which maximizes the signal-to-noise ratio of the flux measurement \citep{10.1117/12.857625,2020ksci.rept....7S}. Besides, COA estimates the fraction of flux inside the photometric aperture that comes from the target star, and uses it to correct for contamination within the PDCSAP light curve. This fraction (also called CROWDSAP) is 0.9996 for GJ 1018. In Fig. \ref{fig:tpfplotter}, we plot the selected aperture over a TPF of GJ 1018, together with all the nearby stars detected in the \textit{Gaia} Data Release 3 \citep[DR3;][]{2022arXiv220800211G}. There are no additional sources within the aperture, and the nearby sources surrounding the aperture have a magnitude difference $\Delta G$ > 6 mag with GJ 1018 in the \textit{Gaia} passband. These large contrasts ensure negligible contamination \citep[e.g.,][]{2018AJ....156..277L,2022MNRAS.509.1075C}.

In October 2018, the SPOC pipeline identified a periodic flux decrease (known as threshold crossing event; TCE) of 7.4 days through the Transiting Planet Search (TPS) algorithm. The algorithm first characterizes the power spectral density of the observation noise, and then estimates the likelihood of the existence of a transit-like signal over a wide range of trial transit durations and orbital periods \citep{2002ApJ...575..493J,10.1117/12.856764,2020ksci.rept....9J}. A transit model fit was performed \citep{Li:DVmodelFit2019} and a suite of diagnostic tests were carried out to help making or breaking confidence in the planetary hypothesis \citep{Twicken:DVdiagnostics2018}, all of which the candidate signal passed. Finally, the TESS Science Office examined the light curve and additional information to designate TIC 118327550.01 as a TESS Object of Interest (TOI-244.01) that would benefit from follow-up observations \citep{2021ApJS..254...39G}.

We downloaded from MAST the PDCSAP light curves and removed all data points with a quality flag different from zero. As a result, we removed one data point in S2 (1359.6474 TJD) due to an argabrightening event (bit 5, value 16). In sector S29, we removed 3005 data points (located between 2098.7624 TJD and 2101.3791 TJD, and between 2112.6041 TJD and 2114.4374 TJD) due to scattered light caused by the Earth or Moon (bit 13, value 4096), and we also removed four data points cataloged as impulsive outliers (bit 10, value 512) which have fluxes deviated 5.2$\sigma$ (2097.5194 TJD), 6.5$\sigma$ (2103.2500 TJD), 7.4$\sigma$ (2103.3000 TJD), and 7.1$\sigma$ (2105.0583 TJD) from the PDCSAP light curve. The final TESS light curves are presented in Table \ref{tab:tess_phot}.

We computed the generalized Lomb-Scargle periodogram \citep[\texttt{GLS},][]{2009A&A...496..577Z} of the TESS light curve (for S2 and S29 separately and jointly) and found no significant periodicities. We also computed the transit least squares periodogram \citep[\texttt{TLS},][]{2019A&A...623A..39H} in order to unveil the significance of the TOI-244.01 signal and to determine whether there are additional transit signals. To do so, we first produced a flattened version of the PDCSAP light curve in order to mitigate possible trends caused by the stellar activity or uncorrected systematics. For that matter, we filtered the photometric data of each sector separately using the robust time-windowed biweight method implemented within the \texttt{wotan} package \citep{2019AJ....158..143H} with a 0.5-day window length. We detrended the PDCSAP and joined the two flattened light curves to create a long time series. We present the detrended light curve in Table~\ref{tab:tess_phot}. In Fig.~\ref{fig:tls_periodograms} (left panel), we plot the \texttt{TLS} periodogram of the flattened light curve, which shows a strong peak at 7.4 days with a signal detection efficiency (SDE) of 36.6. In the right panel, we plot the periodogram over the same light curve with the TOI-244.01 transits masked. This periodogram shows no prominent peaks. However, the six highest peaks have SDEs between 6 and 10, which are slightly above some empirical thresholds for transit detection\footnote{Empirical thresholds for transit detection range from SDE > 6 \citep{2015ApJ...807...45D}, SDE > 6.5 \citep{2018AJ....156...78L}, SDE > 7 \citep{2012ApJ...761..123S}, to SDE > 10 \citep{2018MNRAS.473L.131W}.}. To check the reliability of those peaks, we folded the flattened light curve to the corresponding periodicities and inspected a binned version of the folded data. We found no hints of any transit signal, so we conclude those peaks are spurious.

\subsection{ESPRESSO spectroscopy}
\label{sec:espresso_obs}

\begin{figure*}
    \centering
    \includegraphics[width=\textwidth]{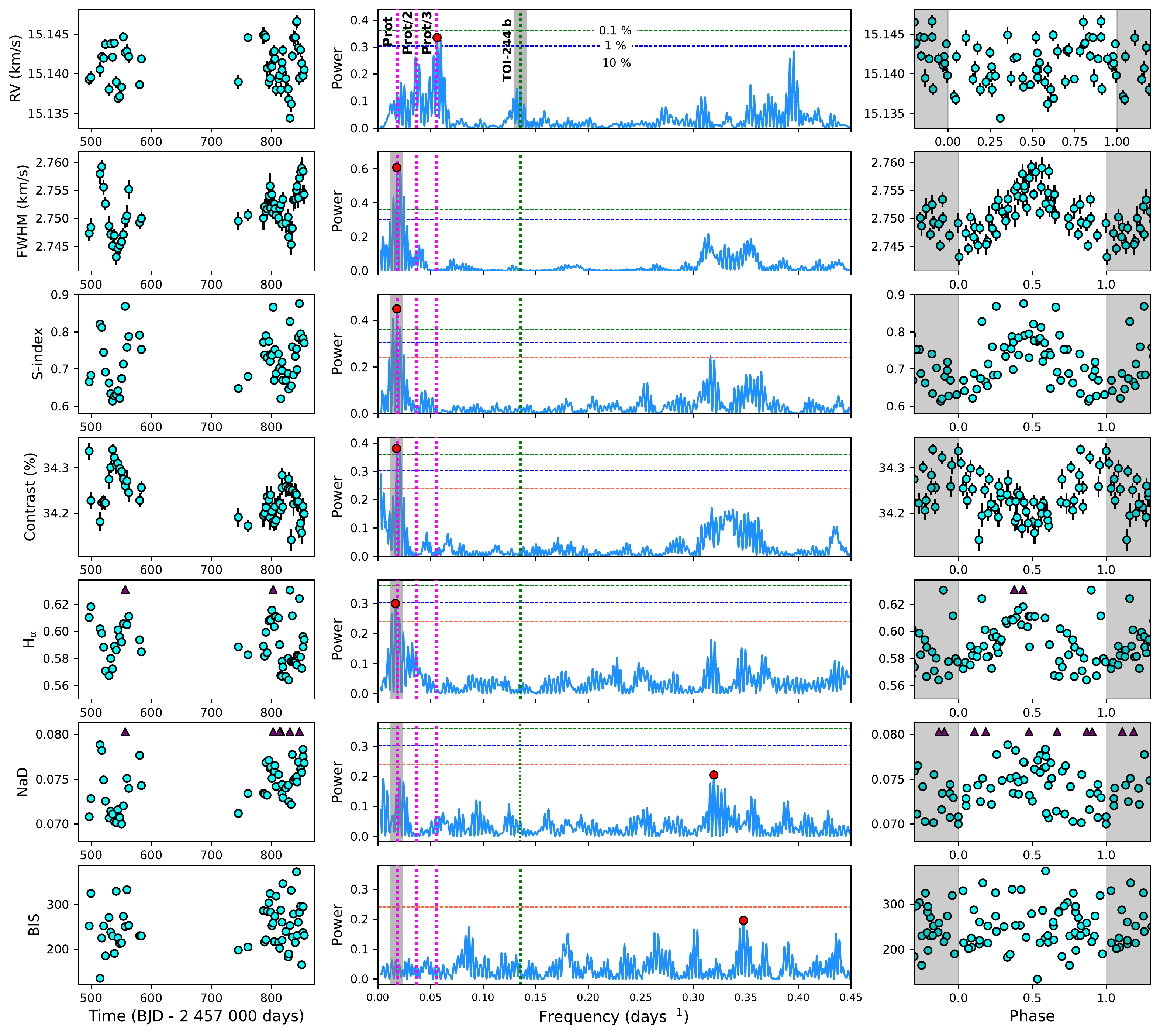}
    \caption{\textit{Left panel:} Time series of the ESPRESSO RVs and activity indicators. \textit{Center panel:} \texttt{GLS} periodograms of the corresponding time series. The red circles highlight the maximum power frequencies. The green dotted vertical lines indicate location of the orbital period of TOI-244~b ($P_{\rm orb}$~=~7.4 days). The magenta dotted vertical lines indicate the rotation period of the star identified in the activity indicators ($P_{\rm rot}$ $\sim$ 56 days) and its second and third harmonic. The gray vertical bands indicate the periods by which the time series in the right panel are folded.  The horizontal dotted lines correspond to the 10$\%$ (orange), 1$\%$ (blue), and 0.1$\%$ (green) FAP levels.  \textit{Right Panel:} ESPRESSO time series folded to the gray bands periods. The triangle markers within the left and right panels indicate the location of data points outside the boundaries of the plot.}
    \label{fig:gls_espresso}
\end{figure*}

We observed GJ 1018 with the ESPRESSO high resolution echelle spectrograph \citep{2021A&A...645A..96P}, which is mounted on the Very Large Telescope (VLT) located at ESO's Paranal Observatory (Chile) and has been operational since 2018. The observations were performed in the course of the ESPRESSO Guaranteed Time Observations (GTOs) under the programs IDs 108.2254.002, 108.2254.005, and 108.2254.006, whose main aim is to determine precise mass measurements of transiting planet candidates. We obtained a total of 57 spectra between 9 October 2021 and 2 October 2022 with a typical cadence of 2-3 days and a typical exposure time of 1200 s, resulting in a mean signal-to-noise ratio (S/N) of 57 at 650 nm. All observations were made in the slow-readout single UT high resolution mode (HR21; 2$\, \times \,$1 binning), which has a spectral resolution power of 140$\,$000 and embraces a wavelength range from 380 nm to 788~nm. For each exposure, we simultaneously illuminated the second fiber with the Fabry-Pérot interferometer, which allows the calibration of the instrumental drift with a precision better than 10 cm$\cdot \rm s^{-1}$ \citep{2010SPIE.7735E..4XW}.

We reduced the data through the ESPRESSO Data Reduction Software (DRS) pipeline version 3.0.0\footnote{The ESPRESSO Data Reduction Software is publicly available at \url{https://www.eso.org/sci/software/pipelines/index.html}.} \citep{2021A&A...645A..96P}. The DRS also extracts the radial velocities (RVs) based on a modified implementation of the original cross-correlation technique presented by \citet{1996A&AS..119..373B}, in which the different spectral lines of a numerical mask are weighed as a function of their RV information \citep{2002A&A...388..632P}. In particular, we used an M3 mask to obtain the cross-correlation function (CCF) of each observation, which was later fitted to a Gaussian profile in order to compute the RVs (center of the Gaussian). Finally, all the RVs were corrected from secular acceleration. The standard deviation of the RV observations is 2.7 m/s and the median uncertainty per data point is 0.7 m/s. The DRS also computes activity indicators such as the full width at half maximum (FWHM) and the amplitude or contrast of the CCF, the bisector span (BIS), the $\rm H_{\alpha}$ and Sodium doublet (NaD) line depths, and the S-index.  We present the complete ESPRESSO data set in Table \ref{tab:espresso_rvs}.

In Fig.~\ref{fig:gls_espresso}, we show the \texttt{GLS} periodogram of the ESPRESSO RVs and activity indicators time series. The FWHM, S-index, Contrast of the CCF, and $\rm H_{\alpha}$ show maximum power periods of 56.0, 55.9, 56.8, and 60.6 days, with false alarm probabilities (FAPs)\footnote{All the FAPs in this work have been computed analytically following Eq. 24 of \citet{2009A&A...496..577Z}.}  of  $5 \times 10^{-6}~\%$, $1.8 \times 10^{-3}~\%$, $1.2 \times 10^{-1}~\%$, and 1.1$~\%$ respectively, thus unveiling the presence of a significant activity-related signal that most likely corresponds to the rotation period of the star. The remaining two indicators BIS and NaD show no significant peaks within the periodogram. Their maximum power periods correspond to 2.9 days (39.0$\%$ FAP), and 3.1 days (30.4$\%$ FAP) respectively, being most likely related to the mean observing cadence of $\sim$2-3 days. It is remarkable that, when folded to the $\sim$56-day periodicity found in most indicators, the NaD time series shows moderate signs of a sinusoidal behavior, suggesting that the activity signal might be manifested in this indicator as well. The RV periodogram shows a peak coinciding with the orbital period of TOI-244.01 that cannot be seen in any of the indicators, which suggests a planetary origin for the signal. However, the signal is not significant. The maximum power period of the RV time series is 17.8 days ($2.9 \times 10^{-1} \% $ FAP), which is compatible with the third harmonic of the $\sim$56-day signal present in most activity indicators. Also, in this region of the periodogram, there is another peak at 28.3 days with a 9.5$\%$ FAP that is compatible with the second harmonic of the $\sim$56-day signal. The presence of these two prominent peaks indicates that the RV time series is significantly affected by the stellar activity. Finally, there is another relevant peak at 2.5 days (2.2$\%$ FAP). This periodicity is compatible with the mean observing cadence, but we also explore the possibility of it being an additional planetary signal in Sect.~\ref{subsec:rv_modeling}.

\subsection{HARPS spectroscopy}
\label{sec:harps_obs}
GJ 1018 was observed by the High Accuracy Radial velocity Planet Searcher (HARPS), which is mounted on the 3.6~m telescope located at ESO's La Silla Observatory (Chile) and has been operational since 2003. HARPS is a fiber-fed cross dispersed echelle espectrograph located in a vacuum vessel that protects the instrument from temperature and refractive index variations. It has a spectral resolution power of 115$\,$000 and covers a wavelength range between 378 and 691 nm. A total of 15 spectra were acquired between 15 December 2018 and 8 January 2019 under the program 1102.C-0339 (PI X. Bonfils). We downloaded the reduced spectra (DRS version 3.8), which are publicly available in the ESO's Science Portal\footnote{\url{http://archive.eso.org/scienceportal/home}}. Eleven spectra were acquired with 1800 s exposure time, resulting in a median S/N of 14.6 per resolution element at 650 nm, and the remaining four spectra were acquired with 1500 s exposure time, resulting in a median S/N of 11.9.   We present the complete HARPS data set in Table~\ref{tab:harps_rvs}. The standard deviation of the RV observations is 4.3 m/s and the median uncertainty per data point is 2.1 m/s. The S/N values obtained with relatively long exposures are remarkably low, leading to a photon-noise limited RV precision (i.e., three times worse than ESPRESSO).

\subsection{ASAS-SN photometry}
\label{sec:asas_sn}

\begin{table*}[]
\caption{Summary of the ASAS-SN observations of GJ 1018.}
\renewcommand{\arraystretch}{1.3}
\begin{tabular}{lllllll}
\hline
Start date & End date & Camera & Band & Station & Location & $\#$Obs \\ \hline
11 May 2014 & 15 Sep 2018 & bf & V & Cassius & Cerro Tololo International Observatory (Chile) & 263 \\
18 Sep 2017 & Operational & bj & g & Henrietta Leavitt & McDonald Observatory (USA) & 385 \\
10 Nov 2017 & Operational & bn & g & Cecilia Payne & South African Astrophysical Observatory & 196 \\
7 Oct 2018 & Operational & bF & g & Cassius & Cerro Tololo International Observatory (Chile) & 266 \\ \hline
\end{tabular}
\label{tab:asas_sn}
\end{table*}

\begin{figure*}
    \centering
    \includegraphics[width=\textwidth]{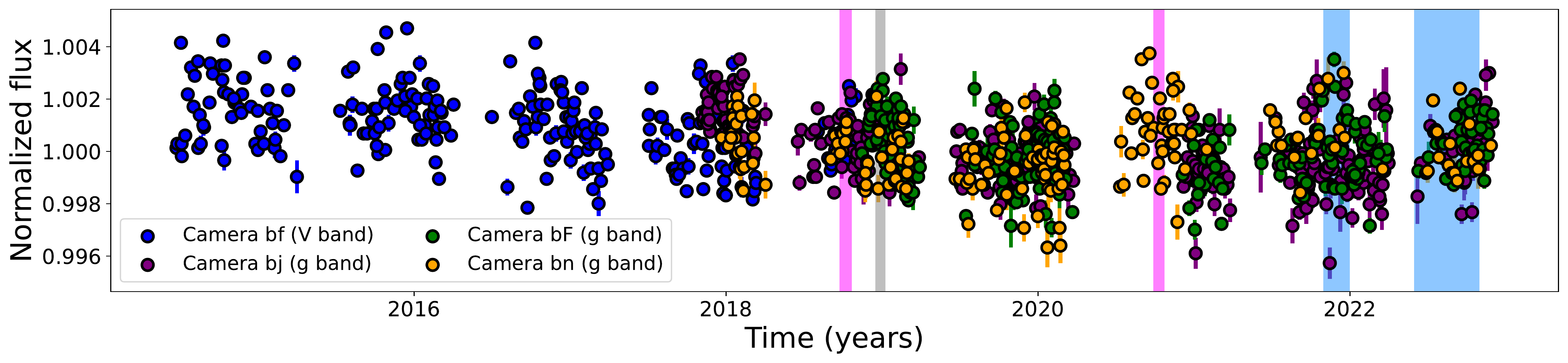}
    \includegraphics[width=\textwidth]{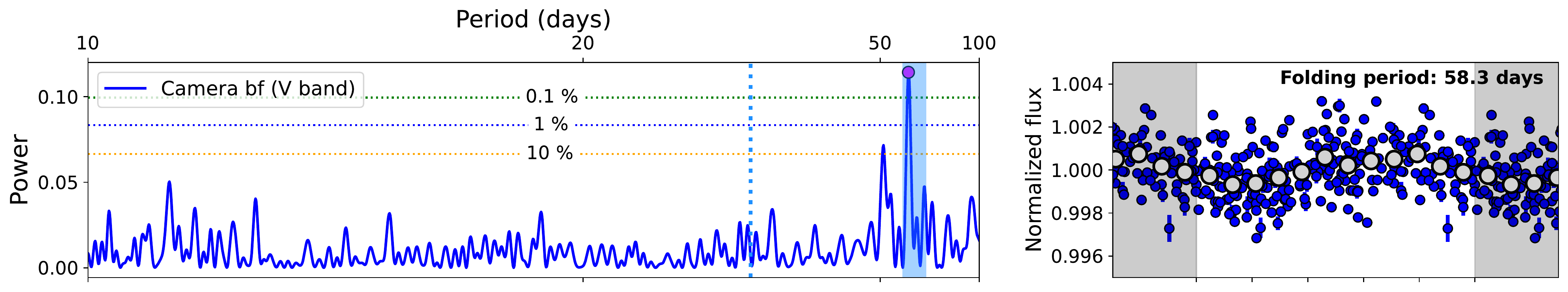}
    \includegraphics[width=\textwidth]{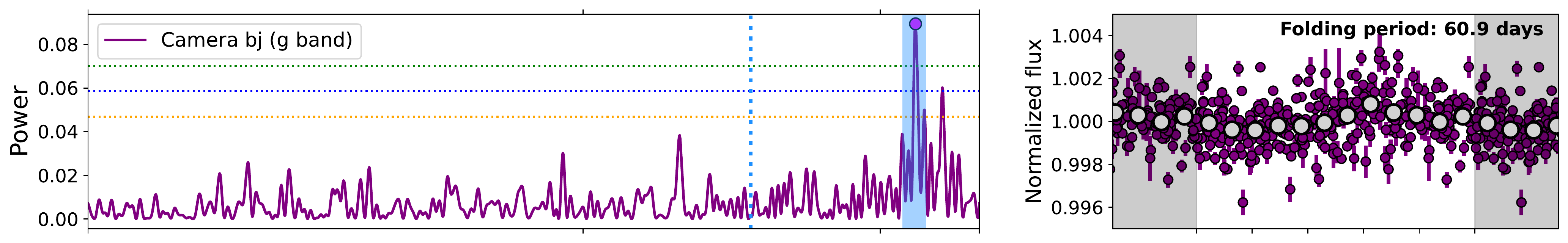}
    \includegraphics[width=\textwidth]{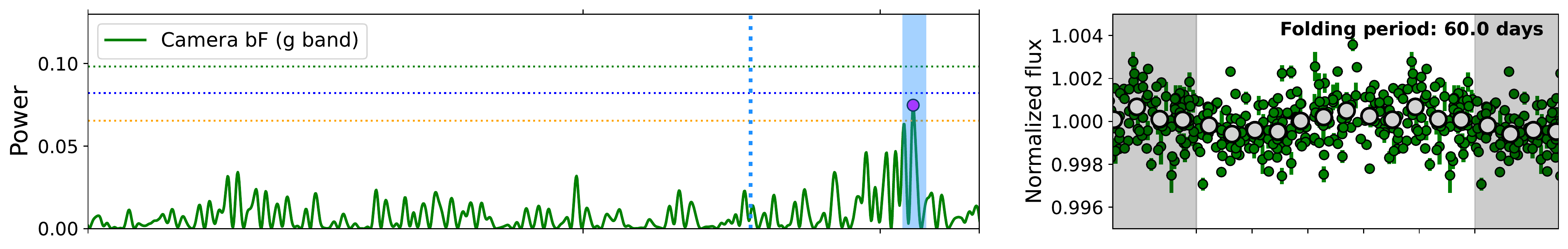}
    \includegraphics[width=\textwidth]{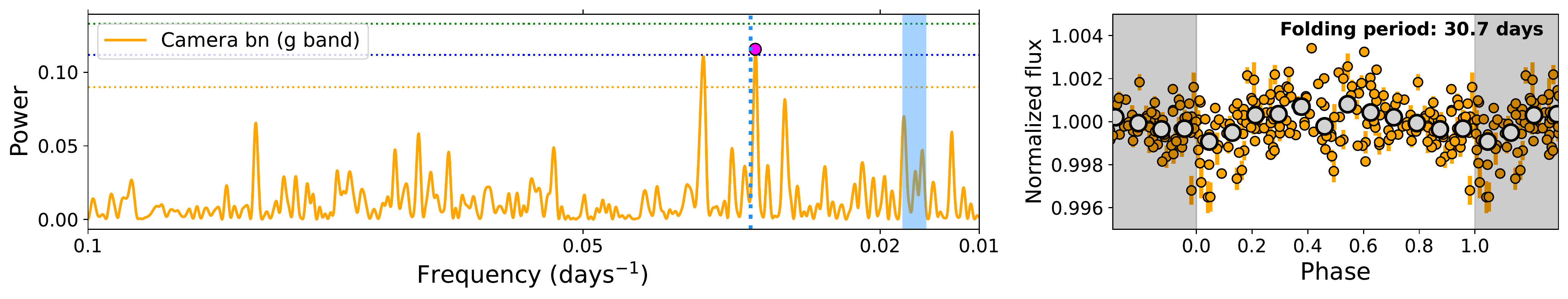}
    \caption{\textit{Top panel:} ASAS-SN photometric time series of GJ 1018. The magenta, gray, and blue vertical lines correspond to the TESS, HARPS, and ESPRESSO observing windows, respectively. \textit{Left panels}: \texttt{GLS} periodograms of the ASAS-SN photometry corrected for the long-term trend. The vertical blue solid line indicates the median rotation modulation period obtained by the \texttt{GLS}, and the vertical blue dashed line indicate its second harmonic. The horizontal dotted lines indicate the 10$\%$ (orange), 1$\%$ (blue), and 0.1$\%$ (green) FAP levels. \textit{Right panels:} Phase-folded light curves to the maximum power period obtained by the \texttt{GLS} periodogram. The gray data points correspond to a 5-day binning.}
    \label{fig:asassn}
\end{figure*}

The sky region encompassing GJ 1018 is being observed by three different stations of the All-Sky Survey for Supernovae \citep[ASAS-SN,][]{2014ApJ...788...48S}. In Table \ref{tab:asas_sn}, we include the observing periods, the number of observations, the observing bands, and the location of the different stations. Each station consists of four Nikon telephoto lenses of 14 cm aperture equipped with a 2048 $\times$ 2048 pixels CCD camera. The pixel scale is 8.0 arc seconds, which corresponds to a 4.5 $\rm deg^{2}$ field of view per camera. Every night, the cameras take three consecutive exposures of 90 s exposure time each, which are combined afterward to increase the signal-to-noise of the light curves by a factor of $\sqrt{3}$. Images obtained in poor weather conditions, that are out of focus (FWHM > 2.5 pixels), or where the studied source is near the detector edge (at a distance < 0.2 deg) are discarded by the survey pipeline. The pipeline performs aperture photometry at a selected location through \texttt{IRAF} \citep{1986SPIE..627..733T} considering a 2 pixel radius aperture and 7-10 pixel radius annulus for the target star and for the reference stars, which are selected from the Photometric All-Sky Survey \citep[APASS,][]{2012JAVSO..40..430H,2019JAVSO..47..130H} following the procedure from \citet{2017PASP..129j4502K}.

We computed the light curves through the ASAS-SN Sky Patrol web interface\footnote{\url{https://asas-sn.osu.edu/}}. Given that GJ 1018 is a high proper motion star ($\mu_{\alpha}$ = $-154.88 \pm 0.02$ $\rm mas/yr$, $\mu_{\delta}$ = $45.63 \pm 0.03$ $\rm mas/ yr$) and the typical FWHMs are comparable to the radius of the aperture ($\sim$ 2 pixels), we shifted the aperture location over time in order to ensure the target centering and thus avoid flux loses. GJ 1018 is visible from the three observatories from mid-May to mid-February of the following year, so we computed the photometry on a year-by-year basis, introducing the coordinates corrected for proper motion corresponding to the central time of each observing window (1 October). After the light curves computation, we discarded those epochs in which the flux is below the estimated 5-$\sigma$ detection limit for the target location, as well as those data points with a deviation greater than 5-$\sigma$ of a flattened version of the photometric time series. In Table \ref{tab:asassn_phot}, we present the complete photometric data set.

In Fig.~\ref{fig:asassn}, we show the ASAS-SN photometric time series of GJ 1018. The complete data set shows a long-term trend that could be caused by instrumental systematics, or more likely by the magnetic cycle of the star, given the similarity of the trends in each individual camera. We also show the \texttt{GLS} periodograms of the photometry corrected for the abovementioned trends; that is, we divided the time series of each camera by a degree three polynomial that was previously fit to the data. For cameras bf, bj, and bF, we obtain maximum power periods of 58.3, 60.9, and 60.0 days, with FAPs of 0.011$\%$, 0.0019$\%$, and 2.79$\%$ respectively. These periodicities are consistent with the ones observed in most of the ESPRESSO activity indicators (Sect.~\ref{sec:espresso_obs}). Hence, they most likely reflect the rotation period of GJ 1018 (see Sect.~\ref{subsec:joint_analysis} for an accurate determination). The maximum power period for camera bn is 30.7 days (FAP = 0.70$\%$), which coincides with the second harmonic of the aforementioned rotation period.

\section{Stellar characterization}
\label{sec:stellar_charact}

\subsection{General description of GJ 1018}
\label{sec:general_description}

%%%% stellar params table %%%%%%%
\begin{table}[]
\caption{Stellar properties of GJ 1018.}
\label{tab:stellar_params}
\renewcommand{\arraystretch}{1.05}
\setlength{\tabcolsep}{1.5pt}
\begin{tabular}{llc}
\hline \hline
Parameter & Value & Reference \\ \hline
\multicolumn{3}{l}{Identifiers} \\ \hline
TOI, TIC & 244, 118327550 & [1], [2] \\
GJ & 1018 & [3] \\
LP & 937-95 & [4] \\
2MASS & J00421695-3643053 & [5] \\
Gaia DR3 & 5001098681543159040 & [6] \\ \hline
\multicolumn{3}{l}{Coordintates, parallax and kinematics} \\ \hline
RA, DEC & 00:42:16.74, -36:43:04.71 & [6] \\
$\rm \mu_{\alpha}$ (mas/yr) & -154.880 $\pm$ 0.020 & [6] \\
$\rm \mu_{\delta}$ (mas/yr) & 45.632 $\pm$ 0.029 & [6] \\
Parallax (mas) & 45.300 $\pm$ 0.027 & [6] \\
Distance (pc) & 22.075 $\pm$ 0.013 & [6] \\
RV (km/s) & 15.1410 $\pm$ 0.013 & [7], Sect. \ref{subsec:joint_analysis} \\
U (km/s) & -23.30 $\pm$ 0.02 & [7], Sect. \ref{sec:abundances} \\
V (km/s) & 17.09 $\pm$ 0.02 & [7], Sect. \ref{sec:abundances} \\
W (km/s) & -8.62 $\pm$ 0.03 & [7], Sect. \ref{sec:abundances} \\
Gal. population & Thin disk & [7], Sect. \ref{sec:abundances} \\ \hline
\multicolumn{3}{l}{Atmospheric parameters and spectral type} \\ \hline
$T_{\rm eff}$ (K) & 3433 $\pm$ 100 & [7], Sect. \ref{sec:steparsyn} \\
log $g$ (dex) & 4.66 $\pm$ 0.07 & [7], Sect. \ref{sec:steparsyn} \\
$\rm [Fe/H]$ (dex) (1) & -0.39 $\pm$ 0.07 & [7], Sect. \ref{sec:steparsyn} \\
$\rm [Fe/H]$ (dex) (2) & -0.03 $\pm$ 0.11 & [7], Sect. \ref{sec:steparsyn} \\
SpT & M2.5 V & [7], Sect. \ref{sec:steparsyn} \\ \hline
Physicall parameters &  &  \\ \hline
$\rm R_{\star}$ ($\rm R_{\odot}$) & 0.428 $\pm$ 0.025 & [7], Sect. \ref{sec:stellar_radius_mass} \\
$\rm M_{\star}$  ($\rm M_{\odot}$) & 0.427 $\pm$ 0.029 & [7], Sect. \ref{sec:stellar_radius_mass} \\
L ($\rm L_{\odot} \times 10^{-2}$) & $2.277\pm 0.042$ & [7], Sect. \ref{sec:luminosity} \\
$\rm P_{rot}$ (days) & $53.3^{+1.2}_{-1.1}$ & [7], Sect. \ref{subsec:joint_analysis} \\
Age (Gyr) & 7 $\pm$ 4 & [8] \\ \hline
Chemical abundances &  &  \\ \hline
$\rm [Mg/H]$ (dex) (1) & -0.28 $\pm$ 0.07 & [7], Sect. \ref{sec:abundances} \\
$\rm [Si/H]$ (dex) (1) & -0.33 $\pm$ 0.07 & [7], Sect. \ref{sec:abundances} \\
$\rm [Mg/H]$ (dex) (2) & 0.00 $\pm$ 0.10 & [7], Sect. \ref{sec:abundances} \\
$\rm [Si/H]$ (dex) (2) & -0.01 $\pm$ 0.10 & [7], Sect. \ref{sec:abundances} \\ \hline
Magnitudes &  &  \\ \hline
$\rm M_{bol}$ (mag) & 8.847 $\pm$ 0.020 & [7], Sect. \ref{sec:luminosity} \\
TESS (mag) & 10.3475 $\pm$ 0.0073 & [2] \\
G (mag) & 11.553 $\pm$ 0.001 & [6] \\
B (mag) & 14.199 $\pm$ 0.033 & [9] \\
V (mag) & 12.684 $\pm$ 0.047 & [9] \\
g' (mag) & 13.426 $\pm$ 0.063 & [9] \\
r' (mag) & 12.097 $\pm$ 0.045 & [9] \\
i' (mag) & 10.831 $\pm$ 0.062 & [9] \\
J (mag) & 8.827 $\pm$ 0.023 & [5] \\
H (mag) & 8.251 $\pm$ 0.033 & [5] \\
$\rm K_{s}$ (mag) & 7.970 $\pm$ 0.029 & [5] \\ 
\hline \hline
%\noalign{\vskip 1mm} 
\end{tabular}
\tablefoot{Based on (1) \texttt{SteParSyn} (2) \texttt{ODUSSEAS} analysis. \textbf{References.} [1] \citet{2021ApJS..254...39G}; [2] \citet{2019AJ....158..138S}; [3] \citet{1991adc..rept.....G}; [4] \citet{1979nlcs.book.....L}; [5] \citet{skrutskie2006}; [6] \citet{2022arXiv220800211G}; [7] This work; [8] \citet{2019AJ....158..173A};  [9] \citet{2019JAVSO..47..130H}.}

\end{table}

% (1) Derived from or based on {\sc SteParSyn} results; (2) Derived from or based on ODUSSEAS results.
%--------------up----------------

%\renewcommand{\arraystretch}{1.2}
%\setlength{\tabcolsep}{1.5pt}

%-------------down----------------

%\noalign{\vskip 1mm} 
%\tablefootnote{\textbf{References.} [1] \citet{2021ApJS..254...39G}; [2] \citet{2019AJ....158..138S}; [3] \citet{1991adc..rept.....G}; [4] \citet{1979nlcs.book.....L}; [5] \citet{skrutskie2006}; [6] \citet{2022arXiv220800211G}; [7] This work; [8] \citet{2019AJ....158..173A};  [9] \citet{2019JAVSO..47..130H}. }
%%%%%%%%%%%%%%%%%%%%%%%%%%%%%%%%%

GJ 1018 is a bright ($K$ = 7.97 mag) early-type M-dwarf star located in the solar neighborhood. The \textit{Gaia} DR3 provides a  parallax of $\pi$ =  45.300 $\pm$ 0.027 mas, which corresponds to a distance of $d$ = 22.075 $\pm$ 0.013 pc. The photometry-based TESS Input Catalog \citep[TIC v8.0.1,][]{2019AJ....158..138S} estimates an effective temperature of $T_{\rm eff}$ = 3407 $\pm$ 157 K, surface gravity of log $g$ = 4.820 $\pm$ 0.004 dex, stellar radius of $R$ = 0.41 $\pm$ 0.01 $\rm R_{\odot}$, and stellar mass of $M$ = 0.40 $\pm$ 0.02 $\rm M_{\odot}$. In the next sections, we describe our stellar characterization based on precise photometry and a high-resolution, high S/N spectrum obtained from the combination of all ESPRESSO spectra. In Table \ref{tab:stellar_params}, we summarize the general properties of GJ 1018 as well as all our derived parameters.

\subsection{Stellar atmospheric parameters}
\label{sec:steparsyn}

We computed the stellar atmospheric parameters of GJ 1018 by means of the {\sc SteParSyn} code\footnote{\url{https://github.com/hmtabernero/SteParSyn/}} \citep{tab22}. The code implements the spectral synthesis method with an MCMC sampler to retrieve the stellar atmospheric parameters. We employed a grid of synthetic spectra computed with the {\sc Turbospectrum} code \citep{ple12} alongside BT-Settl stellar atmospheric models \citep{all12} and atomic and molecular data of the Vienna atomic line database \citep[VALD3;][]{rya15}. We considered a selection of \ion{Fe}{i}, \ion{Ti}{i} lines, and TiO molecular bands that are well-suited to analyze M-dwarf stars \citep{mar21}. In all, {\sc SteParSyn} has allowed us to compute the following stellar parameters: $T_{\rm eff}$~$=$~3433~$\pm$~10~K, $\log{g}$~$=$~4.66~$\pm$~0.07 dex, and [Fe/H]~$=$~$-0.39$~$\pm$~0.07~dex. In order to account for systematics associated with $T_{\rm eff}$ when it is determined by spectral fitting techniques (e.g., \citealt{marfil21} found a dispersion between 40 and 100 K among different model atmospheres valid for M-dwarfs), we increased the $T_{\rm eff}$ uncertainty up to 100 K.

We performed an independent computation of the atmospheric parameters through the newly machine learning tool \texttt{ODUSSEAS}\footnote{\url{https://github.com/AlexandrosAntoniadis/ODUSSEAS/}} \citep{Antoniadis-Karnavas2020}, which has been designed to compute the $T_{\rm eff}$ and [Fe/H] of M-dwarfs. The method is based on measuring the pseudo Equivalent Widths (pEWs) of specific absorption  and blended lines in the wavelength range 5300\,{\AA}-6900\,{\AA} \citep{neves14}. Briefly, \texttt{ODUSSEAS} compares the measured EWs to the machine learning models generated from reference HARPS spectra, convolved to the resolution of the observed  spectrum. Although \texttt{ODUSSEAS} was initially developed for spectra with resolution powers from 48\,000 to 115\,000, it has been applied successfully to ESPRESSO spectra with a resolution power of 140\,000. For the very high resolutions of ESPRESSO and HARPS, the results derived by \texttt{ODUSSEAS} are essentially the same either by using directly the original highest-resolution spectra of ESPRESSO, or by convolving first the ESPRESSO spectra to the lower, but still high, resolution of HARPS. We ran \texttt{ODUSSEAS} by considering a reference data set composed of 47 stars with interferometry-based $T_{\mathrm{eff}}$ \citep{khata21, rabus19}, and [Fe/H] derived through the photometric calibration by \citet{neves12} using the \textit{Gaia} DR3 parallaxes. We obtain $T_{\mathrm{eff}}$~=~3419~$\pm$~92~K and [Fe/H]~=~-0.03~$\pm$~0.11 dex.

The \texttt{ODUSSEAS} effective temperature is in very good agreement with the value computed by {\sc SteParSyn}. However, there exist a strong discrepancy between the metallicities. Obtaining accurate metallicities for M-dwarfs is a complicated task, given the strong blending of spectral lines and molecular bands. For example, \citet{2022A&A...658A.194P} found mean deviations of around 0.1-0.3 dex between different techniques, showing that any uncertainty below those values is underestimated. Overall, we adopt the $T_{\rm eff}$ and $\log{g}$ estimations from the spectral syntesis method. However, given the strong discrepancy in [Fe/H], we decide not to adopt either estimate, and instead discuss both results independently in the subsequent sections.

\subsection{Stellar bolometric luminosity}
\label{sec:luminosity}

\begin{figure}
    \centering
    \includegraphics[scale=0.32]{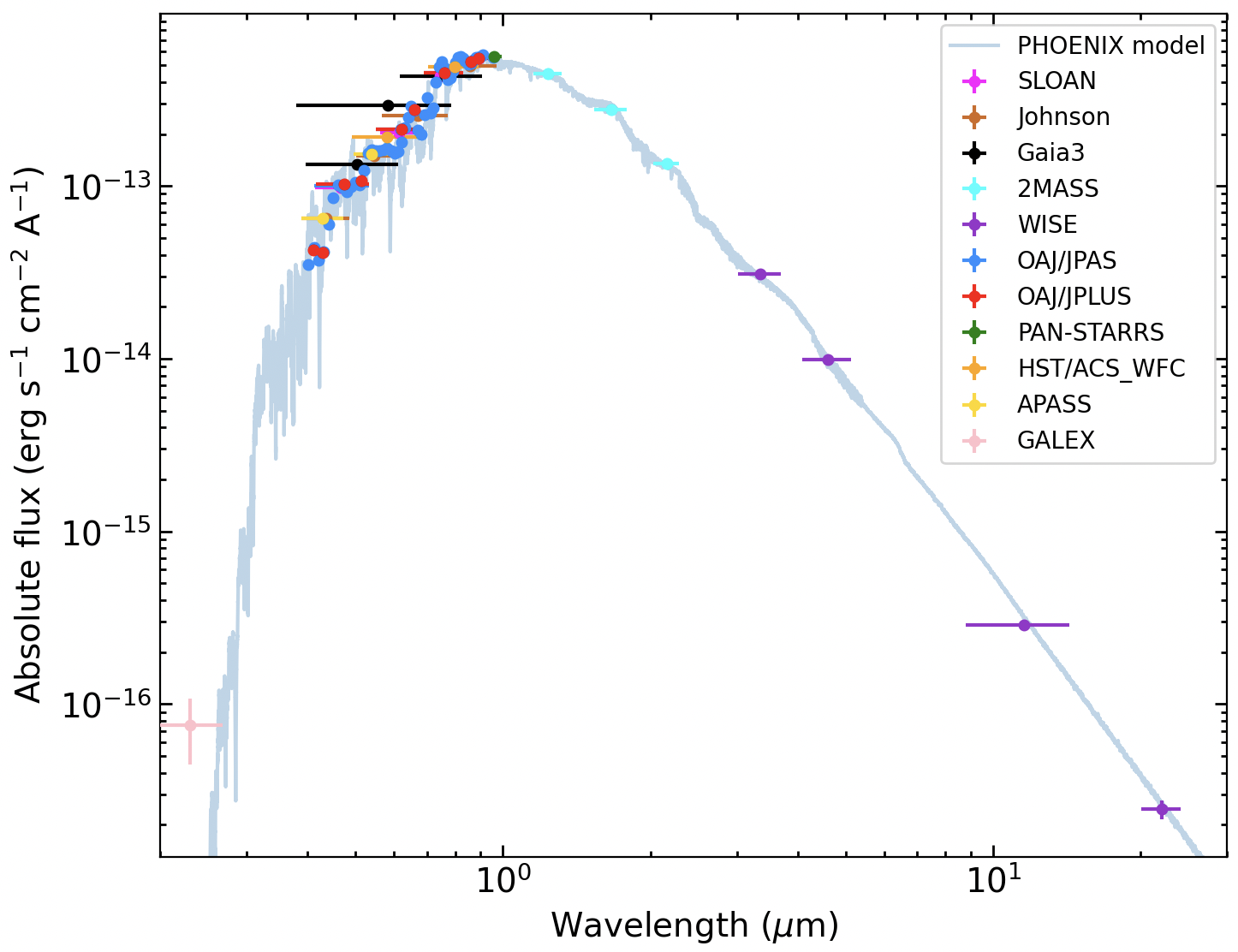}
    \caption{Photometric spectral energy distribution of GJ 1018 from 0.23 to $\sim$25 $\mu$m. The PHOENIX model corresponding to 3300 K, solar metallicity and high gravity is added \citep{2003IAUS..211..325A} in order to show that most fluxes are photospheric in origin and that there is no infrared flux excesses at long wavelengths. Vertical error bars denote flux uncertainties and horizontal error bars account for the width of the passbands. The fluxes, effective wavelengths and widths of all passbands were taken from the Virtual Observatory SED Analyzer database.}
    \label{fig:phot_sed}
\end{figure}

To determine the bolometric luminosity of GJ 1018, we first built the photometric spectral energy distribution (SED) of the star using broadband and narrowband photometry from the literature. The stellar SED is shown in Fig.~\ref{fig:phot_sed}, which includes the Galaxy Evolution Explorer (GALEX) near-ultraviolet photometry \citep{2017ApJS..230...24B}, the Johnson $BVRI$ photometry \citep{2002yCat.2237....0D}, the $BV$ data from the AAVSO Photometric All-sky Survey \citep[APASS,][]{2019JAVSO..47..130H}, the $griz$ data from the Sloan Digital Sky Survey \citep{2000AJ....120.1579Y}, the {\sl Gaia} Early Data Release 3 photometry \citep{2021A&A...649A...1G}, the $y$ data from the Panoramic Survey Telescope and Rapid Response System (Pan-STARRS, \citealt{2020ApJS..251....7F}), the Two Micron All Sky Survey (2MASS) near-infrared $JHK_s$ photometry \citep{skrutskie2006}, the Wide-field Infrared Survey Explorer ({\sl WISE}) $W1$, $W2$, $W3$, and $W4$ data \citep{wright2010}, and the optical multiphotometry of the Observatorio Astrofísico de Javalambre (OAJ) Physics of the Accelerating Universe Astrophysical Survey (JPAS) and Photometric Local Universe Survey (JPLUS) catalogs accessible through the Spanish Virtual Observatory \citep{bayo2008}. In total, there are 89 photometric data points defining the SED of GJ 1018 between 0.23 and $\sim$25 $\mu$m. The OAJ/JPAS data cover very nicely the optical region in the interval 0.40--0.95 $\mu$m with a cadence of one measurement per 0.01 $\mu$m. We used the {\sl Gaia} trigonometric parallax to convert all observed photometry and fluxes into absolute fluxes, which we present in Table \ref{tab:svo_phot}. The SED of GJ 1018 clearly indicates its photospheric origin for wavelengths longer than 0.4 $\mu$m; there are no mid-infrared flux excesses up to 25 $\mu$m.

We integrated the SED over wavelength to obtain the absolute bolometric flux ($F_{\rm bol}$) using the trapezoidal rule. We did not include the $\sl Gaia$ $G$-band flux in the computations because the filter large passband width encompasses various redder and bluer filters. We then applied $M_{\rm bol} = -2.5~{\rm log}\,F_{\rm bol} - 18.988$ \citep{cushing2005}, where $F_{\rm bol}$ is in units of W$\cdot$m$^{-2}$, to derive the absolute bolometric magnitude $M_{\rm bol}$ = 8.8465 $^{+0.0203}_{-0.0200}$ mag for GJ 1018, from which we obtain a bolometric luminosity of $L$ = $2.277 \, \pm \, 0.042 \times 10^{-2}$ L$_{\odot}$. The quoted error bar accounts for the photometric uncertainties in all observed bands and the trigonometric distance error.

\subsection{Stellar radius and mass}
\label{sec:stellar_radius_mass}
We determined the radius of GJ 1018 from the well-known Stefan-Boltzmann law. Based on the effective temperature from Sect.~\ref{sec:steparsyn} and the bolometric luminosity from Sect.~\ref{sec:luminosity}, we obtained a radius of $R_{\star} = 0.428 \pm 0.025$ R$_\odot$. This radius determination is independent of any evolutionary model and depends only on distance (well known from ${\sl Gaia}$), bolometric luminosity (well determined from the SED, see Sect.~\ref{sec:luminosity}), and the model atmospheres used to fit the observed spectra.

The mass of GJ 1018 can be derived following different approaches. One is given by the empirical mass-luminosity relationship of \citet{mann19}, which is derived from 62 nearby, late-type binaries with known orbits. The authors calibrated the stellar mass as a function of the absolute $K$-band magnitude of the stars, finding that metallicity has little impact on the mass of the stars for approximately solar composition. According to the \citet{mann19} relationship, GJ 1018 has a mass of $M_{\star}$ = 0.400 $\pm$ 0.025 M$_\odot$, where the error budget includes the photometric error and the dispersion of the mass-luminosity relation. Another widely used relation valid for M-dwarfs is the mass-radius equation given in \citet{2019A&A...625A..68S}, obtained from 55 detached, double-lined, doubled-eclipsing, and main-sequence M-dwarf binaries from the literature. Following this relation, we obtain a mass of $M_{\star}$ = $0.427\pm0.029$~$\rm M_{\odot}$. Both mass determinations are compatible at the 1$\sigma$ level. Given that the relation by \citet{2019A&A...625A..68S} makes use of the Stefan-Boltzmann law, its mass derivation is consistent with our method for obtaining the stellar radius, thus we adopted $R_{\star} = 0.428 \pm 0.025$ R$_\odot$ and $M_{\star} = 0.427 \pm 0.029$ M$_\odot$.

\subsection{Galactic membership and Mg and Si abundances}
\label{sec:abundances}

Because of heavy line blending, the determination of the individual elemental abundances of M-dwarfs from the visible spectra is very hard  \citep[e.g.,][]{2020A&A...644A..68M}. In this work, we estimated the abundance of Mg and Si closely following the procedure presented in \citet{2021A&A...653A..41D}. In brief, we used the systemic RV, parallax, ra/dec coordinates and proper motions from \textit{Gaia} DR3 to derive the Galactic space velocity UVW of GJ~1018. We obtain U  = -23.30 $\pm$ 0.02 km/s, V = 17.09 $\pm$ 0.02 km/s, and W = -8.62 $\pm$ 0.03 km/s with respect to the local standard of rest (LSR) adopting the solar peculiar motion from \citet{2003A&A...409..523R}. Based on these velocities, we used the widely-known kinematic approach from \citet{2003A&A...410..527B} and the kinematic characteristics for stellar components in the Solar neighborhood from \citet{2003A&A...409..523R} to estimate the probability that GJ 1018 belongs to the thin disk (D), the thick disk (TD), and the halo (H), obtaining 99.11$\%$, 0.889$\%$, and 0.001$\%$ respectively. Hence, it is very likely that GJ 1018 is a member of
the Galactic thin-disk population. Then, from the APOGEE DR17 \citep{2022ApJS..259...35A} we selected cool stars with metallicities similar to that of GJ 1018 and belonging to the chemically defined Galactic thin disk. We obtained a sample of several thousand stars for which we calculated the mean abundances of Mg and Si and their standard deviation (star-to-star scatter). Considering the {\sc SteParSyn} metallicity, we obtain [Mg/H] = -0.28 $\pm$ 0.07 dex and [Si/H] = -0.33 $\pm$ 0.07 dex. Considering the \texttt{ODUSSEAS} metallicity, we obtain [Mg/H] = 0.00 $\pm$ 0.10 dex and [Si/H] = -0.01 $\pm$ 0.10 dex. 

\section{Analysis and results}
\label{sec:analysis_results}

\subsection{TESS photometry analysis}
\label{subsec:TESS_analysis}

We first analyzed the TESS PDCSAP photometry described in Sect.~\ref{sec:obs_tess} through a model that consists of two components: a transit model, and a Gaussian process (GP) that models the correlated photometric noise \citep{2006gpml.book.....R,2012RSPTA.37110550R}. 

We implemented the \citet{2002ApJ...580L.171M} quadratic limb darkened transit model through \texttt{batman} \citep{2015PASP..127.1161K}. The model is defined by the orbital period of the planet ($P_{\rm orb}$), the time of inferior conjunction ($T_{\rm 0}$), the orbital inclination ($i$), the quadratic limb darkening (LD) coefficients $u_{\rm 1}$ and $u_{\rm 2}$, the planet-to-star radius ratio ($R_{\rm p}/R_{\star}$), and the semimajor axis scaled to the stellar radius, which we parametrized through $P_{\rm orb}$ and the stellar mass ($M_{\rm \star}$) and radius ($R_{\rm \star}$) following the Kepler's Third Law. We modeled the TESS correlated noise through a GP with an approximate Matérn-3/2 kernel \citep{2017AJ....154..220F,2018RNAAS...2...31F}, which has been successfully used to model the unknown
mixture of stellar variability and residual systematics of TESS SPOC photometry \citep[e.g.,][]{2022arXiv220113274M,2022AJ....163..298M,2023arXiv230409220M}. This kernel is especially suitable to model TESS light curves where the stellar rotation modulation is not observable (as is the case of GJ 1018, see Sect.~\ref{sec:obs_tess}), and instead, residual systematics are a significant component of the photometric variability, given that it has covariance properties that are especially well matched to short-term instrumental red-noise structures \citep{2017AJ....153..177P,2020AJ....160..259S}. The approximate Matérn-3/2 kernel can be written in terms of the temporal separation between two data points $\tau = t_{i} - t_{j}$ as

\begin{equation}
    K_{3/2} = \eta_{\sigma}^{2} \left[ \left(1 + \frac{1}{\epsilon} \right) e^{-(1-\epsilon) \sqrt{3} \tau / \eta_{\rho}} \cdot \left(1 - \frac{1}{\epsilon} \right) e^{-(1+\epsilon) \sqrt{3} \tau / \eta_{\rho}} \right],
\end{equation}

\noindent where the hyperparameters $\eta_{\sigma}$ and $\eta_{\rho}$ are the characteristic amplitude and timescale of the correlated variations, respectively, and $\epsilon$ controls the approximation to the exact Matérn-3/2 kernel. Given that the amplitudes and timescales of the TESS systematics can vary from one sector to another, we fit those parameters independently ($\eta_{\sigma_{i}}$ and $\eta_{\rho_{i}}$, where \textit{i} denotes the sector), while $\epsilon$ was fixed to its default value of $10^{-2}$ \citep{2017AJ....154..220F}. We also included a jitter term for each sector, which we added quadratically to the TESS flux uncertainties in order to model the uncorrelated noise not taken into account in our model. 

We used a Markov Chain Monte Carlo (MCMC) affine-invariant ensemble sampler \citep{2010CAMCS...5...65G} as implemented in \texttt{emcee} \citep{2013PASP..125..306F} in order to sample the posterior probability density function of the different parameters involved in our model. To do so, we used four times as many walkers as the number of parameters, and performed two consecutive runs. The first run (or burn-in) consisted of 200 000 iterations. After this run, we reset the sampler and initialized the second run (or production) with 100 000 iterations while considering the initial values from the last iteration of the burn-in phase. To ensure the convergence of the chains, we estimated the autocorrelation time for each parameter and checked that it is at least 30 times smaller than the chain length. 

We ran an MCMC fit starting from wide uniform priors for all the parameters involved in the model except for those for which we have prior information, which we constrained through Gaussian priors. These parameters are the $T_{0}$ and $P$ of TOI-244.01 (from the \texttt{TLS} periodogram, Sect. \ref{sec:obs_tess}), the stellar radius and mass (from our spectroscopic analysis, Sect. \ref{sec:stellar_radius_mass}), and the quadratic LD coefficients, which we computed from the \texttt{ldtk} package \citep{2015MNRAS.453.3821P}. The package infers the coefficients of a given LD law relying on the \citet{2013A&A...553A...6H} synthetic spectra library, spectroscopic $T_{\rm eff}$, log $g$, and [Fe/H], and the instrument transmission curve. We used the \texttt{ldtk} uncertainties as the widths of the Gaussian distributions. In order to account for possible systematics in the estimated LD coefficients \citep[e.g.,][]{2022AJ....163..228P}, we also ran the MCMC fit by considering conservative widths of 0.2, obtaining identical results for the planetary parameters. We show the prior distributions for this analysis and subsequent ones in Table~\ref{table:priors}.

\begin{figure*}
    \centering
    \includegraphics[scale = 0.28]{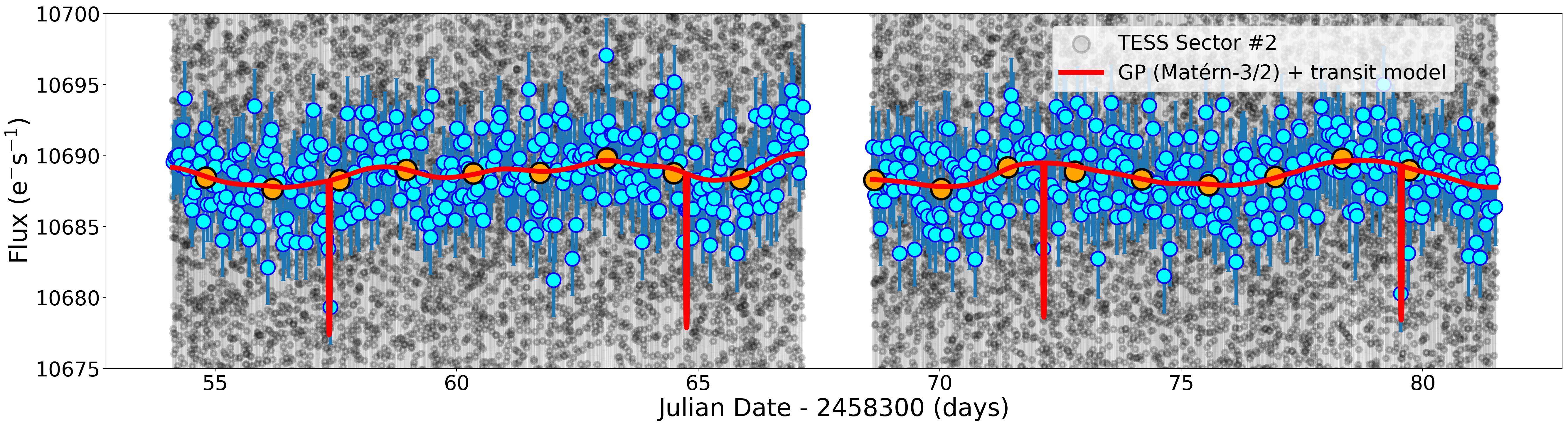}
    \includegraphics[scale = 0.28]{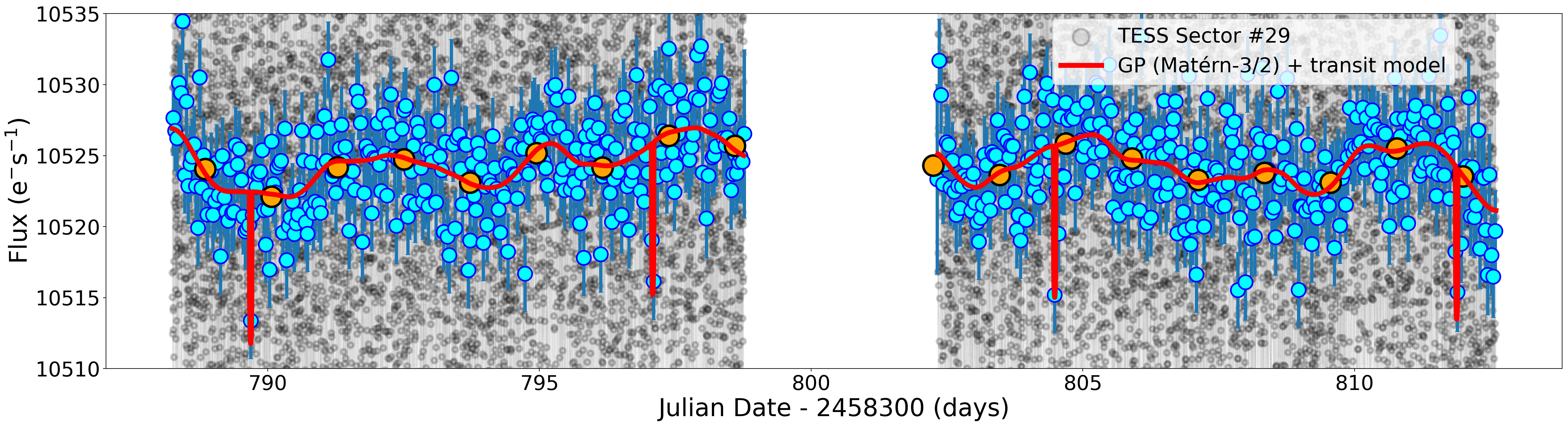}
    \caption{TESS light curve of GJ 1018 with the median posterior global model (transit + GP) superimposed, in which we can see the different locations of the TOI-244.01 transit events over the smooth photometric modulation. The gray data points correspond to the SPOC 2-minute cadence PDCSAP photometry, and the blue and orange data correspond to 50-minute and 1.5-day binned data respectively. }
    \label{fig:tess_photometry}
\end{figure*}

In Table \ref{tab:final_derived_params}, we include the median and 1$\sigma$ (68.3$\%$ credible intervals) of the posterior distributions of the fit parameters. In Fig.~\ref{fig:tess_photometry}, we show the complete TESS light curve together with the global model (transit + GP) evaluated on the fit parameters.

\begin{table}[]
\small
\renewcommand{\arraystretch}{1.175}
\setlength{\tabcolsep}{2pt}
\caption{Prior distribution of the parameters used in our models.}
\label{table:priors}
\begin{tabular}{llc}
\hline \hline
Symbol                             & Parameter         & Distribution                                       \\ \hline 
\multicolumn{3}{l}{Orbital parameters}                                                                                                 \\ \hline
$P$ {[}days{]}                              & Orbital period             & $\mathcal{G} \, (7.397, 0.002)$                                 \\
$T_{\rm 0}$ {[}RJD{]}                       & Time of inf. conj.         & $\mathcal{G} \, (58357.36, 0.02)$                         \\
$i$ {[}degrees{]}                           & Orbital inclination        & $\mathcal{U} \, (50, 90)$                                   \\
$cos(w) \sqrt{e}$                           & Ecc. parametrization       & $\mathcal{U} \, (- \frac{\sqrt{2}}{2}, \frac{\sqrt{2}}{2})$ \\
$sin(w) \sqrt{e}$                           & Ecc. parametrization       & $\mathcal{U} \, (- \frac{\sqrt{2}}{2}, \frac{\sqrt{2}}{2})$ \\ \hline
\multicolumn{3}{l}{Planet parameters}                                                                                                  \\ \hline
$R_{\rm p}$/$R_{\rm \star}$                 & Planet-to-star radius ratio & $\mathcal{U} \, (0.0, 0.1)$                                 \\
$K$ {[}$\rm m/s${]}                        & RV semi-amplitude          & $\mathcal{U} \, (1, 10)$                         \\ \hline
\multicolumn{3}{l}{Stellar parameters}                                                                                                  \\ \hline
$M_{\rm \star}$ $[\rm M_{\odot}]$           & Stellar Mass               & $\mathcal{ZTG} \, (0.427, 0.029)$                           \\
$R_{\rm \star}$ $[\rm R_{\odot}]$           & Stellar Radius             & $\mathcal{ZTG} \, (0.428, 0.025)$                           \\
$u1$                                        & LD coefficient             & $\mathcal{ZTG} \, (0.2204, 0.0011)$                         \\
$u2$                                        & LD coefficient             & $\mathcal{ZTG} \, (0.4112, 0.0017)$                         \\ \hline
\multicolumn{3}{l}{Matérn-3/2 GP hyperparameters}                                                                                      \\ \hline
$\eta_{\sigma_{S2}}$ {[}$\rm e^{-}/s${]}    & S2 amplitude               & $\mathcal{U} \, (0, 10^{2})$                                \\
$\eta_{\rho_{S2}}$ {[}days{]}               & S2 timescale               & $\mathcal{U} \, (0, 10^{2})$                                \\
$\eta_{\sigma_{S29}}$ {[}$\rm e^{-}/s${]}   & S29 amplitude              & $\mathcal{U} \, (0, 10^{2})$                                \\
$\eta_{\rho_{S29}}$ {[}days{]}              & S29 amplitude              & $\mathcal{U} \, (0, 10^{2})$                                \\ \hline
\multicolumn{3}{l}{Quasiperiodic GP hyperparameters}                                                                                  \\ \hline
$\eta_{\rm 1,RV}$ {[}$\rm m/s${]}          & Amplitude (RV)             & $\mathcal{LU} \, (10^{-2}, 10)$                        \\
$\eta_{\rm 2,RV}$ {[}days{]}                & Decay timescale            & $\mathcal{U} \, (0, 5 \cdot 10^{2})$                        \\
$\eta_{\rm 3,RV}$ {[}days{]}                & Correlation period         & $\mathcal{U} \, (40, 80)$                                   \\
$\eta_{\rm 4,RV}$                           & Periodic scale             & $\mathcal{LU} \, (10^{-3}, 10)$                             \\
$\eta_{\rm 1,FWHM}$ {[}$\rm m/s${]}        & Amplitude (FWHM)           & $\mathcal{LU} \, (10^{-2}, 10)$                        \\ \hline
\multicolumn{3}{l}{Instrument-dependent parameters (TESS)}                                                                             \\ \hline
$F_{\rm 0,S2}$ {[}$\rm e^{-}/s${]}          & TESS S2 flux offset        & $\mathcal{U} \, (-10^{2}, 10^{2})$                          \\
$F_{\rm 0,S29}$ {[}$\rm e^{-}/s${]}         & TESS S29 flux offset       & $\mathcal{U} \, (-10^{2}, 10^{2})$                          \\
$\sigma_{\rm TESS,S2}$ {[}$\rm e^{-}/s${]}  & TESS S2 jitter             & $\mathcal{U} \, (0, 10^{2})$                                \\
$\sigma_{\rm TESS,S29}$ {[}$\rm e^{-}/s${]} & TESS S29 jitter            & $\mathcal{U} \, (0, 10^{2})$                                \\ \hline
\multicolumn{3}{l}{Instrument-dependent parameters (ESPRESSO and HARPS)}                                                               \\ \hline
$\gamma_{\rm ESPR,RV}$ {[}$\rm m/s${]}     & ESPR. sys. RV              & $\mathcal{U} \, (13000, 16000)$                                   \\
$\gamma_{\rm HAR,RV}$ {[}$\rm m/s${]}      & HARPS sys. RV              & $\mathcal{U} \, (13\cdot10^{3}, 16\cdot10^{3})$                                   \\
$\gamma_{\rm ESPR,FWHM}$ {[}$\rm m/s${]}   & ESPR. sys. FWHM            & $\mathcal{U} \, (0, 10^{4})$                                    \\
$\gamma_{\rm HAR,FWHM}$ {[}$\rm m/s${]}    & HARPS sys. FWHM            & $\mathcal{U} \, (0, 10^{4})$                                    \\
$\sigma_{\rm ESPR,RV}$ {[}$\rm m/s${]}     & ESPR. RV jitter            & $\mathcal{U} \, (0, 10)$                               \\
$\sigma_{\rm HAR,RV}$ {[}$\rm m/s${]}      & HARPS RV jitter            & $\mathcal{U} \, (0, 10)$                               \\
$\sigma_{\rm ESPR,FWHM}$ {[}$\rm m/s${]}   & ESPR. FWHM jitter          & $\mathcal{U} \, (0, 10)$                               \\
$\sigma_{\rm HAR,FWHM}$ {[}$\rm m/s${]}    & HARPS FWHM jitter          & $\mathcal{U} \, (0, 10)$                               \\ \hline
\multicolumn{3}{l}{Orbital and planet parameters of an additional Keplerian}                                                           \\ \hline
$P_{\rm 2}$ {[}days{]}                      & Orbital period             & $\mathcal{U} \, (0, 5 \cdot 10^{2})$                        \\
$T_{\rm 0,2}$ {[}RJD{]}                     & Time of inf. conj.         & $\mathcal{U} \, (58467, 59855)$                             \\
$cos(w_{2}) \sqrt{e_{2}}$                   & Ecc. parametrization       & $\mathcal{U} \, (- \frac{\sqrt{2}}{2}, \frac{\sqrt{2}}{2})$ \\
$sin(w_{2}) \sqrt{e_{2}}$                   & Ecc. parametrization       & $\mathcal{U} \, (- \frac{\sqrt{2}}{2}, \frac{\sqrt{2}}{2})$ \\
$K_{2}$ {[}$\rm m/s${]}                    & RV semi-amplitude          & $\mathcal{U} \, (1, 10)$                         \\ \hline \hline
%\noalign{\vskip 1mm}
\end{tabular}
%\noalign{\vskip 1mm} 
\tablefoot{The $\mathcal{U}(a,b)$ and $\mathcal{LU}(a,b)$ symbols indicate uniform and log-uniform distributions, being $a$ and $b$ the lower and upper limits. The $\mathcal{G}(\mu, \sigma)$ and $\mathcal{ZTG}(\mu, \sigma)$ symbols indicate Gaussian and zero-truncated Gaussian distributions, being $\mu$ and $\sigma$ the mean and width of the distributions. RJD refers to the Reduced Julian Day (JD $-$ 2\,400\,000).}

\end{table}

\subsection{ESPRESSO and HARPS radial velocity analysis}
\label{subsec:rv_modeling}

We analyzed the ESPRESSO and HARPS RV data sets described in Sects.~\ref{sec:espresso_obs} and \ref{sec:harps_obs} through a model composed of three components: a Keplerian, which models planetary-induced RV signals, an instrumental component, which models the systemic velocity as measured by each instrument, and a GP, which models the RV correlated noise induced by the stellar activity. 

We implemented the Keplerian component through the Python package \texttt{radvel} \citep{2018PASP..130d4504F} by using the parametrization $\left\lbrace P, T_{0}, K, \sqrt{e} cos(w), \sqrt{e} sin(w)\right\rbrace$, being $P$ the orbital period of the planet, $T_{0}$ the time of inferior conjunction, $K$ the semi-amplitude, $e$ the orbital eccentricity, and $w$ the planetary argument of periastron. The instrumental component of our model consists of an offset that corresponds to the systemic radial velocity of the star as measured by each instrument ($\gamma_{ins})$. The periodograms of the spectroscopic data show that the stellar rotation induces significant activity-related RVs (Sect.~\ref{sec:espresso_obs}). We modeled this correlated noise through a GP with a quasiperiodic kernel \citep{2015ITPAM..38..252A,2016A&A...588A..31F}. The choice of this kernel is motivated by the fact that although stellar rotation is a periodic phenomenon, the activity-induced RV signals are quasiperiodic, given that active regions evolve; that is, they move on the stellar surface and appear and disappear throughout the magnetic cycle timescale. The quasiperiodic kernel depends on four hyperparameters ($\eta_{1}$, $\eta_{2}$, $\eta_{3}$, and $\eta_{4}$) and can be written in terms of the separation between data points $\tau = t_{i} - t_{j}$ as

\begin{equation}
    K_{QP} (\tau) = \eta_{1}^{2} \rm{exp} \left[ - \frac{\tau^{2}}{2\eta_{2}^{2}} - \frac{2sin^{2} \left(  \frac{\pi \tau}{\eta_{3}} \right)}{\eta_{4}^{2}}  \right].
\end{equation}

\noindent The hyperparameter $\eta_{1}$ scales with the amplitude of the stellar activity signal. $\eta_{3}$ corresponds to the main periodicity of the signal and it is considered a measure of the stellar rotation period \citep[e.g.,][]{2018MNRAS.474.2094A}. $\eta_{2}$ is the lengthscale of exponential decay. For two data points in the X-axis far from each other, the larger $\eta_{2}$ is, the more closely correlated those data points are \citep[e.g.,][]{2018MNRAS.474.2094A}; therefore, $\eta_{2}$ is considered as a measure of the timescale of growth and decline of the active regions \citep{2014MNRAS.443.2517H,2016A&A...588A..31F}. $\eta_{4}$ controls the amplitude of the $\rm sin^{2}$ term. The smaller $\eta_{4}$ is, data points separated by one orbital period will be much more closely correlated than those data points separated by a different period of time; thus, $\eta_{4}$ indicates the complexity of the harmonic content of the activity signal. Finally, in order to model the white noise not taken into account in our model, we included a jitter term per instrument ($\sigma_{ins}$) that we added quadratically to the uncertainties of our RV measurements. 

In order to obtain a more robust estimation of the model parameters, we included to our data set an activity indicator to be modeled jointly with the HARPS and ESPRESSO RVs. This approach consists of using two GP kernels with shared hyperparameters except the signal amplitude (hereafter we differentiate between $\eta_{\rm 1,RV}$ and $\eta_{\rm 1,indicator}$). This approach has been used in previous works and relies on the assumption that the variations on the activity indicators are caused by the stellar activity alone, and that their periodicity and coherence are the same as those of the activity component of the RV \citep[e.g.,][]{2020A&A...639A..77S,2020A&A...642A.121L,2022A&A...665A.154B}. From the time series and periodograms of Fig.~\ref{fig:gls_espresso}, we can perceive a certain correlation between the RVs and the activity indicators. We computed the strength of these correlations by means of the Pearson product-moment correlation coefficient \citep{doi:10.1080/00031305.1988.10475524}, obtaining the following results: 0.35 for the correlation with the FWHM, 0.20 with S-index, -0.07 with $\rm H_{\alpha}$, -0.22 with BIS, 0.02 with NaD, and 0.21 with the contrast of the CCF. Showing a moderate degree of correlation, we decided to use the FWHM of the CCF as the activity indicator in our modeling.

We explored the constraining capacity of the HARPS data, motivated by their shorter time coverage and smaller precision than ESPRESSO data. To do so, we ran an MCMC fit considering a circular Keplerian model ($e$ = 0, $w$ = 0) to the ESPRESSO and ESPRESSO+HARPS data sets. Besides, for each data set, we performed a complementary fit with zero planets in order to obtain the significance against the null hypothesis. To do so, we used the Perrakis algorithm through the \texttt{bayev} implementation \citep{2016A&A...585A.134D} to compute the logarithm of the Bayesian evidence of the model ln($\mathcal{Z}$) based on the 15$\%$ of the final flattened chain. In the same way as in the previous section, we used wide uniform priors for all the parameters, except for those for which we have prior information (see Table \ref{table:priors} for the detailed prior distributions). As a result, the inferred parameters are compatible in both tested data sets (ESPRESSO and ESPRESSO+HARPS). However, we obtain a larger Bayesian evidence against the null hypothesis when using the ESPRESSO+HARPS data set: $\rm \Delta ln(\mathcal{Z})_{ESPR} $ = +16.5, $\rm \Delta ln(\mathcal{Z})_{ESPR+HARPS} $ = +17.7. This way, the inclusion of HARPS data slightly increases the significance of the detection. Consequently, due to the higher evidence, together with the advantage of having a longer time span to search for additional signals, we used the ESPRESSO+HARPS data set for the subsequent analysis. 

\begin{figure}
    \centering
    \includegraphics[width=\columnwidth]{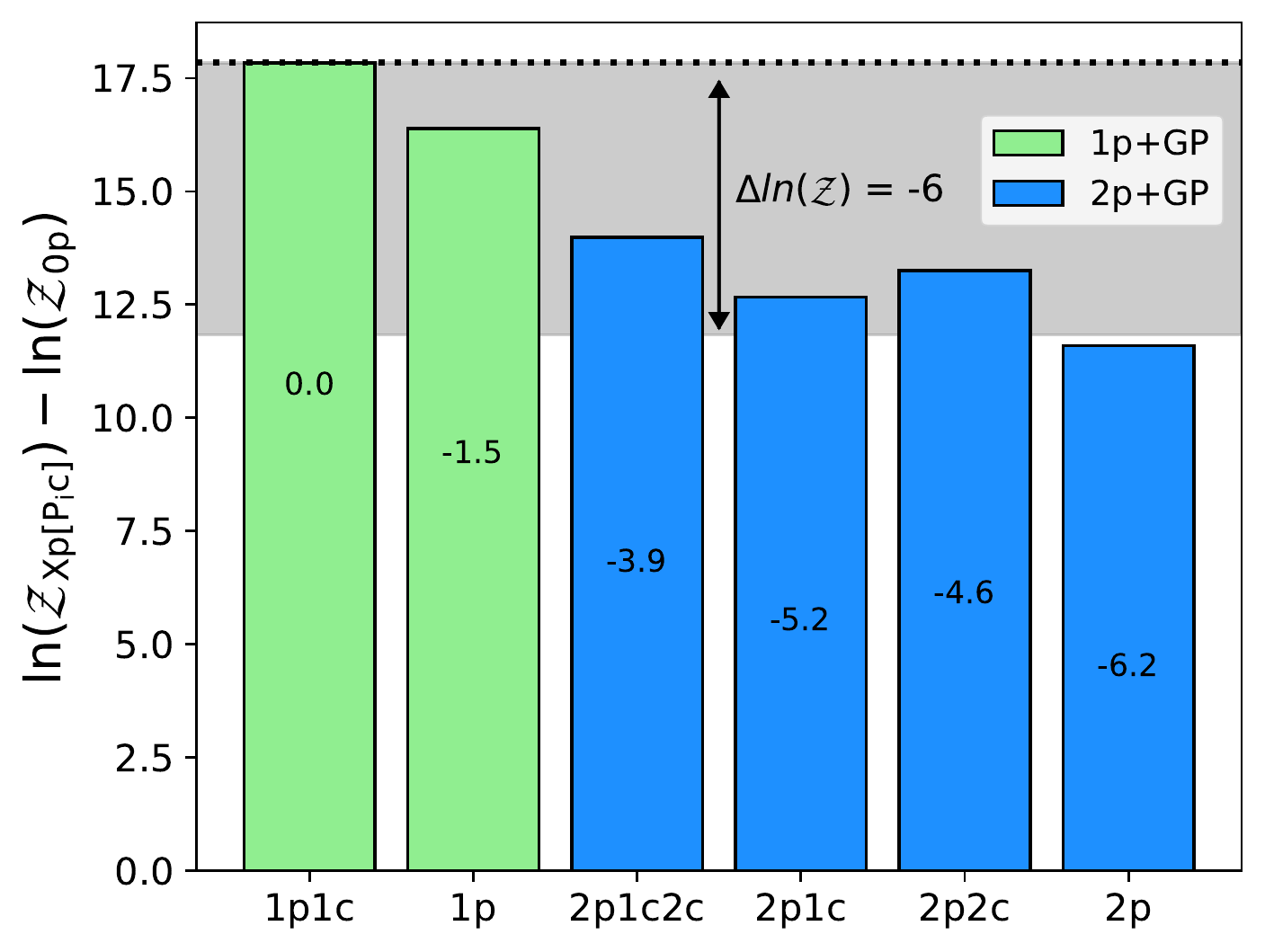}
    \caption{Differences of the log-evidences of different models (labeled in the X axis) and the 0 planet model, tested on the ESPRESSO+HARPS data set. The gray shaded region indicates the $\rm 0 \geq \Delta ln \mathcal{Z} \geq$ -6 region from the largest evidence model (1p1c).} 
    \label{fig:log-evidenes}
\end{figure}

We tested six different models in order to assess whether we can detect additional planetary signals, and to select the simplest model that best represents our data. The models involve one and two planets in all the possible combinations of circular and eccentric orbits. To identify the models, we use the nomenclature Xp[$\rm P_{i}c$], where X is the total number of planets considered in the system, and $\rm P_{i}$ indicates which planets have assumed circular orbits. In this work, planet "1" corresponds to TOI-244.01, and planet "2" to an additional planet without prior orbital information (see Table \ref{table:priors}). This way, we have tested the following models: 1p1c, 1p, 2p1c2c, 2p, 2p1c, and 2p2c. In Fig.~\ref{fig:log-evidenes}, we compare the obtained difference of the log-evidence of each model and the log-evidence of the 0 planet model $\rm  ln(\mathcal{Z}_{\rm Xp[\rm P_{i}c]}) - ln(\mathcal{Z}_{0p})$. We assume that a difference of +6 (i.e., $\rm \Delta ln \mathcal{Z}$ > 6) indicates strong evidence in favor of the largest evidence model \citep{2008ConPh..49...71T}. For our data set, there is no model meeting the condition when compared to a simpler model,  so we chose the simplest model as the one that best represents our data set, which, in this case, coincides with the one with the largest evidence: the 1 planet model in a circular orbit (1p1c). 

\subsection{Joint analysis}
\label{subsec:joint_analysis}

\begin{figure*}
    \centering
    \includegraphics[scale = 0.45]{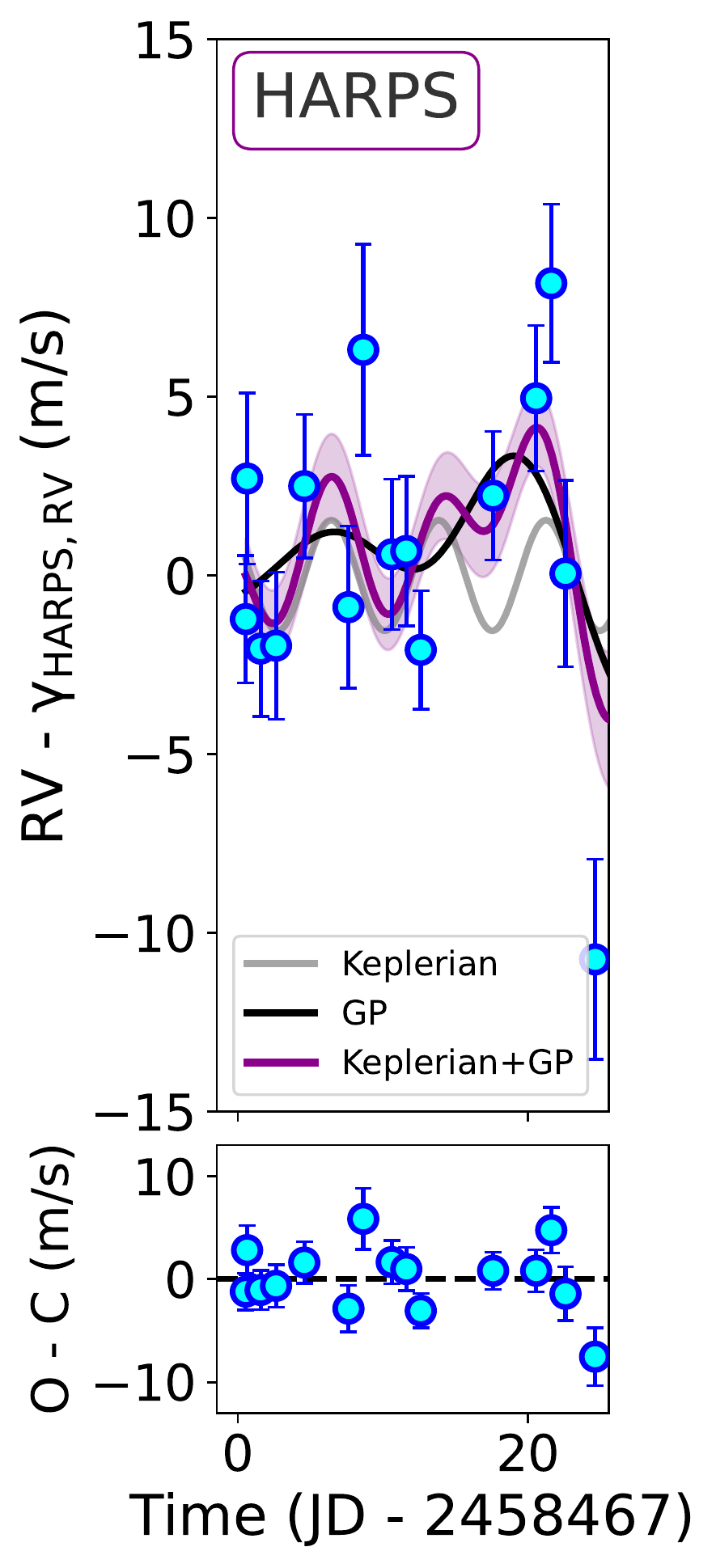}
    \includegraphics[scale = 0.45]{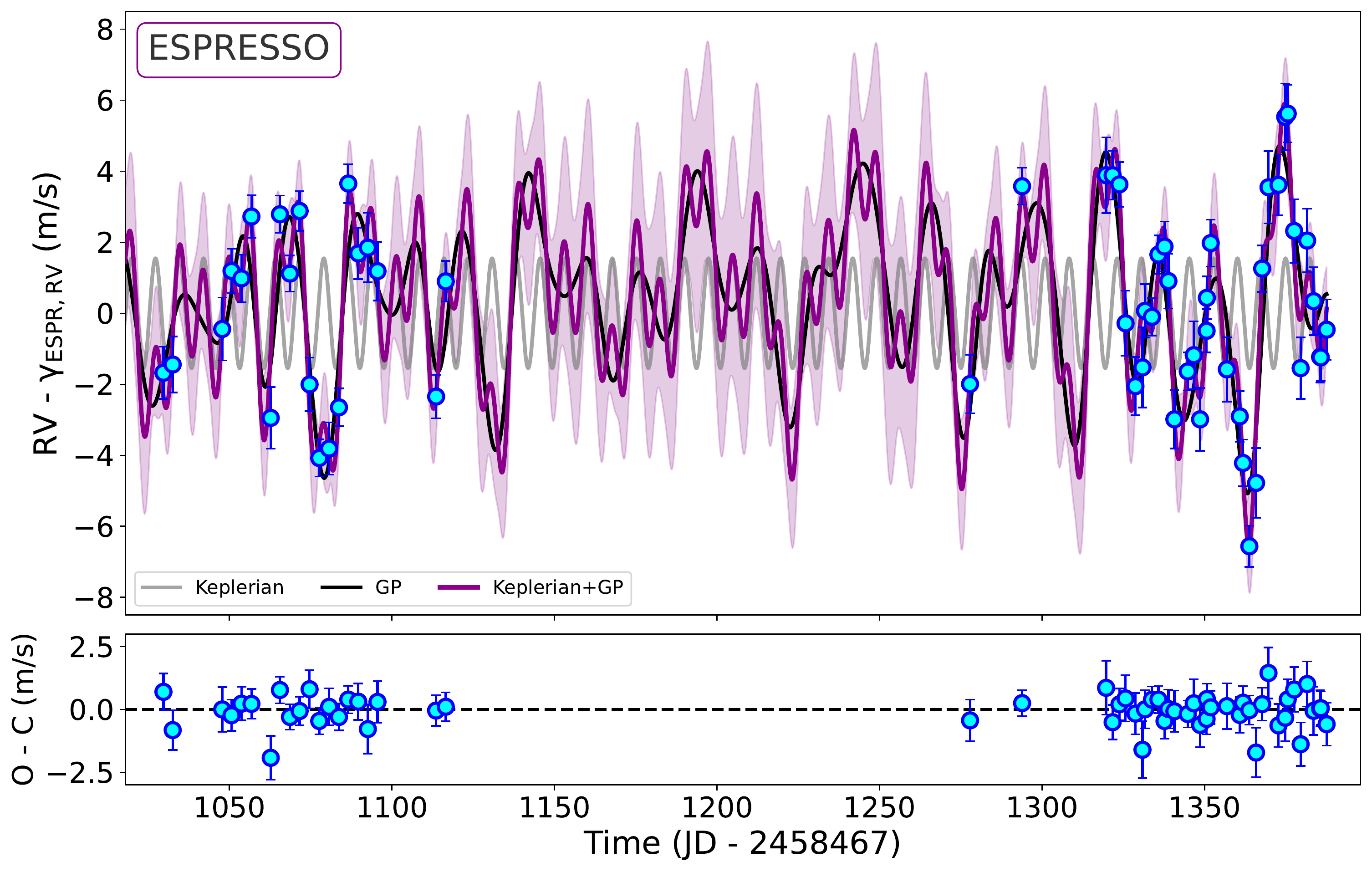}
    \includegraphics[scale=0.63]{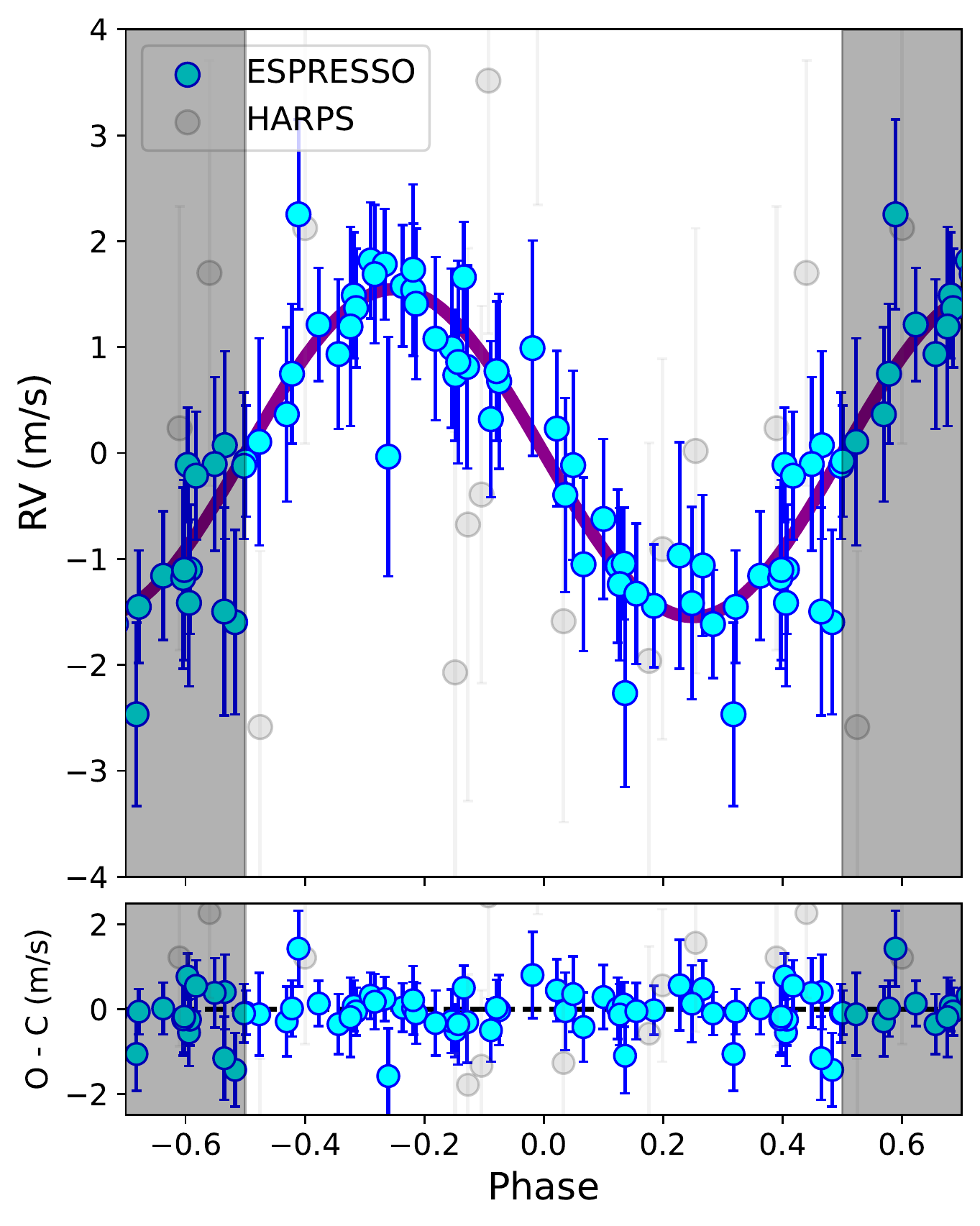}
    \includegraphics[scale=0.63]{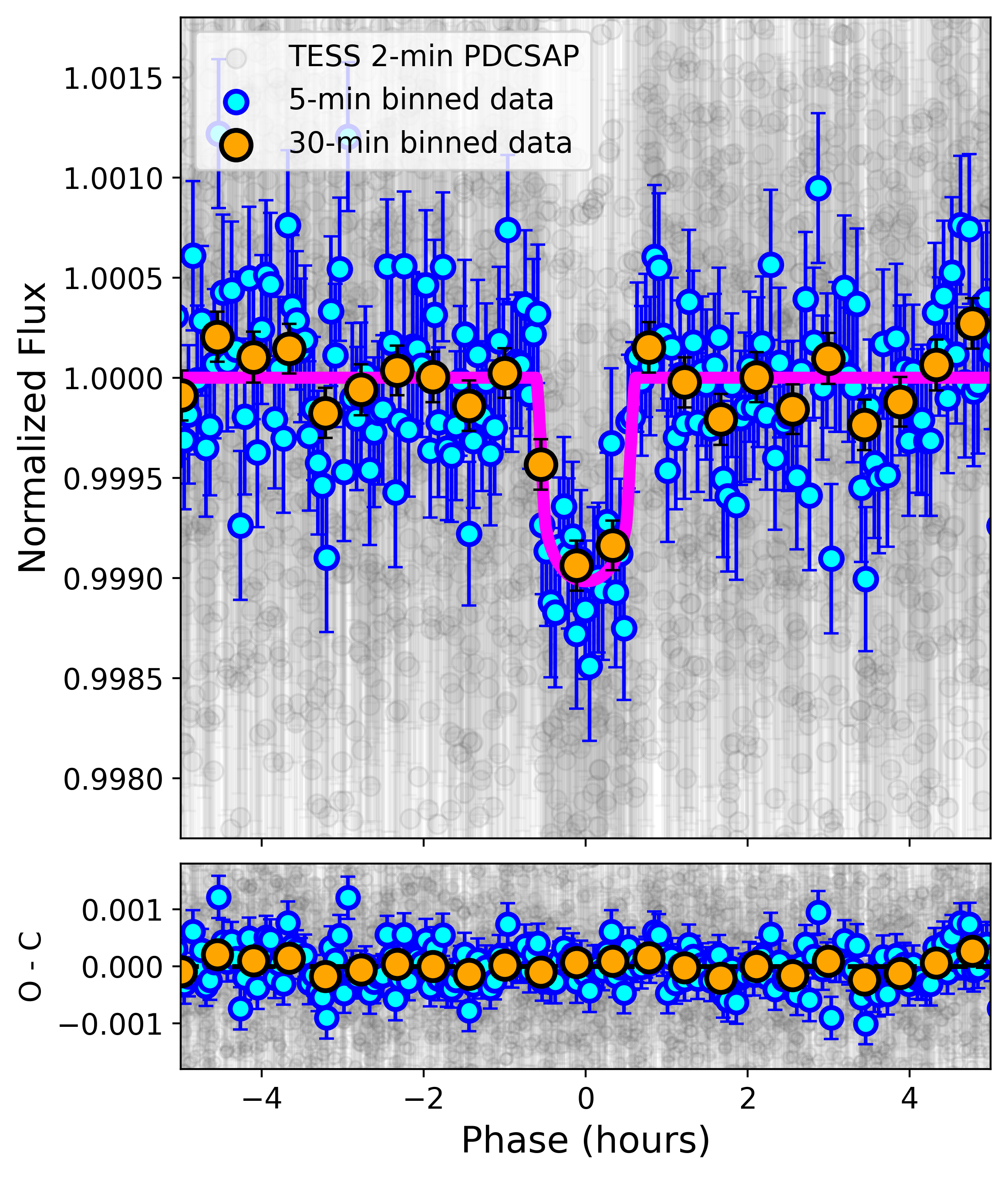}
    \caption{\textit{Top panel:} ESPRESSO and HARPS radial velocities of GJ 1018. The purple solid line indicates the median posterior global model, and the black and gray solid lines indicate the median posterior GP and Keplerian models, respectively. The shaded purple region indicates the 68.7$\%$ confidence interval of the global model. \textit{Bottom panel:} ESPRESSO and HARPS RVs (\textit{left}) and TESS photometry (\textit{right}) subtracted from their corresponding GP components and folded to the orbital period of TOI-244 b. The solid lines indicate the median posterior models.}
    \label{fig:final_plots}
\end{figure*}

We inferred the final parameters of the system by modeling jointly the TESS photometry (same way as in Sect.~\ref{subsec:TESS_analysis}) and ESPRESSO+HARPS radial velocities (same way as in Sect.~\ref{subsec:rv_modeling}; considering the 1p1c model). In Fig.~\ref{fig:final_plots} (top panels), we show the complete RV data set together with the median posterior Keplerian+GP model. In the lower panels, we include the ESPRESSO+HARPS RVs and TESS photometry folded to the inferred orbital period together with the median posterior model, being both subtracted from their corresponding GP components.  In Table \ref{tab:planet_params}, we present the main parameters of TOI-244 b. The complete list of fitted parameters can be found in Table \ref{tab:final_derived_params}. 

\begin{table}[]
\caption{TOI-244 b main parameters obtained from the fitted parameters in the joint analysis (Sect. \ref{sec:analysis_results}, Table \ref{tab:final_derived_params}) and the computed stellar parameters (Sect. \ref{sec:stellar_charact}, Table \ref{tab:stellar_params}).}
\label{tab:planet_params}
\renewcommand{\arraystretch}{1.3}
\setlength{\tabcolsep}{2pt}
\begin{tabular}{lll}
\hline \hline
Symbol                            & Parameter          & Value                     \\ \hline
$P$ {[}days{]}                             & Orbital period              & $7.397225^{+0.000026}_{-0.000023}$ \\
$T_{0}$ {[}RJD{]}                           & Time of inf. conj.    & 58357.3627 $\pm$ 0.0020          \\
$R_{\rm p}$ {[}$\rm R_{\oplus}${]}         & Planet radius               & 1.52 $\pm$ 0.12                    \\
$M_{\rm p}$ {[}$\rm M_{\oplus}${]}         & Planet mass                 & 2.68 $\pm$ 0.30                    \\
$\rho_{\rm p}$ {[}$\rm g \cdot cm^{-3}${]} & Planet density              & 4.2 $\pm$ 1.1                      \\
$a / R_{\star}$                            & $a$ relative to $R_{\star}$ & 28.1 $\pm$ 1.8                     \\
$a$ {[}AU{]}                               & Orbit semi-major axis       & 0.0559 $\pm$ 0.0013                \\
$S$ {[}$\rm S_{\oplus}${]}                 & Insolation flux             & 7.3 $\pm$ 0.4                      \\
$T_{\rm eq}$ {[}K{]}                       & Equilibrium temperature     & 458 $\pm$ 20                       \\
$g$ {[}$\rm m \cdot s^{-2}${]}             & Surface gravity             & 11.3 $\pm$ 2.2                     \\
$b$                                        & Impact parameter            & 0.61 $\pm$ 0.31                    \\
$\delta$ {[}ppt{]}                         & Transit depth               & 1.06 $\pm$ 0.12                    \\
$T_{\rm 14}$ {[}hours{]}                   & Transit duration            & 1.7 $\pm$ 0.5                      \\ \hline \hline
\end{tabular}
\end{table}

\section{Discussion}
\label{sec:discussion}

\subsection{Detectability of additional planets in the RV data set}

\begin{figure}
    \centering
    \includegraphics[width = \columnwidth]{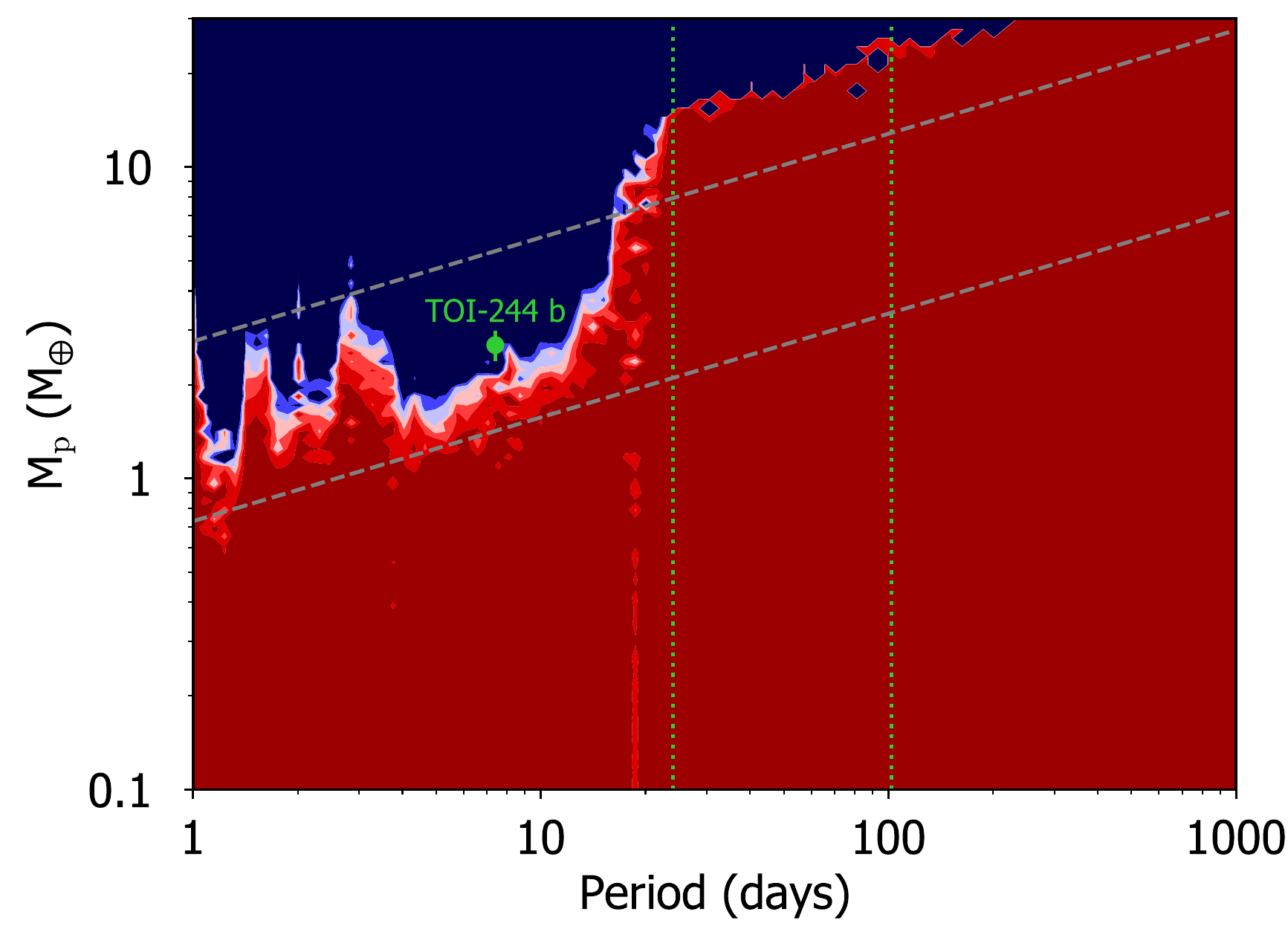}
    \caption{Detectability map for the ESPRESSO+HARPS data set obtained from the injection-recovery test. The redder values correspond to false alarm probabilities (FAPs) larger than 1\% while the blue color indicates that the signal is recovered with FAP $<1$\%. The location of TOI-244\,b is indicated by the green symbol. The green dotted vertical lines indicate the optimistic HZ around GJ 1018 according to \citet{2013ApJ...765..131K}. The bottom dashed line indicates the limit imposed by the median radial velocity uncertainty of the original data set (i.e., before removing the planet and activity contribution) while the upper dashed line indicates the limit corresponding to the rms of the data.}
    \label{fig:IR_RV}
\end{figure}

We constrained the possibility of additional planets in the radial velocity data set by following the injection-recovery procedure described in \citet{2022arXiv221207332S}. In brief, we first removed the radial velocity contribution from the confirmed planet TOI-244\,b. The residuals from this include the activity of the star. We then injected a Keplerian model from a grid of periods (from 1  to 1000 days, 100 bins in log-space) and planet masses (from 0.1 to 30 $\rm M_{\oplus}$, 100 bins in log-space) and random phases. We assumed coplanar orbits with TOI-244\,b. From the resulting RVs we computed the activity model using the hyper-parameters determined in the joint analysis section and removed its contribution. We then computed the false alarm probability of the power at the injected period. We consider the injected planet signal is detected if the FAP is below 1\%. Figure~\ref{fig:IR_RV} shows the result of this exercise. All injections implying a root mean square (rms) larger than 1.5 times the original rms are assumed as detected. We repeated the process five times and average all the iterations to obtain the final detectability matrix shown in Fig.~\ref{fig:IR_RV}. The current ESPRESSO+HARPS data set allows us to detect planets down to 2~$\rm M_{\oplus}$ at orbital periods up to around 22 days.

It is of special interest to constrain the presence of planets in the habitable zone (HZ) of GJ 1018, given that late K-dwarfs and early M-dwarfs represent an ideal trade-off between detectability and true habitability of their planets \citep[see the KOBE experiment;][]{2022A&A...667A.102L}. For the HZ of GJ 1018, which is located in the period range 24-102 days \citep{2013ApJ...765..131K}, the current data allows us to discard planets with masses above 20~$\rm M_{\oplus}$.

\subsection{Internal structure of TOI-244 b}
\label{sec:internal_structure}

\subsubsection{TOI-244 b in the mass-radius diagram}
\label{sec:M-R}

\begin{figure*}
    \centering
    \includegraphics[width=\columnwidth]{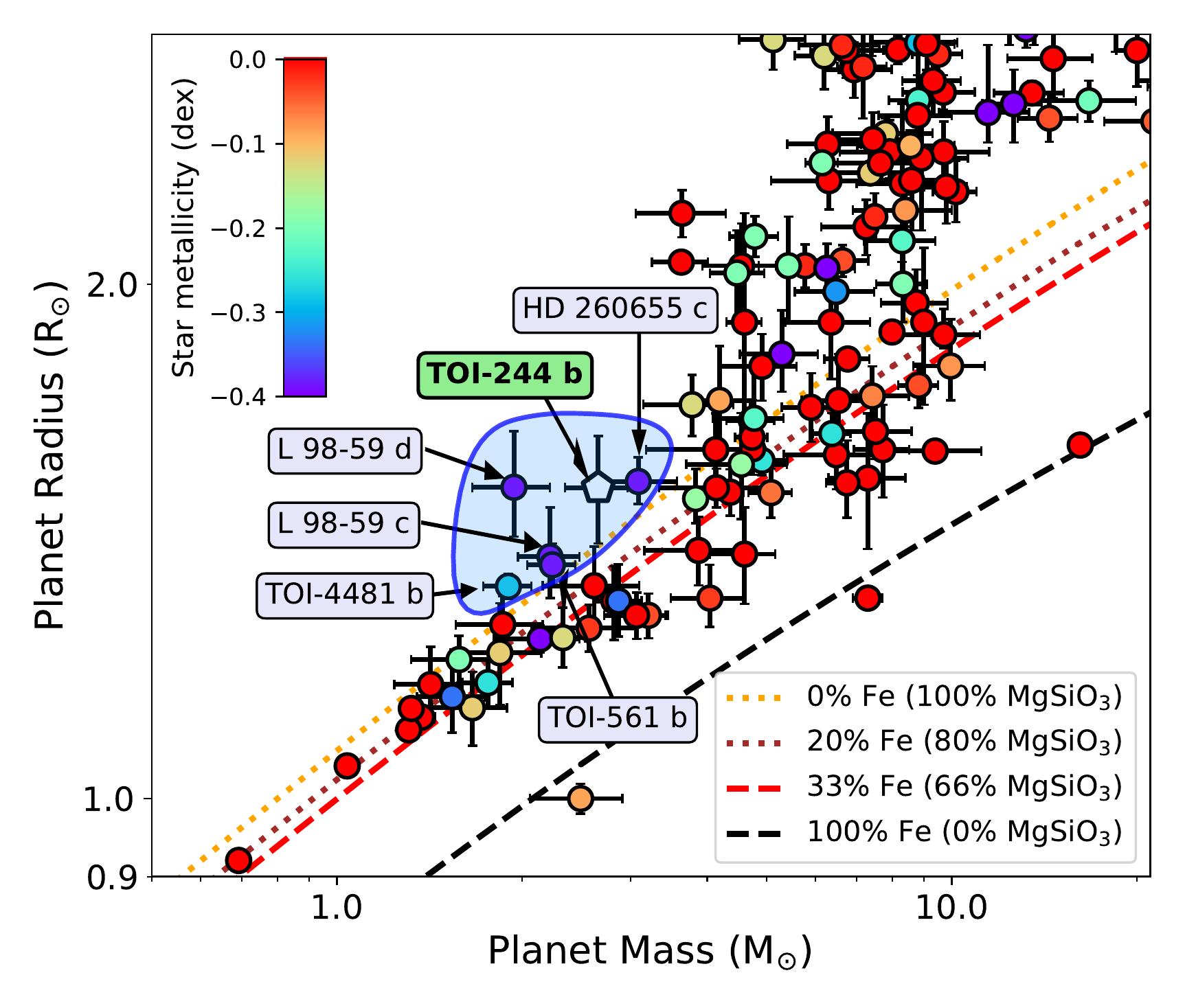}
    \includegraphics[width=\columnwidth]{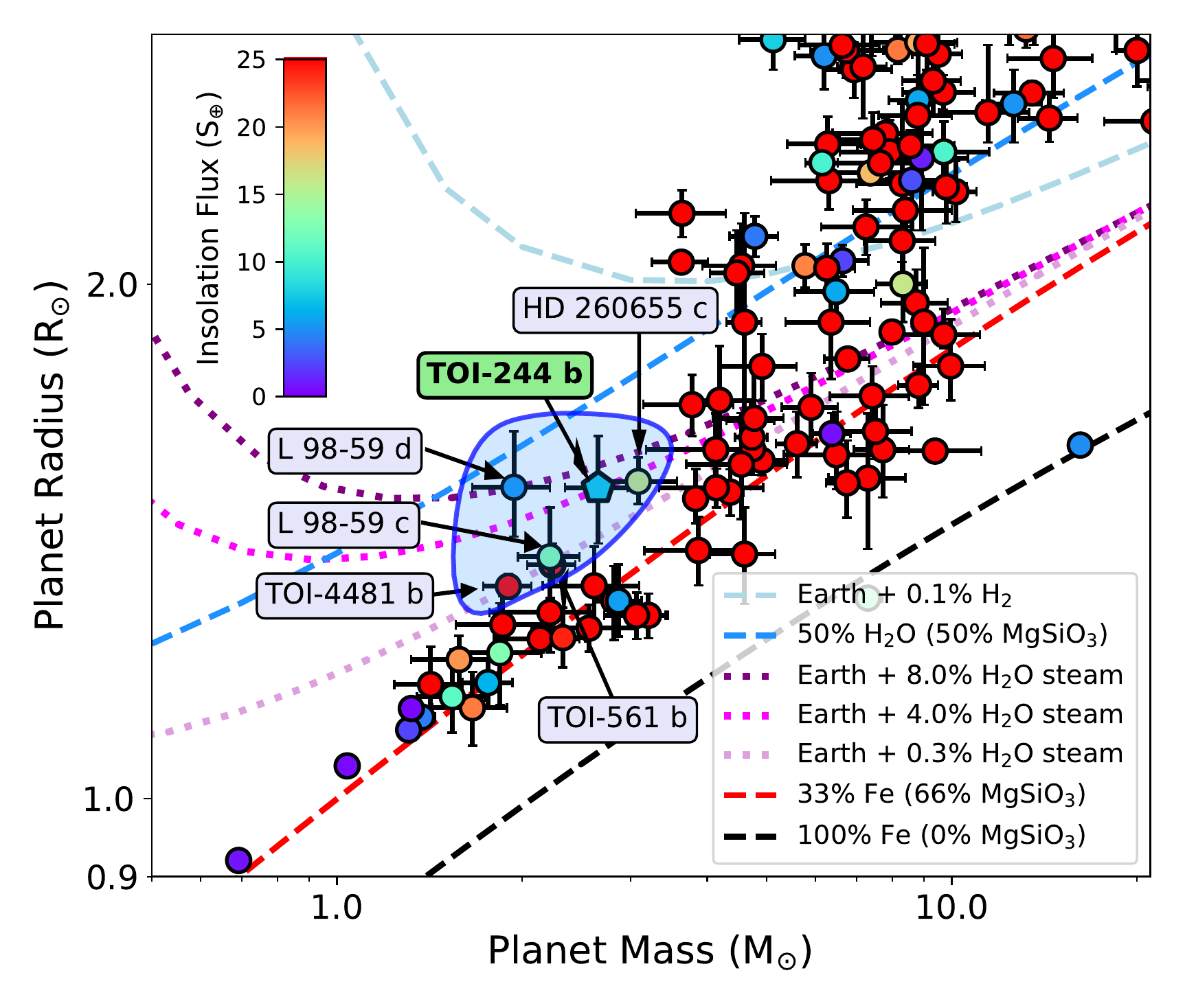}
    \caption{Mass-radius diagrams for the small planet sample with measured dynamical masses with a precision better than 20$\%$. \textit{Left:} Color coding indicates the stellar host metallicities. TOI-244~b is not filled with any color since we get inconsistent metallicities with the different methods used (see Sect.~\ref{sec:steparsyn}). The dashed and dotted lines correspond to theoretical interior models that consider different mass percentages of Fe and $\rm MgSiO_{3}$ \citep{2016ApJ...819..127Z,2019PNAS..116.9723Z}. \textit{Right:} Color coding indicates the stellar insolation received by the planets. The dashed lines correspond to theoretical models from \citet{2019PNAS..116.9723Z}, which consider planets without a significant amount of volatiles (back and red lines), planets with condensed water (dark blue), and planets with $\rm H_{2}$ atmospheric envelopes (light blue). The dotted lines correspond to the \citet{2020A&A...638A..41T} theoretical models for Earth-like planets with $\rm H_{2}O$-dominated atmospheres. These plots have been prepared using \texttt{mr-plotter}, which is available at \url{https://github.com/castro-gzlz/mr-plotter}.}
    \label{fig:M-R}
\end{figure*}

In Fig.~\ref{fig:M-R}, we plot the radius versus the mass for all the known planets from the NASA Exoplanet Archive with measured dynamical masses with a precision better than $20\%$. TOI-244 b is located in an unpopulated region, significantly separated (1.8$\sigma$ and 6.7$\sigma$ in radius and mass respectively) from the Earth-like composition curve \citep[33$\%$ Fe  and 66$\%$ $\rm MgSiO_{3}$ in mass;][]{2019PNAS..116.9723Z}, where rocky planets typically reside. Being located above that curve, TOI-244 b has a lower density than expected. We highlight other five planets in this particular region of the parameter space: TOI-561\,b \citep{2021MNRAS.501.4148L,2021AJ....161...56W,2023AJ....165...88B}, L 98-59~c and d \citep{2021A&A...653A..41D}, HD 260655 c \citep{2022A&A...664A.199L}, and TOI-4481\,b \citep{2023arXiv230106873P}. The existence of these planets might be explained by the presence of lighter elements than those expected for Earth-like compositions. This translates into three possible scenarios: planets with a scarcity or complete absence of iron on their cores (being thus practically composed of silicates), planets with a significant amount of volatile elements, or planets in which both scenarios coexist and have a significant effect on the density of the planet. In the following, we discuss each scenario separately.

\subsubsection{TOI-244 b as an iron-free planet}
\label{sec:iron_free}

\begin{figure}
\includegraphics[scale = 0.57]{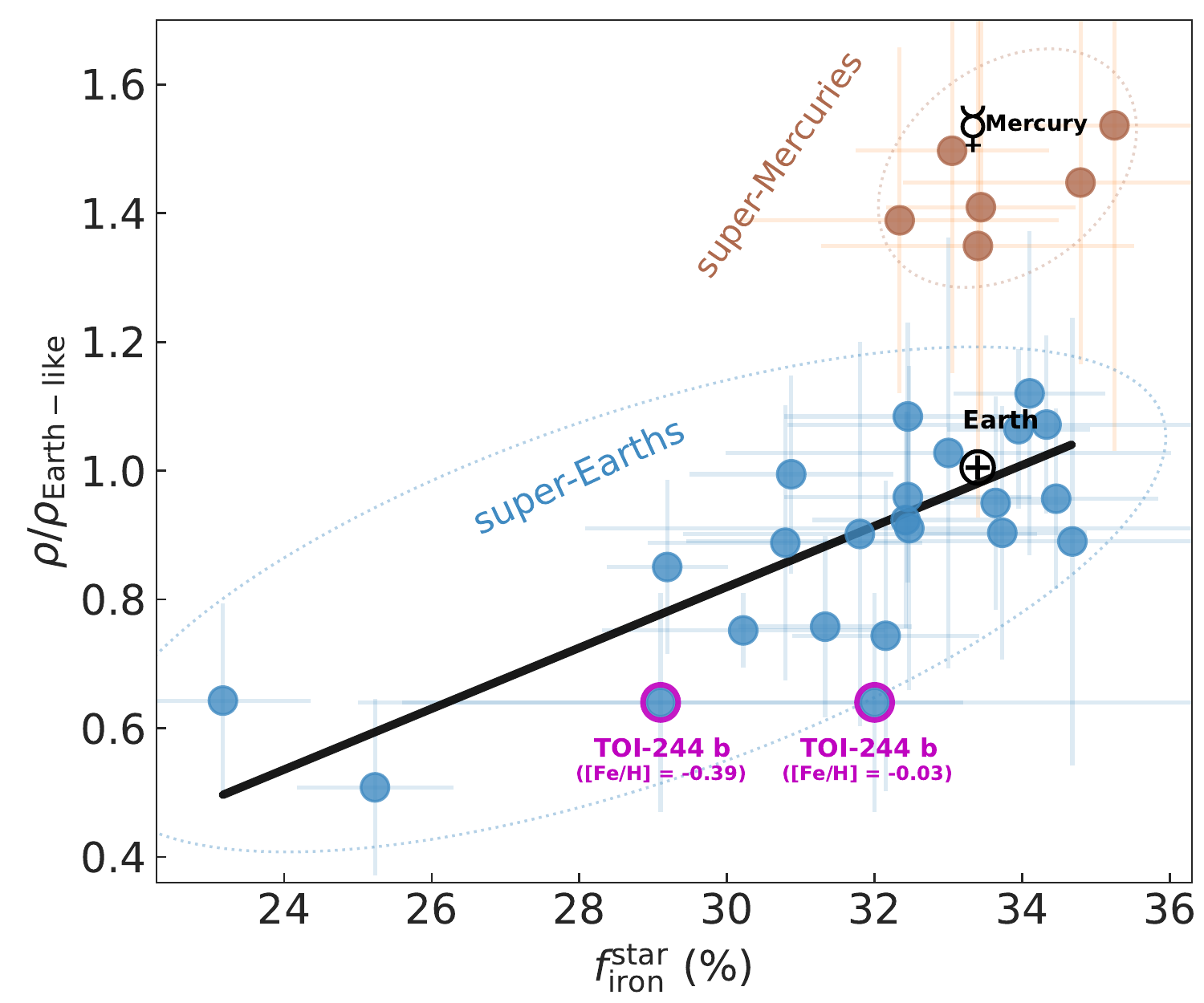}
    \caption{Planet density normalized by the expected density of an Earth-like composition \citep{2017A&A...597A..37D} versus the estimated iron-to-silicate mass fraction of the original protoplanetary disk \citep{2015A&A...580L..13S,2017A&A...608A..94S} for the sample studied in \citet{2021Sci...374..330A} and for TOI-244~b considering the metal-poor and solar metallicity characterizations (see Sect.~\ref{sec:steparsyn}).  The Earth and Mercury are indicated with their respective symbols in black.  All error bars
show 1$\sigma$ uncertainties.}

    \label{fig:density_vs_iron}
\end{figure}

The abundance of refractory
elements such as Mg, Si, and Fe of solar-type stars is considered a proxy of the composition of
the initial protoplanetary disk from which stars and planets were formed  \citep[e.g.,][]{2015A&A...577A..83D,Unterborn_2016}. In a recent study, \citet{2021Sci...374..330A} used stellar abundances of 22 stars to estimate the iron-to-silicate mass fractions of their initial protoplanetary disks ($f^{\rm star}_{\rm iron}$)  based on the stoichiometric model presented by \citet{2015A&A...580L..13S,2017A&A...608A..94S}. The authors found a relationship between $f^{\rm star}_{\rm iron}$ and the densities and iron mass fractions of their hosted planets, disclosing a compositional link between the rocky planets and their host stars. In other words, stars with lower $f^{\rm star}_{\rm iron}$ have lighter planets with lower iron-to-silicate mass fractions\footnote{The sample analyzed in \citet{2021Sci...374..330A} is composed of highly irradiated planets, which are assumed to have negligible atmospheres, similar to the scenario we are discussing.}. In Fig.~\ref{fig:density_vs_iron}, we plot the planet density normalized by the expected density of an Earth-like composition \citep[$\rho$/$\rho_{Earth-like}$;][]{2017A&A...597A..37D} as a function of $f^{\rm star}_{\rm iron}$ for the same planet sample as in \citet{2021Sci...374..330A} and for TOI-244~b. For the metal-poor stellar characterization, we estimate $f^{\rm star}_{\rm iron}$ = 29.2 $\pm$ 4.2 $\%$, which would place TOI-244 b in agreement with the correlation. For the solar metallicity characterization, we estimate $f^{\rm star}_{\rm iron}$ = 32.0 $\pm$ 6.5 $\%$, which would place TOI-244~b slightly more deviated from the trend.

Overall, the low density of the planet together with the possibility of having a relatively low $f^{\rm star}_{\rm iron}$ (in the metal-poor scenario) suggests that TOI-244~b might have an iron mass fraction smaller than that expected for an Earth-like planet. However, according to Fig.~\ref{fig:M-R} (left), TOI-244~b has a density below what is expected for a 100$\%$ silicate composition at the 1$\sigma$ level. Hence, the most likely scenario to explain the low-density of TOI-244~b is that it has a nonnegligible amount of volatile elements in its composition. The color coding of Fig.~\ref{fig:M-R} (left) represents the metallicity of the stellar hosts, showing that the growing population of low-density super-Earths tend to be formed around metal-poor stars (see Sect.~\ref{sec:density_metallicity} for a more extended discussion). Interestingly, the presence of volatiles in planets orbiting metal-poor stars is very expected, given that the building blocks of the original protoplanetary disks that form metal-poor stars are expected to have lower iron and higher water content than the expected for the disks that form stars with solar metallicities \citep{2017A&A...608A..94S}.

\subsubsection{TOI-244 b as a volatile-rich planet}
\label{volatile_rich}

The most abundant volatiles on protoplanetary disks are $\rm H_{2}$/He and $\rm H_{2}O$  \citep[e.g.,][]{2003ApJ...591.1220L,2007ApJ...667..303T}. In interior modeling, it is common to consider a hydrogen-rich envelope, which is representative of a primordial atmosphere. However, water is commonly considered in condensed form (either liquid or solid), based on traditional mass-radius relationships \citep[e.g.,][]{2007ApJ...669.1279S, 2012ApJ...744...59S,2016ApJ...819..127Z}. In recent work, \citet{2020A&A...638A..41T} computed new mass-radius relationships taking into account that for $\rm H_{2}O$-rich rocky planets more irradiated than the runaway greenhouse limit (i.e $S$ > 1.1 $\rm S_{\oplus}$), water is unstable in condensed form, so it would form a thick $\rm H_{2}O$-dominated atmosphere. Hence, given the much lower density of water vapor than liquid or solid water, the water content of highly irradiated $\rm H_{2}O$-rich planets is expected to be dramatically lower than the commonly computed ones \citep{2020ApJ...896L..22M}. In the right-hand mass-radius diagram of Fig.~\ref{fig:M-R}, we include a theoretical composition model consistent with Earth-like planets with $\rm H_{2}$-dominated atmospheres contributing 0.1$\%$ to the total planetary mass \citep{2019PNAS..116.9723Z}, showing that a small amount of hydrogen would significantly increase the radius of a rocky planet. We also include three theoretical models of Earth-like planets with an additional $\rm H_{2}O$-dominated atmosphere \citep{2020A&A...638A..41T}. In this case, a 4$\%$ water mass fraction would reproduce the observed density of TOI-244 b.  In addition, we include the 50$\%$ $\rm H_{2}O$-condensed and 50$\%$ $\rm MgSiO_{3}$ theoretical model from \citet{2019PNAS..116.9723Z}, which is widely used to describe planets in the higher mode of the bimodal radius distribution (2 $\rm R_{\oplus}$ < $R_{\rm p}$ < 4 $\rm R_{\oplus}$). This model allows us to illustrate the huge amount of condensed water ($\sim$20$\%$ in mass) that would be needed to explain the low density of TOI-244 b, in contrast with the 4$\%$ mass fraction when considering steam. In this diagram, the color coding indicates the isolation fluxes received by the planets. Interestingly, TOI-244~b receives a lower insolation than the typically received by planets over the Earth-like composition curve, which could be interpreted as a hint that it could have maintained an atmospheric envelope (we discuss a possible density-insolation relation on a population level in Sect.~\ref{sec:density_insolation}). To infer whether the atmosphere of TOI-244~b could be a primary \ce{H2}-dominated atmosphere, or a secondary outgassed potentially water-dominated atmosphere \citep{Kite2020pnas}, we estimated the amount of material that could be removed by atmospheric loss processes such as photoevaporation and Jeans escape.

Planets in close proximity to their host stars are exposed to high levels of X-ray and extreme UV radiation (XUV), which can lead to partial or total loss of a hydrogen envelope by photoevaporation. If the escape is energy-limited, the atmospheric mass loss rate by photoevaporation can be written as \citep{Erkaev2007,2013ApJ...775..105O}
\begin{equation}
    \dot{M} = \epsilon \frac{\pi F_\mathrm{XUV} R_\mathrm{p}^3}{G M_\mathrm{p}}, \label{eq:atmospheric-loss}
\end{equation}
where $F_\mathrm{XUV}$ is the XUV flux received by the planet, $G$ the gravitational constant and $\epsilon$ 
is an efficiency parameter. Stars emit the most XUV at a young age, during the so-called saturation regime \citep{Pizzolato2003}, which is particularly long for M-dwarfs. This makes photoevaporation extremely efficient in the vicinity of M-dwarfs at a young age. Using the analytical fits of \cite{Sanz-Forcada2011}, we estimated that GJ 1018 emitted $3.9\times 10^{29}$ erg$\cdot$s\textsuperscript{-1} of XUV during its saturation regime, that lasted for $\tau_\mathrm{sat}=350$ Myr. This translates into a mass loss rate during the saturation regime of \rm $\rm 0.074~M_{\oplus}$/Gyr, during which the planet lost $\rm 0.026~M_{\oplus}$ of hydrogen. Following the approach of \cite{Aguichine2021}, we integrated Eq. \ref{eq:atmospheric-loss} in time assuming that, at first order, the mass and radius of the planet remain roughly constant, and only $F_{\mathrm{XUV}}$ varies in time, following the fit of \cite{Sanz-Forcada2011}. We then find that, in total, TOI-244 b could have lost $\rm 0.1~M_{\oplus}$ of hydrogen by photoevaporation. This value is most likely a lower estimate, since an initially greater \ce{H2} content would produce a larger planetary radius, greatly increasing the mass loss rate (see Eq. \ref{eq:atmospheric-loss}). This estimate is consistent with more refined models of hydrogen mass loss. For example, the work by \cite{Rogers2023} also predicts that planets with core masses $\rm <3~M_{\oplus}$ at $T_\mathrm{eq}=500$ K are entirely stripped of their envelopes, if the latter are made of pure \ce{H2}.

Despite the efficient loss of hydrogen, recent studies indicate that secondary atmospheres, made of a mixture of volatile gases, can be restored by outgassing from the magma after the photoevaporation phase \citep{Kite2020pnas,Tian2023}. The equilibrium temperature of TOI-244 b is greater than that of the Earth, but so is its surface gravity. The characteristics of TOI-244 b are such that its Jeans escape parameter $\Lambda_\mathrm{p} = GM_\mathrm{p}m_\mathrm{H}/(k_\mathrm{B}T_\mathrm{eq}R_\mathrm{p})$ \citep{2017A&A...598A..90F} has a value consistent with the Jeans escape parameter of the Earth within the measurements uncertainty range: $\Lambda_\mathrm{p}/\Lambda_\oplus = 1.07^{+0.28}_{-0.23}$. 

Our computations suggest that even if hydrogen was initially present in the secondary atmosphere of TOI-244 b, it has possibly been removed by Jeans escape, as it happened on Earth \citep[see][]{Catling2017book}. Both Jeans escape and photoevaporation appear to be efficient mechanisms to entirely remove hydrogen from the atmosphere of TOI-244~b. This favors the modeling of its interior with an atmosphere of high mean molecular weight volatiles, such as water vapor. However, we cannot rule out the possibility that its atmosphere contains other molecules, for instance carbon-rich species such as \ce{CH4}, CO or \ce{CO2}. Spectroscopic measurements are necessary to break the compositional degeneracy, but the bulk water content can be used as a proxy for all heavy volatile species (\ce{CH4}, \ce{O2}, CO, \ce{CO2}, etc).

\subsubsection{Internal structure modeling}
\label{sec:final_internal_strucutre}

We performed a retrieval on the mass and radius data shown in Table \ref{tab:planet_params} with a 1D interior-atmosphere model to estimate TOI-244~b's compositional parameters. Our interior model is stratified in three layers: a Fe-rich core, a silicate-dominated mantle \citep{Brugger16,Brugger17}, and a water layer (hydrosphere). Given the irradiation conditions of TOI-244 b, volatiles such as water cannot condense out, and is therefore in gaseous and supercritical phases \citep{Mousis2020}. Our interior structure model takes into account the low density of these phases of water by using an equation of state adequate in this region of the water phase diagram. In addition, we establish the coupling interface between the interior and the atmosphere at 300 bar. The temperature at this pressure constitutes the boundary condition for our interior model, and it is calculated self-consistently by a k-correlated atmospheric model. Furthermore, the atmospheric thickness is calculated by the atmospheric model by integrating the hydrostatic equilibrium equation, and then added to the interior model's radius to obtain the total planetary radius \citep{Acuna21,2023arXiv230501250A}.

The two compositional parameters that are free in our MCMC retrieval are the core mass fraction (CMF) and water mass fraction (WMF). The mean and uncertainties of their 1$\sigma$ intervals are shown in Table \ref{tab:mcmc_interior}. Our observable parameters are the mass and the radius. The Fe/Si mole ratio is not considered as an observable, not only because of the difficulty of obtaining an accurate metallicity for GJ 1018, but also because the assumption that the composition of rocky planets reflects the stellar one
is being questioned \citep{2021Sci...374..330A}. Our CMF is compatible within uncertainties with that of the Earth, as well as with the derived CMF distribution of a sample of Super-Earths, between 0.1 and 0.5 \citep{Plotnykov20}. The 1$\sigma$ confidence interval of the WMF of TOI-244 b spans from 0.04 to 0.20, being in the transition between super-Earths (WMF < 0.05) and sub-Neptunes (WMF > 0.20). Overall, our internal structure modeling suggests that TOI-244~b has a $479^{+128}_{-96}$ km thick hydrosphere over a 1.17~$\pm$~0.09~$\rm R_{\oplus}$ solid structure composed of a Fe-rich core and a silicate-dominated mantle compatible with that of the Earth.

\begin{figure}
    \centering
    \includegraphics[width = 0.5\textwidth]{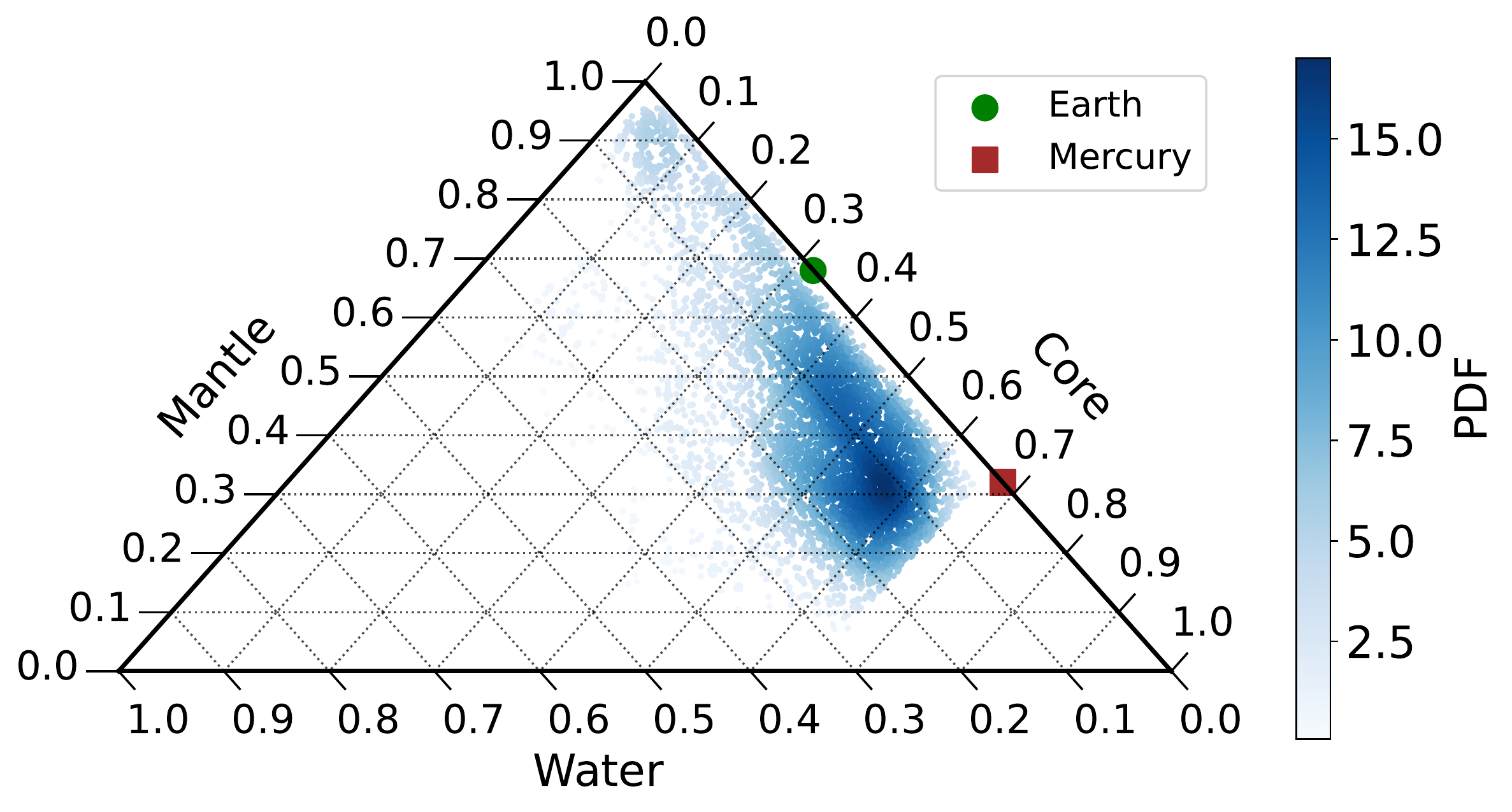}
    \caption{Ternary diagram of the sampled 2D marginal posterior distribution for the CMF and WMF of TOI-244 b in our interior structure retrieval. The color code displays the probability density function (PDF).}
    \label{fig:ternary_interior}
\end{figure}

\begin{table}
\centering
\renewcommand{\arraystretch}{1.3}
\setlength{\tabcolsep}{20pt}
\caption{\label{compo} Composition of TOI~244 b assuming a Fe-rich core, a silicate-dominated mantle, and a water layer (hydrosphere). Errors are the 1$\sigma$ confidence intervals of the interior and atmosphere output parameters.}
\begin{tabular}{lc}\hline \hline
Parameters &  \\
\hline
Core Mass fraction (CMF)  & 0.43 $\pm$ 0.16 \\ 
Water mass fraction (WMF) & 0.12 $\pm$ 0.08  \\
Temperature at 300 bar [K] &  2427$\pm$37 \\ 
Thickness at 300 bar [km] & 479$_{-96}^{+128}$  \\ 
Albedo & 0.28 $\pm$ 0.27 \\
Core + Mantle radius [$\rm R_{\oplus}$]  & 1.17 $\pm$ 0.09 \\
\hline \hline
\end{tabular}
\label{tab:mcmc_interior}
\end{table}

\subsection{Emerging trends in the growing population of low-density super-Earths}

\begin{table*}[]
\caption{\label{tab:low_density} Normalized density to the density expected for a planet composed 100$\%$ of silicates \citep[$\rho$/$\rho_{rock}$;][]{2019PNAS..116.9723Z}, stellar host metallicity ([Fe/H]), insolation flux ($S$), deviation from the expected density of an Earth-like composition ($\sigma_{Earth}$), spectral type (SpT), and reference papers for all the known planets with $R_{\rm p}<2\,\rm R_{\oplus}$, $M_{p}<3.5\,\rm M_{\oplus}$, and a mass precision better than~30$\%$.}
\setlength{\tabcolsep}{14.3pt}
\renewcommand{\arraystretch}{1.3}
\begin{tabular}{lclcccc}
\hline \hline
Planet & $\rho$/$\rho_{rock}$ & \multicolumn{1}{c}{{[Fe/H] (dex)}} & $S$ ($\rm S_{\oplus}$) & $\sigma_{Earth} \, (\sigma)$ & SpT & Ref.               \\ \hline 
L 98-59 c       & 0.90 $\pm$ 0.20               & -0.46 $\pm$ 0.26                        & $12.8^{+2.6}_{-2.1}$               & 1.4                       & M3           & {[}1{]}-{[}3{]}             \\
L 98-59 d       & 0.61 $\pm$ 0.15               & -0.46 $\pm$ 0.26                        & $5.0^{+1.0}_{-0.8}$                & 3.3                       & M3           & {[}1{]}-{[}3{]}             \\
HD 260655 c     & 0.87 $\pm$ 0.16               & -0.43 $\pm$ 0.04                        & 16.1 $\pm$ 0.3                     & 1.9                       & M0           & {[}4{]}                     \\
TOI-4481 b      & 0.89 $\pm$ 0.12               & -0.28 $\pm$ 0.07                        & 130 $\pm$ 6                        & 3.3                       & M1.5         & {[}5{]}                     \\
HD 23472 c      & 0.52 $\pm$ 0.17               & -0.20 $\pm$ 0.05                        & $8.0^{+0.9}_{-0.8}$                & 3.5                       & K4           & {[}6{]}, {[}7{]}            \\
TOI-561 b       & 0.63 $\pm$ 0.15               & -0.40 $\pm$ 0.05                        & 4745 $\pm$ 269                     & 3.4                       & G9           & {[}8{]}-{[}11{]}            \\
Kepler-138 b    & 0.43 $\pm$ 0.08               & -0.18 $\pm$ 0.10                        & 9.9 $\pm$ 0.7                      & 8.0                       & M1           & {[}12{]}-{[}20{]}           \\
Kepler-138 c    & 0.71 $\pm$ 0.18               & -0.18 $\pm$ 0.10                        & 6.8 $\pm$ 0.5                      & 2.3                       & M1           & {[}12{]}, {[}14{]}-{[}24{]} \\
Kepler-138 d    & 0.66 $\pm$ 0.18               & -0.18 $\pm$ 0.10                        & 3.4 $\pm$ 0.2                      & 2.6                       & M1           & {[}12{]}, {[}15{]}-{[}23{]} \\
GJ 1252 b       & 0.93 $\pm$ 0.23               & \multicolumn{1}{c}{+0.10 $\pm$ 0.10}    & $233^{+48}_{-41}$                  & 1.0                       & M3           & {[}25{]}, {[}26{]}          \\
Kepler-114 c    & 0.71 $\pm$ 0.27               & -0.20 $\pm$ 0.10                        & 29 $\pm$ 7                         & 1.6                       & M0           & {[}27{]}, {[}28{]}          \\
TOI-244 b (1)      & 0.79 $\pm$ 0.19               & \multicolumn{1}{c}{-0.39 $\pm$ 0.07}                  & 7.3 $\pm$ 0.4                      & 1.8                       & M2.5         & This work   \\
TOI-244 b (2)      & 0.79 $\pm$ 0.19               & \multicolumn{1}{c}{-0.03 $\pm$ 0.11}                  & 7.3 $\pm$ 0.4                      & 1.8                       & M2.5         & This work   \\
\hline \hline  
%\noalign{\vskip 1mm} 
\end{tabular}
\tablefoot{Metallicity from (1) \texttt{SteParSyn} (2) \texttt{ODUSSEAS}. \textbf{References.} [1] \citet{2019AJ....158...32K}; [2] \citet{2019A&A...629A.111C}; [3] \citet{2021A&A...653A..41D}; [4] \citet{2022A&A...664A.199L}; [5] \citet{2023arXiv230106873P}; [6] \citet{2019A&A...622L...7T}; [7] \citet{2022A&A...665A.154B}; [8] \citet{2021MNRAS.501.4148L}; [9] \citet{2021AJ....161...56W}; [10] \citet{2022MNRAS.511.4551L}; [11] \citet{2023AJ....165...88B}; [12] \citet{2013ApJ...779..188M}; [13] \citet{2014ApJ...784...45R}; [14] \citet{2014ApJ...787...80H}; [15] \citet{2015Natur.522..321J}; [16] \citet{2016ApJ...822...86M}; [17] \citet{2017AJ....153..267M}; [18] \citet{2018MNRAS.478..460A}; [19] \citet{2018ApJ...866...99B}; [20] \citet{2023NatAs...7..206P}; [21] \citet{2012ApJ...750L..37M}; [22] \citet{2014ApJ...784...28K}; [23] \citet{2016ApJS..225....9H}; [24] \citet{2019RAA....19...41G}; [25] \citet{2020ApJ...890L...7S}; [26] \citet{2022ApJ...937L..17C}; [27] \citet{2014ApJS..210...25X}; [28] \citet{2015ApJ...807...45D}.} 
\end{table*}

\begin{figure*}
    \centering
    \includegraphics[scale=0.52]{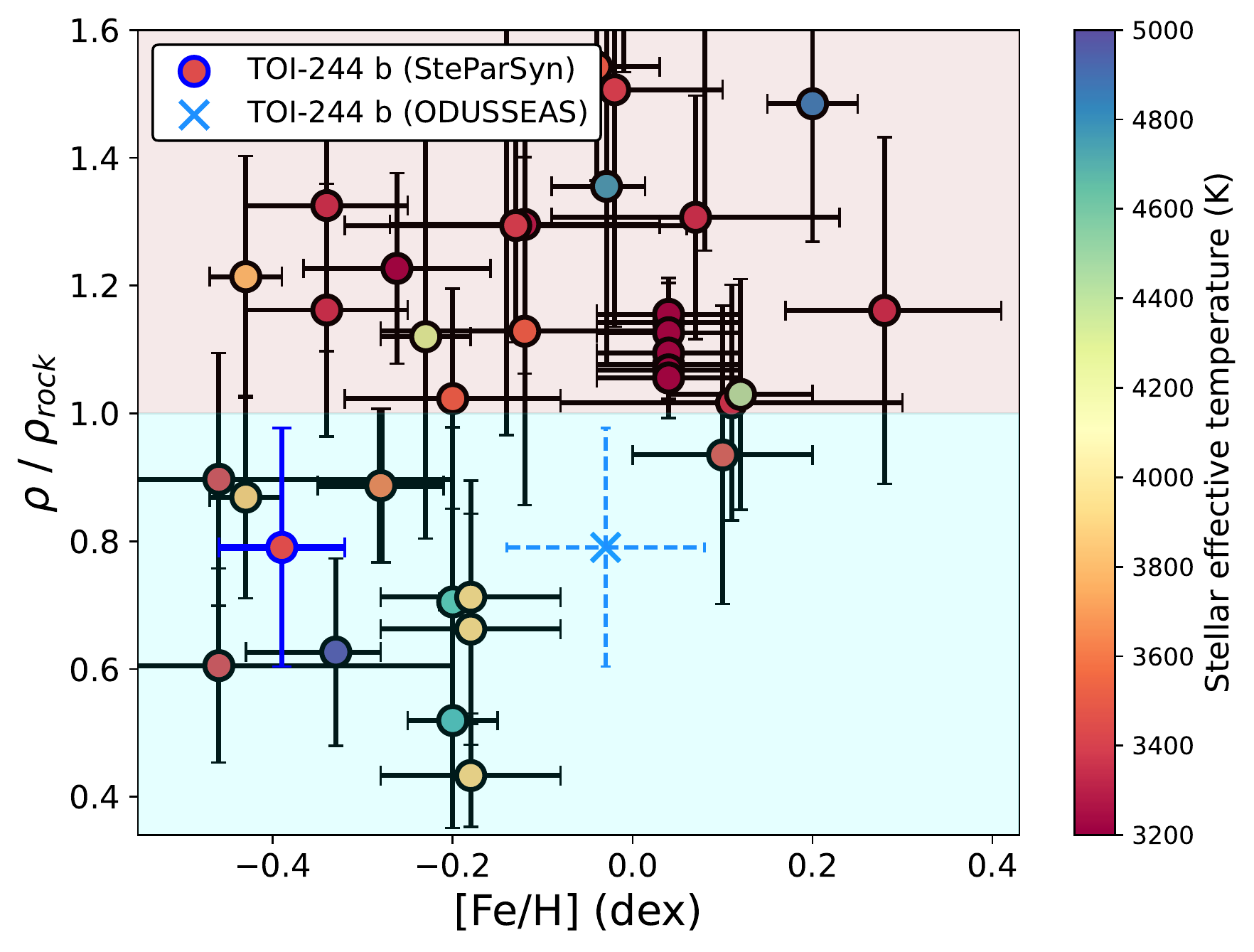}\includegraphics[scale=0.52]{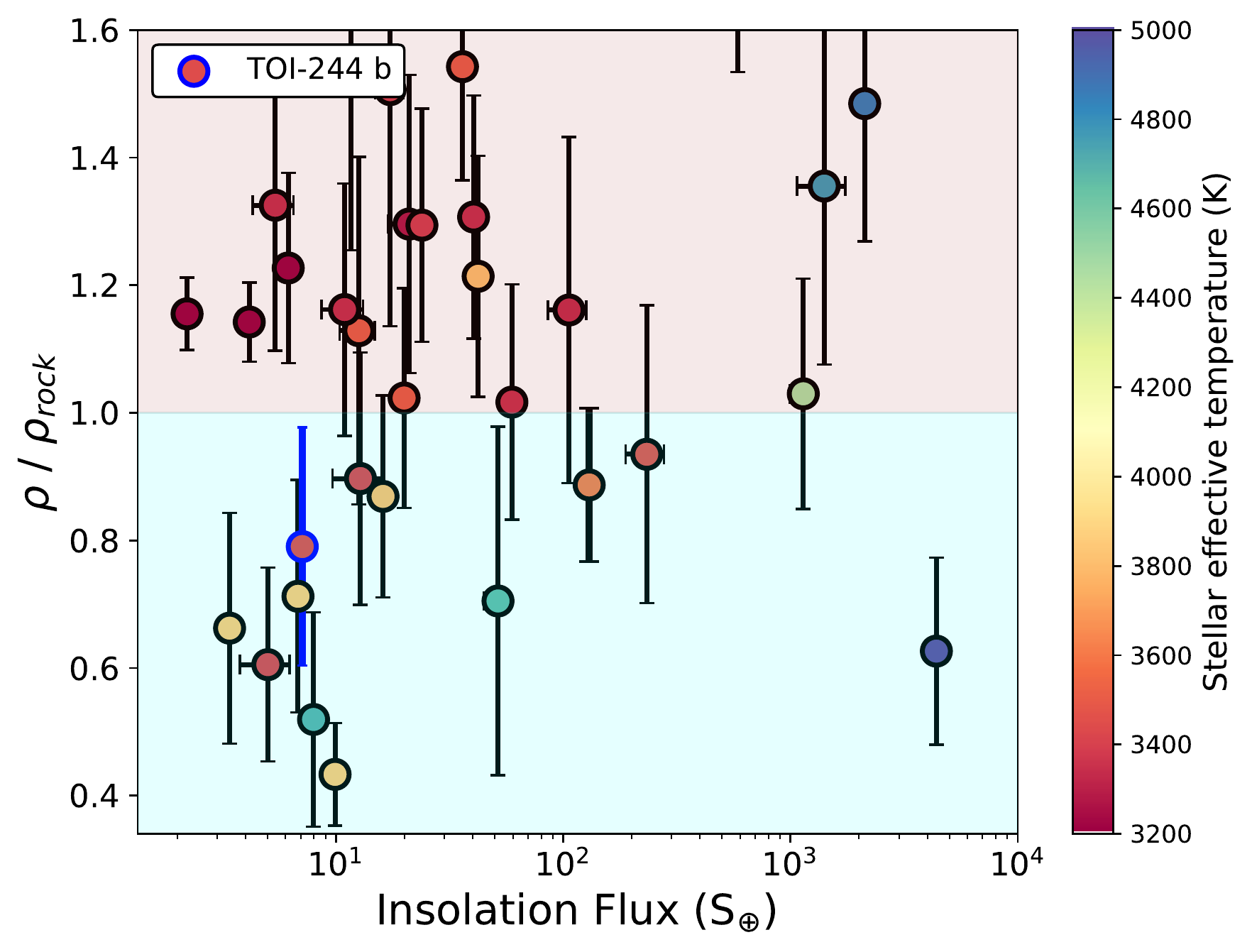}
    \caption{Normalized density to the density expected for a planet composed 100$\%$ of silicates \citep{2019PNAS..116.9723Z} versus the stellar metallicity of the stellar hosts (\textit{left}) and the stellar insolation flux (\textit{right}) for all the confirmed planets with $R_{\rm p}<2\,\rm R_{\oplus}$, $M_{p}<3.5\,\rm M_{\oplus}$, and a mass precision better than~30$\%$. The color coding indicates the effective temperature of the host stars.}
    \label{fig:trends}
\end{figure*}

We now discuss the possible existence of trends in the growing population of low-density super-Earths. In particular, we investigated the possibility of a density-metallicity and a density-insolation relation within the super-Earth population. Our sample consists of all known planets with $R_{\rm p}<2\,\rm R_{\oplus}$, $M_{p}<3.5\,\rm M_{\oplus}$, and a mass precision better than 30$\%$, making a total of 40 well-characterized super-Earths. The threshold in mass aims at discarding puffy sub-Neptunes above the 0.1$\%$~$\rm H_{2}$ theoretical model from \citet{2019PNAS..116.9723Z}. In Table \ref{tab:low_density}, we show the metallicities of the stellar hosts and the received insolation fluxes of the low-density planets in our sample; that is, those with a measured density lower than expected for a planet composed entirely of silicates (i.e., $\rho$/$\rho_{rock}$ $<$ 1).

\subsubsection{Density-metallicity}
\label{sec:density_metallicity}

All the stars hosting published low-density super-Earths are metal-poor ([Fe/H] between -0.20 and -0.45 dex) except GJ~1252, which has a slightly supersolar metallicity ([Fe/H] = +0.10 $\pm$ 0.10 dex). In Fig.~\ref{fig:trends}, we plot $\rho$/$\rho_{rock}$ versus [Fe/H] for all the planets in our sample and for TOI-244~b, considering our two estimated metallicities (see Sect.~\ref{sec:steparsyn}). We see that both Earth-like and low-density planets are found around metal-poor stars. However, there is a scarcity of low-density planets around stars with supersolar metallicities (corresponding to the lower-right parameter space in Fig.~\ref{fig:trends}). In order to quantify this possible trend, we carried out a Kolmogorov–Smirnov (KS) statistical test between the supersolar and subsolar populations, which allowed us to determine how likely is it that we would see both populations if they were drawn from the same probability distribution. We obtain a D statistic of $0.43\pm0.05$, which corresponds to a p-value of $0.0013^{+0.0055}_{-0.0011}$. This result allows us to confidently reject the null-hypothesis (i.e., both samples are drawn from the same distribution). Although statistically significant, we only propose this trend as a possible or tentative trend. This is because the sample of well-characterized rocky planets is still small, and most of those planets orbit M-dwarf stars, whose computed metallicities are less reliable than those of FGK stars.

According to several previous works, this possible trend is expected to exist. Metal-poor stars are known to have an enhancement of Mg and Si \citep[e.g.,][]{2003A&A...410..527B}, and, as discussed in Sect.~\ref{sec:iron_free}, \citet{2021Sci...374..330A} found that stars with higher Mg/Fe and Si/Fe ratios host lighter planets with higher Mg/Fe and Si/Fe ratios. However, the correlation from \citet{2021Sci...374..330A}  would not explain the possible metallicity-density trend for planets with $\rho$/$\rho_{rock}$ < 1, since the densities of those planets must be affected not only by the core and mantle composition, but also by the presence of volatile elements. Interestingly, the water mass fractions of protoplanetary disks forming metal-poor stars are known to be higher than the ones expected for stars with solar metallicities \citep{2017A&A...608A..94S}, so metal-poor stars are expected to host iron-poor and water-rich planets. Overall, both the observed densities and the previous knowledge of stellar composition point toward a possible compositional connection between the volatile content of rocky planets and their host stars. In order to try to understand why metal-poor stars host both low-density and Earth-like super-Earths, in the following we study a possible influence of the insolation flux in the population of low-density super-Earths, which could hint at the capability of those planets of having retained their original abundant water reservoirs.

\subsubsection{Density-insolation}
\label{sec:density_insolation}
In Fig.~\ref{fig:trends} (right-hand panel) we plot $\rho$/$\rho_{rock}$ versus the stellar insolation flux ($S$) received by the planets in our sample. Most of the planets receive insolations lower than 300 $\rm S_{\oplus}$. In this region, we see that the density distribution of planets with $\rho$/$\rho_{rock}$~>~1 is homogeneous across different stellar insolations. This result is to be expected, given that the density of those planets is only determined by the iron and silicates content of their interiors. In contrast, for planets with  $\rho$/$\rho_{rock}$ < 1, their low densities must be explained through the presence of a significant amount of volatile elements. Interestingly, the lowest density planets (i.e., 0.4 < $\rho$/$\rho_{rock}$ < 0.8) tend to be clustered in the low-insolation region of the diagram (i.e., $S$ < 10 $\rm S_{\oplus}$), while the less dense planets near the rocky planet domain tend to receive higher insolations. Hence, this trend might indicate an influence of the stellar irradiation on the volatile content of low-density super-Earths. 

In the highly irradiated region (i.e., 300-10\,000 $\rm S_{\oplus}$), we highlight the presence of the low-density planet TOI-561~b. With $\rho$/$\rho_{rock}$ = 0.63 $\pm$ 0.15, it might require volatile elements in its structure. This planet is a clear outlier in the possible density-insolation relation discussed above. With an insolation flux of $S$~=~4745~$\pm$~269~$\rm S_{\oplus}$, it is unlikely that it has managed to maintain a volatile envelope. Hence, other hypothesis are needed to explain the volatile storage mechanism, such as the magma oceans proposed by \citet{2021ApJ...922L...4D}.  

\subsubsection{Synthesis}

In Sect. \ref{sec:density_metallicity} we have seen that low-density super-Earths tend to be hosted by metal-poor stars, which we argue is to be expected since those stars are already known to host planets with iron-poor cores \citep{2021Sci...374..330A} and are also known to have been formed in water-rich protoplanetary disks \citep{2017A&A...608A..94S}. However, there also exist planets with Earth-like densities (i.e., with a negligible water content in mass) around metal-poor stars, which were also presumably formed in water-rich environments. Interestingly, in Sect. \ref{sec:density_insolation} we found that the lowest-density planets (i.e., 0.4 < $\rho$/$\rho_{rock}$ < 0.8) tend to receive low insolation fluxes (i.e., $S$ < 10 $\rm S_{\oplus}$). These two emerging trends combined indicate that low-irradiated super-Earths around metal-poor stars tend to have the lowest densities, and that such low densities are not typically reached when only one of those two factors is met (there is only the exception of TOI-561, which consists on a metal-poor star with a highly irradiated planet). On a compositional level, the aforementioned trend suggests that while many super-Earths were formed on iron-poor and water-rich protoplantary disks, only the less irradiated ones could have been able to preserve a significant amount of water from their original reservoirs.

\section{Summary and conclusions}
\label{conclusions}

We carried out an intensive radial velocity campaign with ESPRESSO in order to confirm and precisely characterize the close-in ($P$ = 7.4 days) transiting super-Earth-sized planet candidate TOI-244.01, which had been detected by TESS orbiting the bright ($K$ = 7.97 mag) and nearby ($d$ = 22 pc) star GJ 1018. The powerful combination of ESPRESSO and TESS data allowed us to determine a very precise planetary radius and mass (8$\%$ and 11$\%$ relative uncertainties, respectively), thus confirming the planetary nature of TOI-244~b. With a radius of $R_{\rm p}$~=~1.52~$\pm$~0.12~$\rm R_{\oplus}$, TOI-244~b is located amidst the lower mode of the bimodal radius distribution of small planets, where planets are considered to be composed of iron and silicates in a proportion similar to that of the Earth (33$\%$ Fe, and 67$\%$ $\rm MgSiO_{3}$ in mass). However, with a measured dynamical mass of $M_{\rm p}$~=~2.66~$\pm$~0.31~$\rm M_{\oplus}$, TOI-244~b is less dense than the majority of super-Earths of its size ($\rho$ = 4.2 $\pm$ 1.1 $\rm g \cdot cm^{-3}$). 

We investigated the possibility that a scarcity of iron in the core of TOI-244~b could contribute to the planet's low density. To do so, we estimated the iron-to-silicate mass fraction that would have been present in the original protoplanetary disk ($f_{\rm iron}^{\rm star}$) based on our computed stellar metallicity and estimated [Si/H] and [Mg/H] abundances. Unfortunately, depending on the technique used, we found a strong discrepancy between different metallicity computations for this M-dwarf (-0.39 $\pm$ 0.07 dex and -0.03 $\pm$ 0.11 dex). We find $f_{\rm iron}^{\rm star}$ = 29.2 $\pm$ 4.2 $\%$ and $f_{\rm iron}^{\rm star}$ = 32.0 $\pm$ 6.5 $\%$ for the metal-poor and solar metallicity scenarios, respectively, suggesting that TOI-244~b could have a lower amount of iron in its core than the Earth. However, according to the location of TOI-244~b in the mass-radius diagram, it is more likely that its low density is related to the presence of a significant amount of volatile elements. In that sense, we find that atmospheric loss processes may have been very efficient to remove a potential primordial hydrogen envelope, but high
mean molecular weight volatiles such as water could have been retained. Hence, we chose our internal structure model to be composed of an iron-rich core, a silicate-dominated mantle, and an hydrosphere. We ran our model without constraining the Fe/Si mode ratio, not only because of the difficulty of obtaining an accurate metallicity for GJ 1018, but also because the assumption that the composition of rocky planets reflects the stellar one is being questioned. As a result, our internal structure modeling suggests that TOI-244~b has a $479^{+128}_{-96}$ km thick hydrosphere over a 1.17~$\pm$~0.09~$\rm R_{\oplus}$ solid structure composed of a Fe-rich core and a silicate-dominated mantle compatible with that of the Earth.

On a population level, we find that the current population of well-characterized low-density super-Earths tend to be found around stars with subsolar metallicities, and that the lowest dense planets tend to receive lower insolation fluxes than the less dense ones in the $\rho$/$\rho_{rock}$ $<$ 1 regime. However, these possible trends need to be confirmed in the future with a larger sample of well-characterized systems.

Overall, the detection and precise characterization of the super-Earth TOI-244~b thanks to ESPRESSO and TESS data allowed us to make a first detailed description of its composition. Given its unusual properties, the brightness of the host star, and the likely presence of an extended atmosphere of high mean molecular weight volatiles, TOI-244~b will probably become a key target for future atmospheric studies. 

\begin{acknowledgements}
We thank the anonymous referee for the thorough report and quick response. A.C.-G. is funded by the Spanish Ministry of Science through MCIN/AEI/10.13039/501100011033 grant PID2019-107061GB-C61. J.L.-B. is supported by the Spanish Ministry of Science through the Ram\'on y Cajal program with code RYC2021-031640-I (funded through MCIN/AEI/10.13039/501100011033 and the NextGenerationEU/PRTR from the European Union) and the "la Caixa" Foundation (ID 100010434) and from the European Unions Horizon 2020 research and innovation programme under the Marie Sklodowska-Curie grant agreement No 847648, with fellowship code LCF/BQ/PI20/11760023.
YA and JD acknowledge the support of the Swiss National Foundation though grant 192038. This work has been carried out within the framework of the NCCR PlanetS supported by the Swiss National Science Foundation under grants 51NF40$\_$182901 and 51NF40$\_$205606. JIGH, CAP, and ASM acknowledge financial support from the Spanish Ministry of Science and Innovation (MICINN) project PID2020-117493GB-I00. CAP is also thankful for funding from the Spanish government through grants AYA2014-56359-P and AYA2017-86389-P. ASM and JIGH also acknowledge financial support from the Government of the Canary Islands project ProID2020010129. This material is based upon work supported by NASA's Interdisciplinary Consortia for Astrobiology Research (NNH19ZDA001N-ICAR) under award number 19-ICAR19$\_$2-0041. This work was supported by FCT - Funda\c{c}\~ao para a Ci\^encia e a Tecnologia through national funds and by FEDER through COMPETE2020 - Programa Operacional Competitividade e Internacionaliza\c{c}\~ao by these grants: UIDB/04434/2020; UIDP/04434/2020; 2022.06962.PTDC. NCS acknowledges support by the European Union (ERC, FIERCE, 101052347). Views and opinions expressed are however those of the author(s) only and do not necessarily reflect those of the European Union or the European Research Council. Neither the European Union nor the granting authority can be held responsible for them. This work was financed by Portuguese funds through FCT - Funda\c c\~ao para a Ci\^encia e a Tecnologia in the framework of the project 2022.04048.PTDC. CJAPM also acknowledges FCT and POCH/FSE (EC) support through Investigador FCT Contract 2021.01214.CEECIND/CP1658/CT0001. Funding for the DPAC has been provided by national institutions, in particular the institutions participating in the {\it Gaia} Multilateral Agreement. The INAF authors acknowledge financial support of the Italian Ministry of Education, University, and Research with PRIN 201278X4FL and the "Progetti Premiali" funding scheme.  This publication makes use of VOSA, developed under the Spanish Virtual Observatory (\url{https://svo.cab.inta-csic.es}) project funded by MCIN/AEI/10.13039/501100011033/ through grant PID2020-112949GB-I00. This work made use of \texttt{tpfplotter} by J. Lillo-Box (publicly available in \url{www.github.com/jlillo/tpfplotter}), which also made use of the python packages \texttt{astropy}, \texttt{lightkurve}, \texttt{matplotlib} and \texttt{numpy}. This work has made use of data from the European Space Agency (ESA) mission {\it Gaia} (\url{https://www.cosmos.esa.int/gaia}), processed by the {\it Gaia} Data Processing and Analysis Consortium (DPAC, \url{https://www.cosmos.esa.int/web/gaia/dpac/consortium}). We acknowledge the use of public TESS data from pipelines at the TESS Science Office and at the TESS Science Processing Operations Center. Resources supporting this work were provided by the NASA High-End Computing (HEC) Program through the NASA Advanced Supercomputing (NAS) Division at Ames Research Center for the production of the SPOC data products. This research has made use of the Exoplanet Follow-up Observation Program (ExoFOP; DOI: 10.26134/ExoFOP5) website, which is operated by the California Institute of Technology, under contract with the National Aeronautics and Space Administration under the Exoplanet Exploration Program. This research has made use of the NASA Exoplanet Archive,which is operated by the California Institute of Technology, under contract with the National Aeronautics and Space Administration under the Exoplanet Exploration Program. This research has made use of the SIMBAD database \citep{2000A&AS..143....9W}, operated at CDS, Strasbourg, France. This work has made use of the following software: \texttt{astropy} \citep{2022ApJ...935..167A}, \texttt{matplotlib} \citep{2007CSE.....9...90H}, \texttt{numpy} \citep{2020Natur.585..357H}, \texttt{scipy} \citep{2020NatMe..17..261V}, and \texttt{lightkurve} \citep{2018ascl.soft12013L}.

\end{acknowledgements}


\bibliographystyle{aa} % style aa.bst
\bibliography{references} % your references
\begin{appendix}

\section{Additional tables}

%%%%%%%%%%%%%%%%%%
% appendix tables
%%%%%%%%%%%%%%%%%%
%\renewcommand{\arraystretch}{1}
%\setlength{\tabcolsep}{8pt}

\begin{table*}
	\caption{TESS photometry of GJ 1018. The detrended light curve was obtained by dividing the PDCSAP by the trend obtained through the biweight method (see Section \ref{sec:obs_tess}). The complete table is accessible through CDS (see link in the main title caption).}
	\begin{center}
		\begin{tabular}{cccc}
  \hline \hline
			{BJD (days)} & {PDCSAP ($\rm e^{-} s^{-1}$)} & Detrended PDCSAP & Sector \\
			\hline
			2458354.113 & $10687 \pm 14$ & $0.9997 \pm 0.0013$ & TESS02 \\
			2458354.114 & $10699 \pm 14$ & $1.0009 \pm 0.0013$ & TESS02 \\
			... & ... & ... & ... \\
			2459088.244 & $10517 \pm 13$ & $0.9990 \pm 0.0013$ & TESS29 \\
			2459088.246 & $10515 \pm 13$ & $0.9988 \pm 0.0013$ & TESS29 \\
			... & ... & ... & ... \\
   \hline \hline
		\end{tabular}
	\end{center}
 \label{tab:tess_phot}
\end{table*}
\begin{table*}
\caption{Complete ESPRESSO radial velocities and activity indicators of GJ 1018 acquired between 9 October 2021 and 2 October 2022 under the programs IDs 108.2254.002, 108.2254.005, and 108.2254.006. This table is accessible through CDS (see link in the main title caption).}
\renewcommand{\arraystretch}{0.8}
\setlength{\tabcolsep}{7pt}
\small
	\begin{center}
		\begin{tabular}{cccccccc}
            \hline \hline
			{RJD (days)} & RV ($\rm m \, s^{-1}$) & FWHM ($\rm m \, s^{-1}$) & S-index (x1000)& Contrast ($\%$) & $\rm H_{\alpha}$ (x1000) & NaD (x1000) & BIS \\
			\hline
			59496.698 & $15139.30 \pm 0.73$ & $2747.3 \pm 1.5$ & $665.4 \pm 3.5$ & $34.337 \pm 0.018$ & $610.36 \pm 0.20$ & $70.79 \pm 0.12$ & $252.0 \pm 1.5$ \\
			59499.536 & $15139.54 \pm 0.79$ & $2748.4 \pm 1.6$ & $683.5\pm 4.1$ & $34.228 \pm 0.020$ & $618.31 \pm 0.20$ & $72.83 \pm 0.13$ & $324.6 \pm 1.6$ \\
			59514.774 & $15140.54 \pm 0.88$ & $2758.0 \pm 1.8$ & $820.7 \pm 5.1$ & $34.181 \pm 0.022$ & $601.96 \pm 0.23$ & $78.85 \pm 0.16$ & $135.3 \pm 1.8$ \\
			59517.632 & $15142.17 \pm 0.62$ & $2759.3 \pm 1.2$ & $812.0 \pm 2.6$ & $34.223 \pm 0.015$ & $598.70 \pm 0.15$ & $78.210 \pm 0.098$ & $225.1 \pm 1.2$ \\
			59520.697 & $15141.96 \pm 0.67$ & $2755.6 \pm 1.3$ & $745.0 \pm 3.0$ & $34.225 \pm 0.017$ & $588.32 \pm 0.17$ & $74.92 \pm 0.11$ & $252.0 \pm 1.3$ \\
			59523.770 & $15143.70 \pm 0.60$ & $2752.6 \pm 1.2$ & $691.3 \pm 2.6$ & $34.223 \pm 0.015$ & $570.90 \pm 0.14$ & $72.540 \pm 0.090$ & $185.1 \pm 1.2$ \\
			59529.697 & $15138.04 \pm 0.87$ & $2748.6 \pm 1.7$ & $662.5 \pm 4.8$ & $34.275 \pm 0.022$ & $567.23 \pm 0.23$ & $70.66 \pm 0.16$ & $270.6 \pm 1.7$ \\
			59532.530 & $15143.77 \pm 0.52$ & $2747.1 \pm 1.0$ & $633.8 \pm 1.9$ & $34.301 \pm 0.013$ & $580.01 \pm 0.13$ & $71.428 \pm 0.078$ & $238.2 \pm 1.0$ \\
			59535.617 & $15142.10 \pm 0.51$ & $2745.1 \pm 1.0$ & $613.5 \pm 1.8$ & $34.340 \pm 0.013$ & $572.37 \pm 0.12$ & $71.102 \pm 0.076$ & $229.7 \pm 1.0$ \\
			59538.593 & $15143.86 \pm 0.56$ & $2746.9 \pm 1.1$ & $627.4 \pm 2.1$ & $34.323 \pm 0.014$ & $589.11 \pm 0.14$ & $70.274 \pm 0.085$ & $190.7 \pm 1.1$ \\
			59541.659 & $15138.98 \pm 0.76$ & $2743.1 \pm 1.5$ & $630.8 \pm 3.5$ & $34.306 \pm 0.019$ & $586.31 \pm 0.20$ & $70.14 \pm 0.13$ & $329.2 \pm 1.5$ \\
			59544.621 & $15136.91 \pm 0.52$ & $ 	2744.6 \pm 1.0$ & $641.2 \pm 1.9$ & $34.310 \pm 0.013$ & $601.20 \pm 0.13$ & $71.598 \pm 0.078$ & $225.4 \pm 1.0$ \\
			59547.661 & $15137.17 \pm 0.74$ & $2745.2 \pm 1.5$ & $621.0 \pm 3.5$ & $34.299 \pm 0.018$ & $595.90 \pm 0.19$ & $70.72 \pm 0.12$ & $212.6 \pm 1.5$ \\
			59550.695 & $15138.33 \pm 0.53$ & $ 	2745.8 \pm 1.1$ & $674.3 \pm 2.0$ & $34.292 \pm 0.013$ & $592.22 \pm 0.12$ & $69.985 \pm 0.078$ & $214.5 \pm 1.1$ \\
			59553.568 & $15144.63 \pm 0.55$ & $2747.1 \pm 1.1$ & $713.2 \pm 2.1$ & $34.275 \pm 0.014$ & $605.78 \pm 0.13$ & $72.017 \pm 0.081$ & $273.4 \pm 1.1$ \\
			59556.631 & $15142.67 \pm 0.72$ & $2749.6 \pm 1.4$ & $868.8 \pm 3.5$ & $34.260 \pm 0.018$ & $661.87 \pm 0.19$ & $82.32 \pm 0.12$ & $250.5 \pm 1.4$ \\
			59559.584 & $15142.83 \pm 0.98$ & $2750.4 \pm 2.0$ & $758.5 \pm 5.6$ & $34.271 \pm 0.024$ & $604.98 \pm 0.27$ & $75.09 \pm 0.18$ & $332.5 \pm 2.0$ \\
			59562.558 & $15142.17 \pm 0.83$ & $2755.2 \pm 1.7$ & $787.5 \pm 4.5$ & $34.245 \pm 0.021$ & $611.19 \pm 0.22$ & $73.98 \pm 0.14$ & $253.2 \pm 1.7$ \\
			59580.587 & $15138.64 \pm 0.61$ & $2749.3 \pm 1.2$ & $791.4 \pm 2.6$ & $34.228 \pm 0.015$ & $593.87 \pm 0.15$ & $77.662 \pm 0.095$ & $230.3 \pm 1.2$ \\
			59583.554 & $15141.88 \pm 0.57$ & $2750.0 \pm 1.1$ & $752.6 \pm 2.3$ & $34.257 \pm 0.014$ & $584.92 \pm 0.14$ & $74.288 \pm 0.087$ & $230.0 \pm 1.1$ \\
			59744.857 & $15138.99 \pm 0.82$ & $2749.5 \pm 1.6$ & $647.6 \pm 4.3$ & $34.191 \pm 0.020$ & $588.62 \pm 0.21$ & $71.17 \pm 0.15$ & $198.2 \pm 1.6$ \\
			59760.864 & $15144.56 \pm 0.52$ & $2750.6 \pm 1.0$ & $680.0\pm 1.8$ & $34.172 \pm 0.013$ & $582.72 \pm 0.12$ & $73.414 \pm 0.079$ & $205.0 \pm 1.0$ \\
			59786.712 & $15144.9 \pm 1.1$ & $2750.0 \pm 2.1$ & $771.8 \pm 6.3$ & $34.195 \pm 0.027$ & $589.10 \pm 0.28$ & $73.44 \pm 0.21$ & $286.0 \pm 2.1$ \\
			59788.710 & $15144.87 \pm 0.69$ & $2752.1 \pm 1.4$ & $737.6 \pm 3.0$ & $34.199 \pm 0.017$ & $581.69 \pm 0.17$ & $73.32 \pm 0.11$ & $216.2 \pm 1.4$ \\
			59790.806 & $15144.61 \pm 0.63$ & $2751.2 \pm 1.3$ & $789.5 \pm 2.4$ & $34.213 \pm 0.016$ & $599.45 \pm 0.15$ & $76.86 \pm 0.10$ & $221.3 \pm 1.3$ \\
			59792.691 & $15140.70 \pm 0.92$ & $2751.7 \pm 1.8$ & $730.3 \pm 4.9$ & $34.236 \pm 0.023$ & $584.17 \pm 0.23$ & $73.20 \pm 0.17$ & $284.6 \pm 1.8$ \\
			59795.744 & $15138.93 \pm 0.82$ & $2754.0 \pm 1.6$ & $774.2 \pm 3.8$ & $34.229 \pm 0.020$ & $607.84 \pm 0.22$ & $77.12 \pm 0.15$ & $303.3 \pm 1.6$ \\
			59797.895 & $15139.5 \pm 1.1$ & $2755.8 \pm 2.3$ & $720.3 \pm 6.2$ & $34.190 \pm 0.028$ & $608.82 \pm 0.33$ & $76.21 \pm 0.24$ & $323.9 \pm 2.3$ \\
			59798.682 & $15141.05 \pm 0.75$ & $2751.9 \pm 1.5$ & $738.9 \pm 3.5$ & $34.240 \pm 0.019$ & $608.29 \pm 0.19$ & $76.36 \pm 0.13$ & $282.6 \pm 1.5$ \\
			59800.814 & $15140.88 \pm 0.53$ & $2754.3 \pm 1.1$ & $737.1 \pm 1.7$ & $34.203 \pm 0.013$ & $615.84 \pm 0.13$ & $75.087 \pm 0.082$ & $252.2 \pm 1.1$ \\
			59802.806 & $15142.63 \pm 0.54$ & $2751.0 \pm 1.1$ & $866.4 \pm 1.8$ & $34.181 \pm 0.013$ & $679.96 \pm 0.14$ & $83.944 \pm 0.085$ & $258.2 \pm 1.1$ \\
			59804.676 & $15142.86 \pm 0.71$ & $ 	2752.6 \pm 1.4$ & $669.5 \pm 2.9$ & $34.178 \pm 0.018$ & $603.54 \pm 0.18$ & $75.88 \pm 0.12$ & $217.0 \pm 1.4$ \\
			59805.877 & $15141.89 \pm 0.77$ & $2751.7 \pm 1.5$ & $752.9 \pm 3.3$ & $34.214 \pm 0.019$ & $611.17 \pm 0.21$ & $76.52 \pm 0.14$ & $273.8 \pm 1.5$ \\
			59807.711 & $15137.99 \pm 0.82$ & $2750.6 \pm 1.6$ & $687.6 \pm 3.8$ & $34.184 \pm 0.020$ & $611.05 \pm 0.22$ & $74.17 \pm 0.15$ & $318.6 \pm 1.6$ \\
			59811.828 & $15139.34 \pm 0.53$ & $ 	2750.1 \pm 1.1$ & $740.5 \pm 1.8$ & $34.225 \pm 0.013$ & $610.03 \pm 0.13$ & $75.519 \pm 0.083$ & $215.1 \pm 1.1$ \\
			59813.666 & $15139.80 \pm 0.96$ & $2751.8 \pm 1.9$ & $702.9 \pm 5.3$ & $34.222 \pm 0.024$ & $590.24 \pm 0.26$ & $82.35 \pm 0.18$ & $202.3 \pm 1.9$ \\
			59815.623 & $15137.99 \pm 0.89$ & $2752.5 \pm 1.8$ & $620.0 \pm 4.8$ & $34.206 \pm 0.022$ & $567.41 \pm 0.22$ & $87.95 \pm 0.16$ & $287.2 \pm 1.8$ \\
			59817.629 & $15140.49 \pm 0.61$ & $2749.0 \pm 1.2$ & $696.2 \pm 2.4$ & $34.256 \pm 0.015$ & $566.88 \pm 0.14$ & $73.399 \pm 0.096$ & $260.5 \pm 1.2$ \\
			59817.705 & $15141.42 \pm 0.60$ & $2749.0 \pm 1.2$ & $718.7 \pm 2.2$ & $34.284 \pm 0.015$ & $577.86 \pm 0.15$ & $74.411 \pm 0.099$ & $220.2 \pm 1.2$ \\
			59818.893 & $15142.96 \pm 0.66$ & $ 	2753.4 \pm 1.3$ & $669.7 \pm 2.5$ & $34.214 \pm 0.016$ & $573.71 \pm 0.17$ & $72.94 \pm 0.11$ & $346.4 \pm 1.3$ \\
			59823.850 & $15139.40 \pm 0.91$ & $2749.2 \pm 1.8$ & $669.6 \pm 4.4$ & $34.259 \pm 0.023$ & $566.65 \pm 0.25$ & $74.01 \pm 0.17$ & $239.7 \pm 1.8$ \\
			59827.827 & $15138.08 \pm 0.71$ & $2750.2 \pm 1.4$ & $688.1 \pm 2.9$ & $34.255 \pm 0.018$ & $580.15 \pm 0.19$ & $74.33 \pm 0.12$ & $183.0 \pm 1.4$ \\
			59828.825 & $15136.77 \pm 0.66$ & $2746.6 \pm 1.3$ & $645.8 \pm 2.5$ & $34.275 \pm 0.016$ & $564.14 \pm 0.17$ & $72.49 \pm 0.11$ & $189.7 \pm 1.3$ \\
			59830.775 & $15134.42 \pm 0.58$ & $2748.3 \pm 1.2$ & $827.7 \pm 2.0$ & $34.253 \pm 0.014$ & $630.62 \pm 0.15$ & $81.738 \pm 0.094$ & $253.0 \pm 1.2$ \\
			59832.849 & $15136.20 \pm 0.98$ & $ 	2745.3 \pm 2.0$ & $654.7 \pm 4.9$ & $34.141 \pm 0.024$ & $577.67 \pm 0.27$ & $72.18 \pm 0.19$ & $331.4 \pm 2.0$ \\
			59834.713 & $15142.24 \pm 0.65$ & $2748.2 \pm 1.3$ & $759.9 \pm 2.5$ & $34.251 \pm 0.016$ & $611.63 \pm 0.17$ & $77.76 \pm 0.11$ & $227.2 \pm 1.3$ \\
			59836.668 & $15144.5 \pm 1.0$ & $2753.3 \pm 2.0$ & $684.3 \pm 5.3$ & $34.220 \pm 0.025$ & $577.48 \pm 0.28$ & $74.86 \pm 0.20$ & $278.6 \pm 2.0$ \\
			59839.738 & $15144.60 \pm 0.85$ & $ 	2753.5 \pm 1.7$ & $734.0 \pm 4.0$ & $34.245 \pm 0.021$ & $577.22 \pm 0.23$ & $75.10 \pm 0.15$ & $215.0 \pm 1.7$ \\
			59841.810 & $15146.51 \pm 0.94$ & $2755.1 \pm 1.9$ & $754.1 \pm 4.7$ & $34.241 \pm 0.023$ & $575.28 \pm 0.26$ & $75.27 \pm 0.18$ & $372.5 \pm 1.9$ \\
			59842.586 & $15146.60 \pm 0.81$ & $2755.7 \pm 1.6$ & $698.5 \pm 3.8$ & $34.224 \pm 0.020$ & $582.39 \pm 0.21$ & $74.69 \pm 0.14$ & $231.0 \pm 1.6$ \\
			59844.569 & $15143.30 \pm 0.89$ & $2753.6 \pm 1.8$ & $784.1 \pm 4.7$ & $34.226 \pm 0.022$ & $581.37 \pm 0.23$ & $76.93 \pm 0.16$ & $260.2 \pm 1.8$ \\
			59846.556 & $15139.44 \pm 0.87$ & $2757.2 \pm 1.7$ & $876.1 \pm 4.5$ & $34.166 \pm 0.021$ & $624.29 \pm 0.23$ & $82.24 \pm 0.16$ & $281.8 \pm 1.7$ \\
			59848.563 & $15143.03 \pm 0.90$ & $ 	2758.3 \pm 1.8$ & $794.7 \pm 4.7$ & $34.178 \pm 0.022$ & $581.16 \pm 0.24$ & $76.12 \pm 0.16$ & $295.8 \pm 1.8$ \\
			59850.543 & $15141.32 \pm 0.96$ & $ 	2759.0 \pm 1.9$ & $776.4 \pm 5.4$ & $34.157 \pm 0.024$ & $572.63 \pm 0.25$ & $75.76 \pm 0.17$ & $165.4 \pm 1.9$ \\
			59852.544 & $15139.75 \pm 0.72$ & $ 	2754.6 \pm 1.4$ & $777.6 \pm 3.2$ & $34.215 \pm 0.018$ & $596.29 \pm 0.18$ & $78.35 \pm 0.12$ & $237.3 \pm 1.4$ \\
			59852.749 & $15139.74 \pm 0.66$ & $2758.4 \pm 1.3$ & $782.7 \pm 2.7$ & $34.199 \pm 0.016$ & $588.54 \pm 0.17$ & $77.55 \pm 0.11$ & $294.7 \pm 1.3$ \\
			59854.545 & $15140.52 \pm 0.85$ & $2754.3 \pm 1.7$ & $770.0 \pm 4.2$ & $34.199 \pm 0.021$ & $594.01 \pm 0.23$ & $76.82 \pm 0.15$ & $231.0 \pm 1.7$ \\
   \hline \hline
		\end{tabular}
	\end{center}
 \label{tab:espresso_rvs}
\end{table*}
\begin{table}
\caption{Complete HARPS RVs and FWHMs of GJ 1018 acquired between 15 December 2018 and 8 January 2019 under the program ID 1102.C-0339. This table is accessible through CDS (see link in the main title caption).}
\renewcommand{\arraystretch}{1.04}
\setlength{\tabcolsep}{12pt}
	\begin{center}
		\begin{tabular}{ccc}
  \hline \hline
			{RJD (days)} & {RV ($\rm m \, s^{-1}$)} & {FWHM ($\rm m \, s^{-1}$)} \\
			\hline
			58467.550 & $15127.3 \pm 1.8$ & $2960.4 \pm 1.8$ \\
			58467.635 & $15131.3 \pm 2.4$ & $2956.1 \pm 2.4$ \\
			58468.566 & $15126.5 \pm 1.9$ & $2959.4 \pm 1.9$ \\
			58469.627 & $15126.6 \pm 2.1$ & $2961.8 \pm 2.1$ \\
			58471.576 & $15131.1 \pm 2.0$ & $2968.6 \pm 2.0$ \\
			58474.620 & $15127.7 \pm 2.3$ & $2955.3 \pm 2.3$ \\
			58475.641 & $15134.9 \pm 3.0$ & $2954.2 \pm 3.0$ \\
			58477.600 & $15129.2 \pm 2.1$ & $2951.3 \pm 2.1$ \\
			58478.603 & $15129.2 \pm 2.1$ & $2953.5 \pm 2.1$ \\
			58479.601 & $15126.5 \pm 1.7$ & $2958.9 \pm 1.7$ \\
			58484.586 & $15130.8 \pm 1.8$ & $2969.1 \pm 1.8$ \\
			58487.555 & $15133.5 \pm 2.0$ & $2962.2 \pm 2.0$ \\
			58488.590 & $15136.7 \pm 2.2$ & $2964.9 \pm 2.2$ \\
			58489.573 & $15128.6 \pm 2.6$ & $2964.9 \pm 2.6$ \\
			58491.613 & $15117.8 \pm 2.8$ & $2956.2 \pm 2.8$ \\
   \hline \hline
		\end{tabular}
	\end{center}
 \label{tab:harps_rvs}
\end{table}
\begin{table}
\caption{ASAS-SN photometry \citep{2014ApJ...788...48S} of GJ 1018 computed by shifting the photometric aperture every 1 October in order to ensure the target centering. The complete table is accessible through CDS (see link in the main title caption).}
\renewcommand{\arraystretch}{1.0418}
\setlength{\tabcolsep}{9pt}
	\begin{center}
		\begin{tabular}{cccc}
   \hline \hline
			{HJD (days)} & Flux (mJy) & Camera & Filter \\
			\hline
			2456788.907 & $32.464 \pm 0.080$ & bf & V \\
			2456789.906 & $32.402 \pm 0.075$ & bf & V \\
			2456794.907 & $32.41 \pm 0.13$ & bf & V \\
			2456805.887 & $31.331 \pm 0.066$ & bf & V \\
			2456808.883 & $31.868 \pm 0.049$ & bf & V \\
			2456809.884 & $30.971 \pm 0.072$ & bf & V \\
			2456810.872 & $32.599 \pm 0.085$ & bf & V \\
			... & ... & ... & ... \\
			2458014.683 & $16.229 \pm 0.029$ & bj & g \\
			2458015.680 & $16.303 \pm 0.028$ & bj & g \\
			2458016.678 & $16.282 \pm 0.049$ & bj & g \\
			2458018.673 & $16.214 \pm 0.028$ & bj & g \\
			2458019.670 & $16.314 \pm 0.028$ & bj & g \\
			2458020.663 & $16.236 \pm 0.031$ & bj & g \\
			2458021.656 & $16.289 \pm 0.026$ & bj & g \\
			... & ... & ... & ... \\
			2458068.440 & $16.650 \pm 0.023$ & bn & g \\
			2458076.419 & $16.700 \pm 0.020$ & bn & g \\
			2458080.404 & $16.618 \pm 0.020$ & bn & g \\
			2458081.402 & $16.404 \pm 0.028$ & bn & g \\
			2458082.380 & $16.461 \pm 0.031$ & bn & g \\
			2458084.287 & $16.641 \pm 0.035$ & bn & g \\
			2458085.268 & $16.782 \pm 0.064$ & bn & g \\
			... & ... & ... & ... \\
			2458398.671 & $16.381 \pm 0.027$ & bF & g \\
			2458404.736 & $16.470 \pm 0.033$ & bF & g \\
			2458406.512 & $16.181 \pm 0.043$ & bF & g \\
			2458407.551 & $16.134 \pm 0.034$ & bF & g \\
			2458408.512 & $16.305 \pm 0.038$ & bF & g \\
			2458413.678 & $15.659 \pm 0.065$ & bF & g \\
			2458415.775 & $16.02 \pm 0.12$ & bF & g \\
			... & ... & ... & ... \\
   \hline \hline
		\end{tabular}
	\end{center}
 \label{tab:asassn_phot}
\end{table}
\begin{table}
\caption{Absolute fluxes in different filters that make up the SED of GJ 1018. All the catalogs were accessed through the Spanish Virtual Observatory \citep[SVO;][]{bayo2008}. The complete table is accessible through CDS (see link in the main title caption).}
	\begin{center}
		\begin{tabular}{cc}
  \hline \hline
			Filter ID & Abs. Flux ($\rm erg \, s^{-1} cm^{-2} \AA^{-1}$) \\
			\hline
			GALEX/GALEX.NUV & $\left(7.6 \pm 3.1\right) \times 10^{-17}$ \\
			Misc/APASS.B & $\left(6.48 \pm 0.15\right) \times 10^{-14}$ \\
			Generic/Johnson.B & $\left(6.503 \pm 0.057\right) \times 10^{-14}$ \\
			SLOAN/SDSS.g & $\left(9.85 \pm 0.18\right) \times 10^{-14}$ \\
			GAIA/GAIA3.Gbp & $\left(1.3338 \pm 0.0038\right) \times 10^{-13}$ \\
			Misc/APASS.V & $\left(1.527 \pm 0.039\right) \times 10^{-13}$ \\
			Generic/Johnson.V & $\left(1.5008 \pm 0.0054\right) \times 10^{-13}$ \\
			HST/ACS$\_$WFC.F606W & $\left(1.9170 \pm 0.0036\right) \times 10^{-13}$ \\
			GAIA/GAIA3.G & $\left(2.9298 \pm 0.0076\right) \times 10^{-13}$ \\
			SLOAN/SDSS.r & $\left(2.038 \pm 0.057\right) \times 10^{-13}$ \\
			Generic/Johnson.R & $\left(2.5561 \pm 0.0038\right) \times 10^{-13}$ \\
			SLOAN/SDSS.i & $\left(4.331 \pm 0.084\right) \times 10^{-13}$ \\
			GAIA/GAIA3.Grp & $\left(4.327 \pm 0.015\right) \times 10^{-13}$ \\
			HST/ACS$\_$WFC.F814W & $\left(4.8812 \pm 0.0055\right) \times 10^{-13}$ \\
			Generic/Johnson.I & $\left(4.9304 \pm 0.0062\right) \times 10^{-13}$ \\
			GAIA/GAIA3.Grvs & $\left(5.109 \pm 0.032\right) \times 10^{-13}$ \\
			SLOAN/SDSS.z & $\left(5.5047 \pm 0.0052\right) \times 10^{-13}$ \\
			PAN-STARRS/PS1.y & $\left(5.5937 \pm 0.0075\right) \times 10^{-13}$ \\
			2MASS/2MASS.J & $\left(4.492 \pm 0.095\right) \times 10^{-13}$ \\
			2MASS/2MASS.H & $\left(2.765 \pm 0.084\right) \times 10^{-13}$ \\
			2MASS/2MASS.Ks & $\left(1.354 \pm 0.036\right) \times 10^{-13}$ \\
			WISE/WISE.W1 & $\left(3.013 \pm 0.063\right) \times 10^{-14}$ \\
			WISE/WISE.W2 & $\left(9.75 \pm 0.16\right) \times 10^{-15}$ \\
			WISE/WISE.W3 & $\left(2.880 \pm 0.050\right) \times 10^{-16}$ \\
			WISE/WISE.W4 & $\left(2.46 \pm 0.30\right) \times 10^{-17}$ \\
			OAJ/JPAS.J0400 & $\left(3.494 \pm 0.091\right) \times 10^{-14}$ \\
			OAJ/JPLUS.J0410 & $\left(4.262 \pm 0.066\right) \times 10^{-14}$ \\
			OAJ/JPAS.J0410 & $\left(4.427 \pm 0.077\right) \times 10^{-14}$ \\
			OAJ/JPAS.J0420 & $\left(3.706 \pm 0.077\right) \times 10^{-14}$ \\
			OAJ/JPLUS.J0430 & $\left(4.107 \pm 0.067\right) \times 10^{-14}$ \\
			OAJ/JPAS.J0430 & $\left(4.154 \pm 0.082\right) \times 10^{-14}$ \\
			OAJ/JPAS.J0440 & $\left(5.994 \pm 0.094\right) \times 10^{-14}$ \\
			OAJ/JPAS.J0450 & $\left(8.53 \pm 0.11\right) \times 10^{-14}$ \\
			OAJ/JPAS.J0460 & $\left(1.017 \pm 0.012\right) \times 10^{-13}$ \\
			OAJ/JPAS.J0470 & $\left(9.77 \pm 0.13\right) \times 10^{-14}$ \\
			OAJ/JPAS.gSDSS & $\left(1.0074 \pm 0.0019\right) \times 10^{-13}$ \\
			OAJ/JPLUS.gSDSS & $\left(1.0286 \pm 0.0019\right) \times 10^{-13}$ \\
			OAJ/JPAS.J0480 & $\left(9.29 \pm 0.13\right) \times 10^{-14}$ \\
			OAJ/JPAS.J0490 & $\left(9.90 \pm 0.13\right) \times 10^{-14}$ \\
			OAJ/JPAS.J0500 & $\left(1.052 \pm 0.013\right) \times 10^{-13}$ \\
			OAJ/JPAS.J0510 & $\left(1.009 \pm 0.014\right) \times 10^{-13}$ \\
			OAJ/JPLUS.J0515 & $\left(1.079 \pm 0.0098\right) \times 10^{-13}$ \\
			OAJ/JPAS.J0520 & $\left(1.228 \pm 0.015\right) \times 10^{-13}$ \\
			OAJ/JPAS.J0530 & $\left(1.547 \pm 0.015\right) \times 10^{-13}$ \\
			OAJ/JPAS.J0540 & $\left(1.616 \pm 0.016\right) \times 10^{-13}$ \\
			OAJ/JPAS.J0550 & $\left(1.601 \pm 0.016\right) \times 10^{-13}$ \\
			OAJ/JPAS.J0560 & $\left(1.609 \pm 0.016\right) \times 10^{-13}$ \\
			OAJ/JPAS.J0570 & $\left(1.629 \pm 0.017\right) \times 10^{-13}$ \\
			OAJ/JPAS.J0580 & $\left(1.657 \pm 0.016\right) \times 10^{-13}$ \\
			OAJ/JPAS.J0590 & $\left(1.602 \pm 0.015\right) \times 10^{-13}$ \\
			OAJ/JPAS.J0600 & $\left(1.547 \pm 0.014\right) \times 10^{-13}$ \\
			OAJ/JPAS.J0610 & $\left(1.582 \pm 0.015\right) \times 10^{-13}$ \\
			OAJ/JPAS.rSDSS & $\left(2.1212 \pm 0.0024\right) \times 10^{-13}$ \\
			... & ... \\
   \hline \hline
		\end{tabular}
	\end{center}
 \label{tab:svo_phot}
\end{table}

%\renewcommand{\arraystretch}{1.24}
%\setlength{\tabcolsep}{15pt}
%\caption{Inferred parameters of TOI-244 b obtained from the \textit{TESS} light curve analysis, ESPRESSO and HARPS radial velocities analysis, and the final joint analysis based on the \textit{TESS} light curve, ESPRESSO and HARPS radial velocities. }

\begin{table*}[]
\renewcommand{\arraystretch}{1.6}
\setlength{\tabcolsep}{15pt}
\caption{Inferred parameters of TOI-244 b obtained from the TESS light curve analysis, ESPRESSO and HARPS radial velocities analysis, and the final joint analysis based on the TESS, ESPRESSO and HARPS data. }
\begin{tabular}{lccc}
\hline \hline
Parameter                          & TESS photometry           & ESPRESSO and HARPS RVs & Joint fit                 \\ \hline
\multicolumn{4}{l}{Orbital parameters}                                                                                                                  \\ \hline
$P$ {[}days{]}                              & $7.397225^{+0.000026}_{-0.000024}$ & $7.39714^{+0.00074}_{-0.00075}$   & $7.397225^{+0.000026}_{-0.000023}$ \\
$T_{\rm 0}$ {[}JD{]}                       & 2458357.3627 $\pm$ 0.0020          & 2458357.3622 $\pm$ 0.017        & 2458357.3627 $\pm$ 0.0020          \\
$i$ {[}degrees{]}                           & 88.32 $\pm$ 0.16                   & --                              & 88.32 $\pm$ 0.16                   \\ \hline
\multicolumn{3}{l}{Planet parameters}                                                                              &                                    \\ \hline
$R_{\rm p}$/$R_{\rm \star}$                 & $0.0326^{+0.0017}_{-0.0018}$         & --                              & $0.0326^{+0.0017}_{-0.0018}$         \\
$K$ {[}$\rm m/s${]}                        & --                                 & 1.54 $\pm$ 0.16                 & 1.55 $\pm$ 0.16                    \\ \hline
\multicolumn{3}{l}{Stellar parameters}                                                                              &                                    \\ \hline
$M_{\rm \star}$ $[\rm M_{\odot}]$           & 0.428 $\pm$ 0.029                  & --                              & 0.428 $\pm$ 0.029                  \\
$R_{\rm \star}$ $[\rm R_{\odot}]$           & 0.426 $\pm$ 0.025                  & --                              & 0.426 $\pm$ 0.025                  \\
$u1$                                        & 0.2204 $\pm$ 0.0011                & --                              & 0.2204 $\pm$ 0.0011                \\
$u2$                                        & 0.4112 $\pm$ 0.0017                & --                              & 0.4112 $\pm$ 0.0017                \\ \hline
\multicolumn{3}{l}{Matérn-3/2 GP hyperparameters}                                                                  &                                    \\ \hline
$\eta_{\sigma_{S2}}$ {[}$\rm e^{-}/s${]}    & $0.88^{+0.42}_{-0.22}$             & --                              & $0.87^{+0.36}_{-0.21}$             \\
$\eta_{\rho_{S2}}$ {[}days{]}               & $1.13^{+0.95}_{-0.45}$             & --                              & $1.09^{+0.79}_{-0.45}$             \\
$\eta_{\sigma_{S29}}$ {[}$\rm e^{-}/s${]}   & $1.80^{+0.63}_{-0.34}$             & --                              & $1.79^{+0.63}_{-0.34}$             \\
$\eta_{\rho_{S29}}$ {[}days{]}              & $0.79^{+0.43}_{-0.25}$             & --                              & $1.09^{+0.44}_{-0.24}$             \\ \hline
\multicolumn{3}{l}{Quasiperiodic GP hyperparameters}                                                              &                                    \\ \hline
$\eta_{\rm 1,RV}$ {[}$\rm m/s${]}          & --                                 & $3.5^{+1.0}_{-0.7}$      & $3.5^{+1.0}_{-0.7}$       \\
$\eta_{\rm 2,RV}$ {[}days{]}                & --                                 & $86^{+26}_{-23}$                  & $85^{+26}_{-23}$                     \\
$\eta_{\rm 3,RV}$ {[}days{]}                & --                                 & $53.2^{+1.3}_{-1.1}$              & $53.3^{+1.2}_{-1.1}$                 \\
$\eta_{\rm 4,RV}$                           & --                                 & $0.62^{+0.11}_{-0.09}$            & $0.62^{+0.11}_{-0.09}$               \\
$\eta_{\rm 1,FWHM}$ {[}$\rm m/s${]}        & --                                 & $3.34^{+0.70}_{-0.53}$   & $3.35^{+0.70}_{-0.54}$      \\ \hline
\multicolumn{3}{l}{Instrument-dependent parameters (TESS)}                                                         &                                    \\ \hline
$F_{\rm 0,S2}$ {[}$\rm e^{-}/s${]}          & $0.04^{+0.31}_{-0.32}$             & --                              & $0.05^{+0.29}_{-0.30}$             \\
$F_{\rm 0,S29}$ {[}$\rm e^{-}/s${]}         & $0.17^{+0.56}_{-0.54}$             & --                              & $0.18^{+0.54}_{-0.55}$             \\
$\sigma_{\rm TESS,S2}$ {[}$\rm e^{-}/s${]}  & $0.21^{+0.23}_{-0.15}$             & --                              & $0.20^{+0.23}_{-0.15}$             \\
$\sigma_{\rm TESS,S29}$ {[}$\rm e^{-}/s${]} & $1.90^{+0.53}_{-0.76}$             & --                              & $1.92^{+0.51}_{-0.74}$             \\ \hline
\multicolumn{3}{l}{Instrument-dependent parameters (ESPRESSO and HARPS)}                                           &                                    \\ \hline
$\gamma_{\rm ESPR,RV}$ {[}$\rm m/s${]}     & --                                 & $15141.0^{+1.2}_{-1.3}$   & $15141.0^{+1.2}_{-1.3}$      \\
$\gamma_{\rm HAR,RV}$ {[}$\rm m/s${]}      & --                                 & $15128.5^{+2.5}_{-2.6}$   & $15128.5^{+2.4}_{-2.6}$      \\
$\gamma_{\rm ESPR,FWHM}$ {[}$\rm m/s${]}   & --                                 & 2750.9 $\pm$ 1.2             & 2750.9 $\pm$ 1.2                 \\
$\gamma_{\rm HAR,FWHM}$ {[}$\rm m/s${]}    & --                                 & 2959.9 $\pm$ 2.6              & 2959.9 $\pm$ 2.6                \\
$\sigma_{\rm ESPR,RV}$ {[}$\rm m/s${]}     & --                                 & $0.19^{+0.21}_{-0.13}$ & $0.19^{+0.20}_{-0.13}$    \\
$\sigma_{\rm HAR,RV}$ {[}$\rm m/s${]}      & --                                 & $2.9^{+1.3}_{-1.1}$    & $2.9^{+1.3}_{-1.1}$       \\
$\sigma_{\rm ESPR,FWHM}$ {[}$\rm m/s${]}   & --                                 & $0.55^{+0.39}_{-0.36}$ & $0.56^{+0.39}_{-0.37}$   \\
$\sigma_{\rm HAR,FWHM}$ {[}$\rm m/s${]}    & --                                 & $4.4^{+1.5}_{-1.2}$    & $4.4^{+1.5}_{-1.2}$    \\ \hline \hline
\end{tabular}
\label{tab:final_derived_params}
\end{table*}
%%%%%%%%%%%%%%%%%%%
%%%%%%%%%%%%%%%%%%%

\end{appendix}

\end{document}